\documentclass[fleqn,usenatbib]{mnras}
\usepackage{mathptmx}
\usepackage[T1]{fontenc}
\DeclareRobustCommand{\VAN}[3]{#2}
\let\VANthebibliography\thebibliography
\def\thebibliography{\DeclareRobustCommand{\VAN}[3]{##3}\VANthebibliography}
\usepackage{graphicx}	
\usepackage{amsmath}	
\usepackage{amssymb}	
\usepackage{rotating}
\usepackage{sidecap}
\usepackage{lscape}
\usepackage{multicol, blindtext}

\usepackage{aecompl}
\usepackage{wasysym}
\usepackage{array}
\usepackage{times}
\usepackage{url}
\usepackage{xcolor}
\usepackage{hyperref}
\hypersetup{colorlinks=true,citecolor=blue,linkcolor=blue,filecolor=blue,urlcolor=blue}
\usepackage{soul}

\newcommand{\kms}{\ \text{km}\ \text{s}^{-1}}
\newcommand{\cmg}{\ \text{cm}^{2}\ \text{g}^{-1}}
\newcommand{\zz}{$\left(0,0\right)$ }
\newcommand{\mm}{$\left(-2,-2\right)$ }

\title[Structure Formation with Inelastic 2cDM]{Structure Formation under Inelastic Two-Component Dark Matter: Halo Statistics and Matter Power Spectra in the High-$z$ Universe}

\author[R. Low et al.]
    {Ryan Low$^{1}$\thanks{E-mail: rtlow@ku.edu},
    Rakshak Adhikari$^{2,1}$,
    Jonah C. Rose$^{3,4}$,
    Stephanie O'Neil$^{5,6}$,
    \newauthor
    Mikhail~V. Medvedev$^{1,7}$,
    Paul Torrey$^{8,9,10}$,
    Mark Vogelsberger$^{11,12}$
     \vspace{0.3cm}\\
    $^{1}$Department of Physics and Astronomy, University of Kansas, Lawrence, KS 66045, USA\\
    $^{2}$Center for Relativity and Cosmology, Troy University, Troy, Al, 36081, USA\\
    $^{3}$Department of Astronomy, University of Florida, Gainesville, FL 32611, USA\\
    $^{4}$Center for Computational Astrophysics, Flatiron Institute, 162 5th Avenue, New York, NY 10010, USA\\
    $^{5}$Department of Physics and Astronomy, University of Pennsylvania, Philadelphia, PA 19104, USA\\
    $^{6}$Department of Physics, Princeton University, Princeton, NJ 08544, USA\\
    $^{7}$Laboratory for Nuclear Science, Massachusetts Institute of Technology, Cambridge, MA 02139, USA\\
    $^{8}$Department of Astronomy, University of Virginia, Charlottesville, VA 22904, USA\\
    $^{9}$Virginia Institute for Theoretical Astronomy, University of Virginia, Charlottesville, VA 22904, USA\\
    $^{10}$The NSF-Simons AI Institute for Cosmic Origins, USA\\
    $^{11}$Department of Physics and Kavli Institute for Astrophysics and Space Research,
          Massachusetts Institute of Technology,
          Cambridge, MA 02139, USA\\
    $^{12}$The NSF AI Institute for Artificial Intelligence and Fundamental Interactions, Massachusetts Institute of Technology, Cambridge, MA 02139, USA
    }

\begin{document}
\pagerange{\pageref{firstpage}--\pageref{lastpage}}

\maketitle
\label{firstpage}

\begin{abstract}
We present hydrodynamic simulations of a
flavour-mixed two-component dark matter (2cDM) model
that utilize IllustrisTNG baryonic physics.
The model parameters are explored
for two sets of power laws of the velocity-dependent cross sections, favoured on the basis of previous studies.
The model is shown to suppress the formation
of structures at scales $k\gtrsim 10^2\ h\text{ Mpc}^{-1}$
up to 40\% compared to cold dark matter (CDM) 
at redshifts $z\sim5-2$. We compare our results to structure enhancement
and suppression due to cosmological and astrophysical parameters presented
in the literature and find that 2cDM effects remain relevant
at galactic and subgalactic scales.
The results indicate the robustness of the role of nongravitational dark matter interactions in structure formation and the absence of putative degeneracies introduced by baryonic feedback at high $z$. The predictions made can be further tested with future Ly-$\alpha$ forest observations.
\end{abstract}


\section{Introduction}
With the beginning of new observational missions
alongside the maturation of multi-messenger astronomy
comes the promise of unprecedented tests of the cosmological standard
model, $\Lambda$CDM. 
These new observations will join the decades-long effort
of precision cosmology observations and direct detection experiments
to determine the physical nature of dark matter (DM),
which still remains a mystery.

Historically, it was thought that $N$-body simulations of $\Lambda$CDM produced
major tensions with observations on the dwarf-galaxy scale. Some of the most
notable small-scale problems include (see \cite{dwarfsBaryonReview} for a comprehensive review): 
the core-cusp problem 
\citep{CoreCuspFlores, CoreCuspKlypin, CoreCupspNavarro}, where the innermost density
profiles of dwarf galaxies form isothermal cores as opposed to cusps in $N$-body simulation;
the missing satellites problem \citep{MissingSatellites, DMsubstructure}, where the observed
number of satellites around Milky Way-like systems is far fewer than what $N$-body
simulations predicted; the Too-Big-to-Fail problem \citep{TBTF, TBTFII, TBTFIII}, where $N$-body
simulations predict the formation of massive subhalos around 
Milky Way-like systems that could not have failed to form a significant stellar component;
and the rotation curve diversity problem \citep{CurveDiversity, diversityII}, where observed
dwarf galaxies exhibit a large diversity in rotation
curves despite the $N$-body prediction of a universal density profile \citep{NFWProfile}.

Elastic self-interacting dark matter (SIDM) was introduced as a possible
explanation for the core-cusp problem \citep{SpergelSteinhardt}, and has
since been extended to give potential explanations for additional 
small scale problems.
Notably, elastic SIDM has been shown 
to create Milky Way-like systems
with diverse rotation curves \citep{VdepSIDM, DiversityBaryonSIDM}
(see \cite{sidmreview} for a comprehensive review).

In current literature, the existence and significance of 
the small scale problems is a matter of debate \citep{NoMissing}.
Despite this, there is still good reason to construct alternative DM models
and find their cosmological implications. 
Modern searches for particle DM candidates regularly construct DM models with
non-negligible self-interactions \citep{higgsDM, eRecoilDM, BDMTheoryI}
as plausible detection scenarios for current generation detector
and accelerator experiments \citep{inelDetection, inelDetectionLHC, BDMTheoryII}.
Notably, these models include inelastic self-interactions.
The particle theory behind such self-interactions are generally known and well-studied
\citep{SchutzInelasticDM}, however the full cosmological implications of such
models is still a matter of active study.

A major part of this effort is to classify alternative DM models and parametrize them
in a way amenable to simulation. One such classification and parametrization scheme
is the “effective theory of structure formation” (ETHOS) \citep{ETHOSI, ETHOSII}.
DM self-interactions generally have two significant regimes. 
Interactions in the early universe before the DM thermally decouples
lead to small-scale perturbations in the initial power spectrum.
These effects are most widely studied in models like warm dark matter,
where they are modelled in simulation by generating initial conditions
consistent with these modified power spectra and then evolve
these initial conditions using a standard CDM simulation \citep{WDMICs}.
Late time interactions,
direct particle collisions during galaxy and cluster formation,
lead to the thermalisation of DM halos and, in the inelastic
case, potentially their evaporation \citep{InelasticSIDMEvaporation, 2cDMCosmo}.

How baryons interact with
varied cosmological and astrophysical parameters is now well-understood
with large hydrodynamical simulation suites such as the 
Cosmology and Astrophysics with Machine-learning Simulations
(CAMELS) project \citep{CAMELS}.
Nevertheless, comprehensive simulation suites
incorporating both realistic baryonic physics
and modified DM are still in their infancy.
One such initiative to rectify this discrepancy,
the DaRk mattEr and Astrophysics with Machine learning and Simulations (DREAMS)
project, will soon produce alternative DM simulation suites
comparable to CAMELS. However, this project will begin with
a focus on warm DM models and will take some time to
generalize to more complicated DM models.
A small but growing number of simulations
exist where these baryonic prescriptions are used
with elastic SIDM:
\cite{SIDMbaryoncosnerve, SIDMbaryonbursty, SIDMbaryons}
are some examples utilizing the IllustrisTNG model,
while \cite{SIDMFIREI,SIDMFIREII,SIDMFIREIII,SIDMFIREIV,SIDMFIREV}
implement the Feedback in Realistic Environments (FIRE) model.
Fewer simulations -- N-body or hydrodynamical -- consider 
general DM models with inelastic effects 
\citep{InelasticSIDMEvaporation, 2cDMCosmo, BDMNBody, EndoDM, ADMI, ADMII}.

In this paper, we present the first suite of simulations
that utilize both an inelastic two-component SIDM model and hydrodynamic
baryonic feedback. 
We organize this paper as follows.
In section \ref{sec:Methods} we describe the DM model and our simulation methods.
In section \ref{sec:Results} we present several summary statistics and demonstrate
how the modified DM physics drives structure formation away from the CDM case.
In section \ref{sec:Discussion} we draw comparisons with other hydrodynamical
simulations to show how the modified dark physics produces a unique signature.
In section \ref{sec:Conclusions} we conclude.
\section{Methods}
\label{sec:Methods}
\subsection{2cDM Model\protect\label{subsec:2cDM-Model}}

The two-component dark matter (2cDM) model is the two-flavour case
of a general $N$-component dark matter model motivated by the physics
of flavour mixed particles \cite{2cDMTheory,2cDMCosmo,Med2010a,Med2010b,m00,m01}.
$N$-component flavour-mixed dark matter is a physically motivated model
for both its kinematic behaviour --- to be discussed below --- and
its ability to avoid tight constraints imposed by the
early universe \citep{TodorokiI}.
Typical multicomponent DM models rely on multiple particle
species that decay or relax to some ground state.
For inelastic effects to be significant at sufficiently late times,
then the abundance of these excited states must be sufficiently
large at late times.
However, the same collision processes must also occur
in the early universe, where the DM density is
order of magnitudes higher thereby increasing the reaction rate
and depleting the abundance of excited states at later times.
Thus, a general multicomponent DM model threads a fine line
where the inelastic effect must be simultaneously small enough that the
excited states remain abundant and large enough that
the modified DM physics can modify halo formation.
flavour mixing involves a single stable particle species
whose self-interactions behave as if there are distinct
excited states due to the difference in 
propagation and interaction between flavour and mass
eigenstates.

The state of flavour-mixed DM particles can be represented in terms of flavour
eigenstates or mass eigenstates. These two representations
are related by a unitary transformation
\[
\left(\begin{array}{c}
\left|\alpha\right\rangle \\
\left|\beta\right\rangle 
\end{array}\right)=U\left(\begin{array}{c}
\left|h\right\rangle \\
\left|\ell\right\rangle 
\end{array}\right)\text{,}
\]
where $\left(\alpha,\beta\right)$ denote flavour states,
$\left(h,\ell\right)$ denote mass states (denoting the `heavier' and `lighter' states), and the unitary
transformation $U$ is parametrized by the mixing angle $\theta$ and given by
\[
U=\left(\begin{array}{cc}
\cos\theta & \sin\theta\\
-\sin\theta & \cos\theta
\end{array}\right)\text{.}
\]
Many-particle states, in particular two-particle states,
are the tensor product of single-particle states. 
Evolution for a state $\left|\Psi\right\rangle $ is determined by
the Schr\"{o}dinger equation
\[
i \hslash\partial_{t}\left|\Psi\right\rangle =H\left|\Psi\right\rangle \text{.}
\]
The Hamiltonian is
$H=H_{free}+H_{grav}+H_{int}$,
where $H_{free}$ describes free propagation,
$H_{grav}$ interaction with a gravitational field, 
and $H_{int}$ particle-particle interactions.
$H_{free}$ and $H_{grav}$ are diagonal in the mass
basis (corresponding to states $\left|h\right\rangle $, $\left|\ell\right\rangle $)
while $H_{int}$ is diagonal in the flavour basis.
In the two-particle case,
$H_{int}$ is transformed to the mass basis by the similarity transformation
$H_{int}=U_{2}^{\dagger}\tilde{H_{int}}U_{2}$,
with $U_{2}\equiv U\otimes U$. 
Because of this, $H_{int}$ necessarily contains non-trivial
off-diagonal elements in the mass basis, 
so particle-particle interactions can lead to conversions between
mass eigenstates.

These mass eigenstate conversions are the inelastic interactions in this model.
In principle, there are six inelastic reactions:
$\left|hh\right\rangle \rightarrow\left|h\ell\right\rangle $, $\left|\ell\ell\right\rangle \rightarrow\left|h\ell\right\rangle $,
$\left|hh\right\rangle \rightarrow\left|\ell\ell\right\rangle $,
and their reverses. In this study, we choose a flavour mixing angle
which eliminates the processes 
$\left|hh\right\rangle \rightarrow\left|\ell\ell\right\rangle $, 
maximizing the inelastic interactions \citep{2cDMTheory, TodorokiI}.
This results in the interaction cross sections taking on the form
\begin{equation}
\sigma_{ii\rightarrow ff}=\frac{\sigma_{0}}{m}\left(\frac{V}{V_{0}}\right)^{\alpha}\left(\begin{array}{cccc}
2 & 1 & 1 & 0\\
\Theta & 1+\Theta & 0 & 1\\
\Theta & 0 & 1+\Theta & 1\\
0 & \Theta & \Theta & 2\Theta
\end{array}\right)\text{,}\label{eq:2cdmsigma}
\end{equation}
where $V$ is the relative velocity between the particles, $\Theta=\Theta\left(E_{final}-E_{initial}\right)$
is the Heaviside step function, $\alpha$ is a power law index, and
$\sigma_{0}/m$ is the cross section per particle mass at $V_{0}$,
which we choose to be $V_{0}=100\kms$. The step functions ensure
that the upscattering processes, i.e. $\left|h\ell\right\rangle \rightarrow\left|hh\right\rangle $,
are kinematically allowed. Downscattering is always kinematically
allowed. In principle, the power laws for each type of process can
be different, so Equation \ref{eq:2cdmsigma} can be summarized as
\begin{equation}
\sigma\left(V\right)=\begin{cases}
\sigma_{0}\left(V/V_{0}\right)^{a_{s}} & \text{elastic scattering}\\
\sigma_{0}\left(p_{f}/p_{i}\right)\left(V/V_{0}\right)^{a_{c}} & \text{inelastic conversion}
\end{cases}\text{,}\label{eq:2cdmSigmas}
\end{equation}
with power law indices $\left(a_{s},a_{c}\right)$ for scattering
and conversion respectively. We highlight the extra ratio of state
momenta for the conversion cross section, which provides an extra
inverse power of velocity for conversion cross sections.
That is, the velocity dependence for the conversion cross section
goes as $V^{a_c - 1}$.

\begin{figure}
\includegraphics[width=1\columnwidth]{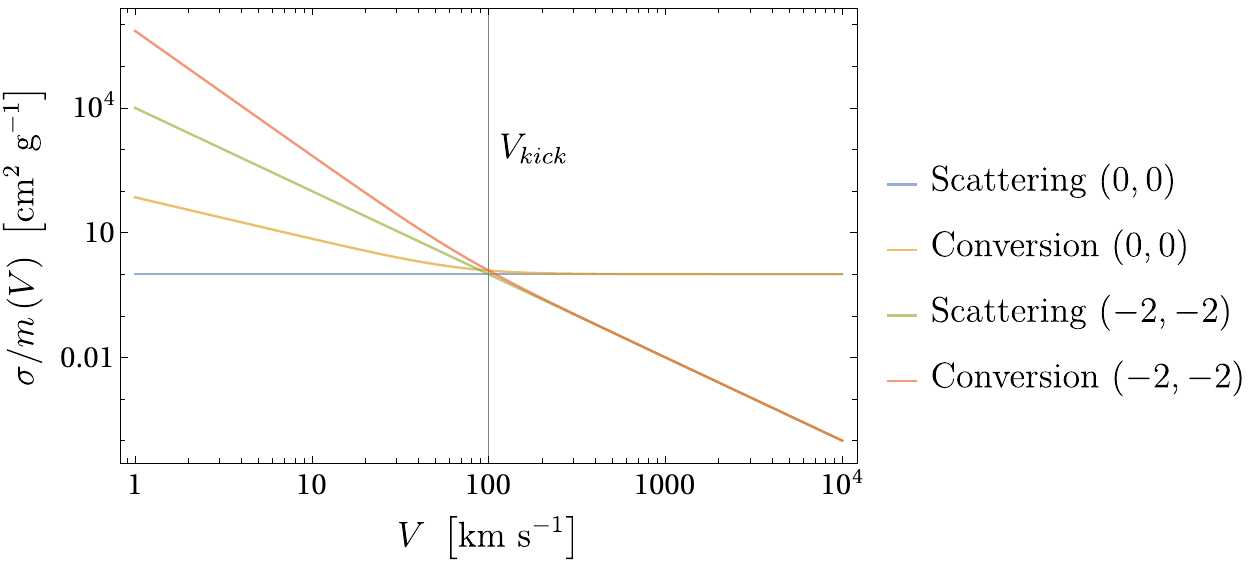}
\caption{
Cross sections for the power law indices
$\left(a_{s},a_{c}\right)=\left(0,0\right)$
and $\left(-2,-2\right)$ with
$\sigma_0=1\cmg$ and $V_{kick}=100\kms$.
For $V>V_{kick}$, $\sigma_c$ behaves as a
simple power law with an extra inverse power,
i.e. while the power law index is $a_c$, the
velocity dependence goes as $V^{a_c - 1}$.
\protect\label{fig:2cDM-cross-sections}}
\end{figure}

\cite{2cDMCosmo} demonstrated that it is sufficient to consider the case
where the masses are highly degenerate, that is $m_{h}\approx m_{\ell}\approx m$.
Nondegenerate models produce
non-physical large scale structure
and halo statistics that wildly disagree with
observational constraints, typically wildly underproducing halo counts
across all scales.
This mass degeneracy, $\Delta m/m$, determines how much kinetic
energy is injected or lost during a conversion. It is useful to discuss
this mass degeneracy in terms of a ``kick velocity'' 
\begin{equation}
V_{kick}=c\sqrt{2\frac{\Delta m}{m}}\text{,}\label{eq:vkick}
\end{equation}
which tells how much kinetic energy a heavy (light) particle gains (loses)
when it converts to the other mass state. More specifically, a heavy particle {\em initially at rest} obtains a velocity equal to $V_{kick}$ when converted into the light state.
We emphasize that $V_{kick}$ is not the exact value of the velocity a moving particle obtains in each collision, but rather an energy-related parameter with the dimension of a velocity. 
The actual velocity change of a particle in each collision depends on the particle's initial state. For highly degenerate particles, the velocity change 
is written in terms of $V_{kick}$ by \citep{2cDMTheory}
\begin{align}
\Delta V = V' - V &\simeq \sqrt{ \left(\Delta m/m\right) c^2 + V^2 } - V \label{eq:deltaV}\\
&\simeq 
\begin{cases}
\left(1/\sqrt{2}\right)V_{kick}, & \text{if } V \ll V_{kick},\\
\left(1/4\right) V_{kick}^2 / V, & \text{if } V \gg V_{kick}.
\nonumber
\end{cases}
\end{align}
Using Equations \ref{eq:2cdmSigmas} and \ref{eq:deltaV}, we plot the
scattering and conversion cross sections for example values of
$\sigma_0$ and $V_{kick}$ in Figure \ref{fig:2cDM-cross-sections}.

Particle interactions are modelled in the code using a Monte-Carlo method.
Using the same implementation as presented in \cite{TodorokiI},
we assume that all collisions are rare and binary.
The probability
of interaction channel $ij\rightarrow i'j'$ within time step $\Delta t$
can therefore be modelled using the pair probability 
\begin{equation}
P_{ij\rightarrow i'j'}=\left(\rho_{j}/m_{j}\right)\sigma_{ij\rightarrow i'j'}\left|{\bf V}_{j}-{\bf V}_{i}\right|\,\Delta t\,\Theta\text{,}
\end{equation}
where, $\rho_{j}/m_{j}$ is the number density of target particles,
$\sigma_{ij\rightarrow i'j'}$ is the cross section for the process
$ij\rightarrow i'j'$, ${\bf V}_{j}-{\bf V}_{i}$ is
the relative velocity of initial particles, and $\Theta$ is the same
Heaviside function from Equation \ref{eq:2cdmsigma}. In every time
step, for each particle, $38\pm10$ nearest neighbours are identified as
potential scattering partners, from which the scattering probabilities
are calculated. A random number is generated to determine which, if
any, scattering channels are chosen. If an interaction occurs, the
final particle energy and momenta magnitude are calculated in the
centre-of-mass frame, with energy injected or removed in an inelastic
interaction if need be. The final directions of momentum are chosen at
random (but still opposite to each other) in the centre-of-mass frame.
Under the rare binary collision approximation, any interactions
within a single time step involving more than two particles
are rejected.
Both elastic and kinematically allowed inelastic interactions use this
process to determine if a scattering event occurs.
The implementation is agnostic to the interaction
type and only uses the scattering probability
for each channel to determine the kind of interaction.
In principle, particle masses must also be adjusted in an interaction.
In practice, the mass degeneracy is much smaller than the mass resolution
of a simulation. Therefore, the simulation particles are of equal mass,
but are labelled with their state so that appropriate interactions
take place.

\subsection{Simulation Suites\protect\label{subsec:Simulation-Suites}}

We present new cosmological simulations utilizing the above 2cDM model.
Two kinds of simulations were performed. $N$-body simulations are `dark
matter only' (DMO) and evolve only under gravitation, while hydrodynamical
simulations implement baryonic physics. Simulations are performed using
the advanced hydrodynamical code \textsc{Arepo} \citep{Arepo}. Gravity
is implemented using the same TreePM/SPH code as in GADGET-3 \citep{GadgetII, Aquarius},
while baryonic physics follow the IllustrisTNG model \citep{IllustrisTNG}.
IllustrisTNG is an improvement to the Illustris model \citep{IllustrisMark, IllustrisPaul},
implementing subgrid baryonic physics processes including star formation and
feedback, black hole formation and feedback, and gas enrichment.

Initial conditions were generated using the \textsc{N-GenIC} code
using cosmological parameters from \cite{PlanckCosmo}, where $\Omega_{m}=0.302$,
$\Omega_{\Lambda}=0.698$, $\Omega_{b}=0.046$, $\sigma_{8}=0.83$,
$n_{s}=0.97$, and $H_{0}=100h\kms\,\text{Mpc}^{-1}$ so that $h=0.6909$.

We note that no modified transfer function was used to generate these
initial conditions -- we only consider the late time
dynamics of 2cDM.
This point requires special attention,
as general SIDM models,
such as those parameterized by \cite{ETHOSI, ETHOSII},
typically include some 
modification to the initial power spectrum.
The dynamics of flavour mixing exclude
significant changes to the initial power spectrum, however.
\cite{2cDMTheory} discusses how flavour eigenstates
evolve in flat spacetime, i.e. in the early universe.
Flavour eigenstates are modelled as a wavepacket
superposition of mass eigenstates.
The mass conversion probability goes as
$\left(1-I\right)^2$, where $I$ is the
wavepacket overlap.
In the strong mass degeneracy case,
wavepackets remain well-overlapped over time
with $I\sim1-\left(\Delta m/m\right)^2$.
Therefore, mass eigenstate conversions
are suppressed in early universe
by a factor of $\left(\Delta m/m\right)^4$.
The least mass degenerate cases we consider
have $\Delta m/m\sim10^{-7}$,
so conversions in the early universe are
less probable by a factor of roughly
$10^{28}$.
Hence, in this regime 2cDM has a
negligible effect in the early universe,
so a modified initial transfer function
is not necessary.
An inelastic DM model that includes
a modified initial transfer function
would be a different model than 2cDM.

In our fiducial set of simulations, we choose a periodic box with
side length $L_{box}=3h^{-1}\,\text{Mpc}$ and a total particle count of $N_{part}=256^{3}$.
For the DMO simulations this yields a mass resolution of $1.38\times10^{5}\mathrm{M_{\odot}}$,
while for the hydrodynamical simulations the mass resolutions are
$1.17\times10^{5}\mathrm{M_{\odot}}$ and $2.11\times10^{4}\mathrm{M_{\odot}}$ for
DM particles and gas cells respectively. All simulations have a gravitational
softening length of $404\,\text{pc}$ at $z=0$, yielding a force
resolution scale of $1.31\,\text{kpc}$. Additional simulations demonstrating
how results converge with $L_{box}$ and $N_{part}$ are presented
in Appendix \ref{sec:Convergence}.

All simulations begin at $z=99$. All DMO simulations are evolved
to $z=0$. We evolve hydrodynamical simulations to $z=2$, due to
their computational cost. A follow-up paper discussing results found
at $z=0$ is in preparation.

Substructures are identified within the simulation box using the Friends-of-Friends (FoF) and \textsc{Subfind} algorithms \citep{Springel2001, Dolang2009}.
The FoF algorithm organizes particles into groups. \textsc{Subfind} further organizes
particles by their gravitational boundedness. 
Each FoF group contains a largest main halo and
can have many smaller subhalos.
\textsc{Subfind} subhalos are what are typically associated with galaxies.
The main halo of a FoF group is itself considered a subhalo of the group;
it is the largest subhalo in the FoF group.
For our analysis, we study the subhalo populations of each simulation.

\subsubsection*{DMO Simulations - Parameter Space Exploration}

For this study, we choose explore the 2cDM parameter space of two
power laws identified by \cite{TodorokiII, TodorokiIII} as being both
physically natural as well as consistent with observations.
These have $\left(a_{s},a_{c}\right)=\left(0,0\right)$
and $\left(a_{s},a_{c}\right)=\left(-2,-2\right)$ respectively. Each
is physically motivated: the \zz model corresponds
to $s$-wave scattering, while the \mm model arises
naturally from maximizing the conversion probability \cite{2cDMTheory}. 
\cite{TodorokiII, TodorokiIII}
also demonstrate that for each model, $\sigma_{0}/m\sim1\cmg$ and
$\Delta m/m\sim10^{-8}$ (corresponding to $V_{k}\approx100\kms$)
produce results consistent with observational constraints.
We explore the 2cDM parameter space by varying $\sigma_{0}/m$ and
$V_{kick}$ about these values for the two power laws using
the same initial conditions.
In particular, we vary $\sigma_{0}/m$ between $0.1$ and $10\cmg$ and $V_{kick}$
between $20$ and $200\kms$ each in logarithmic steps.
We also performed a CDM simulation using this initial condition as
a baseline to compare against.

\subsubsection*{Hydrodynamical Simulations}

\begin{figure*}
\includegraphics[width=1\textwidth]{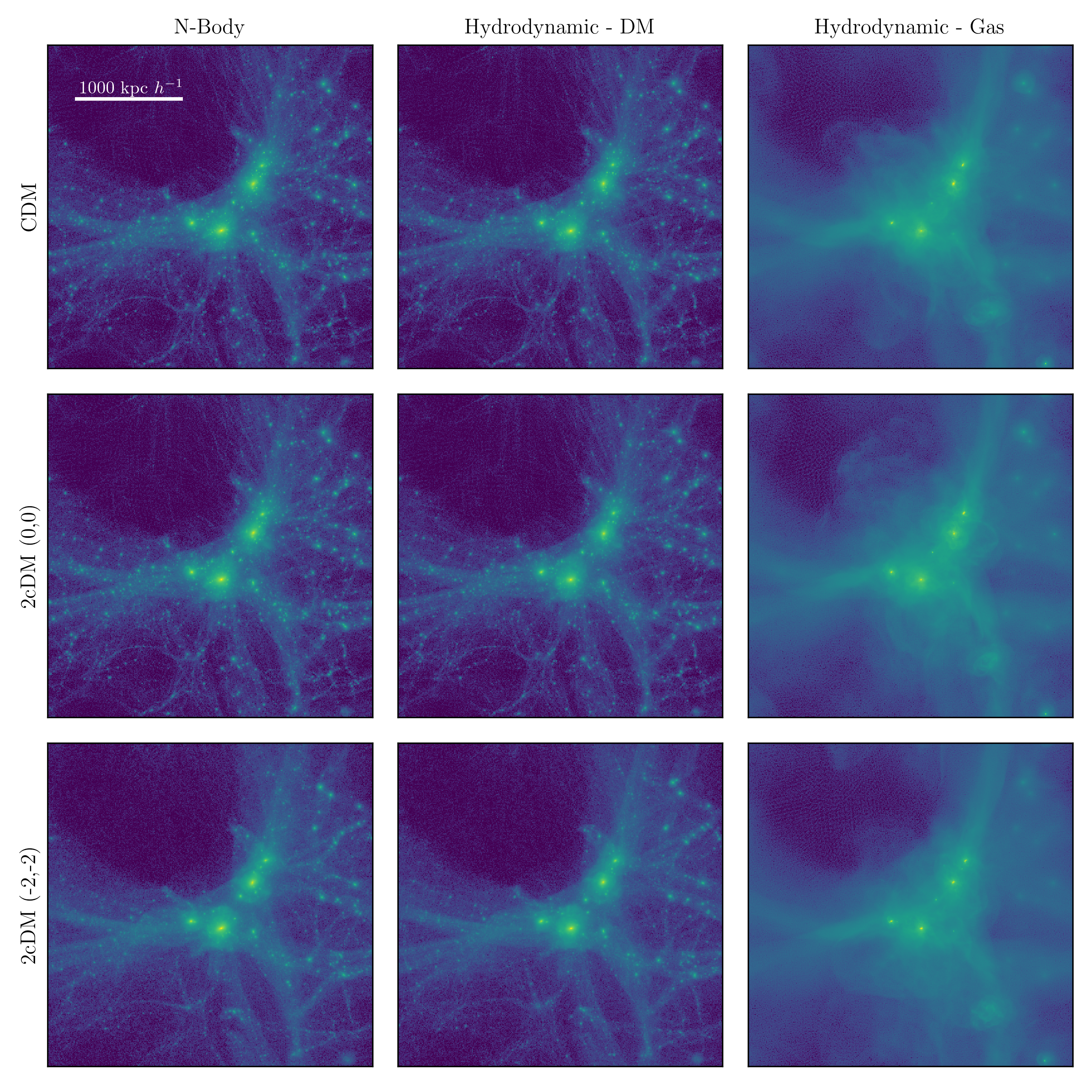}
\caption{
Halo formation within the 2cDM model. Each image
shows the particle density via a histogram in the most dense slice of
the $z$ coordinate.
The top row is CDM, middle is the \zz 2cDM model, and bottom is the \mm 2cDM model.
2cDM simulations have 2cDM parameters $\sigma_0/m=1\cmg$ and $V_{kick}=100\kms$.
The leftmost column is a DMO simulation, the middle column shows the DM particle
density for the corresponding hydrodynamical simulation, and the rightmost column
shows the gas particle density.
Comparison by eye with CDM shows that both 2cDM models suppress small scale halo formation
while keeping the large scale structure intact.
\protect\label{fig:2cDM-halos}}
\end{figure*}

To study how the 2cDM model behaves in the presence of baryons, as well as
to demonstrate the robustness of results, we perform a suite of simulations
using 10 different initial conditions by varying the random seed in \textsc{N-GenIC}.
For each initial condition, we perform a hydrodynamical simulation for each power law
with fixed 2cDM parameters $\sigma_0/m=1\cmg$ and $V_{kick}=100\kms$
as well as a CDM simulation as a baseline to compare against.
This gives a fiducial simulation suite of 30 total simulations.
The CDM simulations share the same initial conditions and random seeds
as the 2cDM simulations to highlight differences solely due to
the altered DM model.
We also performed corresponding
DMO simulations to highlight the effect of the baryons. An overview of structure
formation under the 2cDM model displayed in Figure \ref{fig:2cDM-halos}.

\subsubsection*{Data Products}
For all simulations, we analyse three related summary statistics to
determine the extent at which small scale structures are suppressed.
The halo mass function (HMF) is the cumulative distribution of halo
masses observed within each simulation box. It is related to the maximum
circular velocity function (MCVF) via $V_{max}=\max\left\{ \sqrt{GM\left(>R\right)/R}\right\} $.
In principle, both measure the mass distribution of a halo population.
While the HMF is more instructive in demonstrating the direct effect
of 2cDM interactions, the MCVF is usually more readily measured in
observation, as halo mass estimates are often reliant on direct measurements
of velocity dispersions. To avoid counting spurious non-physical structures
as well as avoid small-scale numerical effects, we only consider \textsc{SubFind}
halos with $>100$ simulation particles, corresponding to halos with
masses $\gtrsim10^{7}\,\mathrm{M_{\odot}}$. In addition, we do not form structures
with masses $\gtrsim10^{11}\,\mathrm{M_{\odot}}$ due to the small box size.

A third metric we analyse is the one-dimensional dimensionless power
spectrum $\Delta^{2}\left(k\right)$, which is a common statistical measure
of density fluctuations with wavenumber $k$ (accordingly with characteristic size
$\lambda=1/k$). $\Delta^{2}\left(k\right)$ is related to the ordinary
power spectrum via $\Delta^{2}\left(k\right)=4\pi\left(2\pi\right)^{-3}k^{3}P\left(k\right)$.
To calculate $P\left(k\right)$, consider fluctuations from the mean density $\bar{\rho}$
\[
\delta\left(x\right) = \frac{\rho\left(x\right) - \bar{\rho} }{\bar{\rho}}.
\]
From this, $P\left(k\right)=\left|\delta\left(k\right)\right|^2$, where 
$\delta\left(k\right)=\mathcal{F}\left\{ \delta\left(\boldsymbol{x}\right)\right\}$
is the Fourier transform of $\delta\left(x\right)$.

Power spectra are common metrics in constraining alternative DM models.
By surveying Lyman-$\alpha$ absorption in quasar spectra,
one can obtain the density fluctuations in neutral Hydrogen at $z\sim2-4$
and thereby estimate the density fluctuations in the
matter density field \citep{FGPAI, FGPAII}.
DM models that suppress or enhance the matter density field at some
characteristic scale will produce a signature on the
Ly-$\alpha$ forest, and therefore the matter power spectrum.
This method has already been used to place large constraints
on warm DM and decaying DM \citep{LymanWDM, LymanDDM, LymanWDMII, LymanConstraints},
and will only become more constraining as next-generation
spectroscopic surveys release more data.

Power spectra were computed using the 
publicly available \textsc{GenPK} code \citep{GenPK}.
Numerically, the smallest scale one can probe before aliasing due
to discretization occurs is the Nyquist wavenumber, $k_{Nyquist}=2\pi N^{1/3}/\left(2L\right)$.
For our simulation suites, $k_{Nyquist}=286\ h\text{Mpc}^{-1}$.
On large scales, accuracy is limited by the small number of modes with
$\lambda\sim L_{box}$.

These metrics give two important pieces of information: a scale
at which suppression occurs and the degree of that suppression. To
more easily discern the degree and scale, we present all quantities
as ratios relative to CDM values in addition to the values themselves.

Unique to hydrodynamic simulations is direct access to the baryonic fields.
From this data, 
track the evolution of the star formation rate (SFR)
for star-forming halos through each simulation.

We also calculate mock Ly-$\alpha$ spectra using the
\textit{fake\_spectra} \citep{fake_spectra} flux extractor package.
We calculate spectra using $5,000$ lines of sight per simulation,
which \cite{TillmanI} found to be sufficient to minimize
variation due to sampling. Along each line of sight,
the optical depth $\tau$ is calculated, from which
the flux $F\left(V\right)=e^\tau$ is obtained.
Due to the smallness of $L_{box}$, each set of spectra
can be considered to be within a single redshift bin.
As such, it is unnecessary for us to construct full light cones.

From the set of spectra, we calculate the 1D flux power spectrum
(P1D), which is obtained from the Fourier transform of the
flux fluctuation
\[
\delta_F\left(V\right) = \frac{F\left(V\right) - \bar{F} }{\bar{F}}.
\]
where $\bar{F}$ is the mean flux in the box.
In our simulations, the Nyquist level for P1D is
$k_{V,Nyquist}\approx2 \ \text{s } \text{km}^{-1}$.
To solely the effect of modified DM on the Ly-$\alpha$ spectra,
we do not add noise to the spectra nor do we attempt to
emulate any specific instrumentation characteristics or
data reduction pipeline.

\section{Results}
\label{sec:Results}
\subsection{Parameter Space Exploration}
\label{subsec:parameter-space}
\subsubsection*{Variation of $V_{kick}$}

\begin{figure*}
\noindent\begin{minipage}[t]{1\columnwidth}
\includegraphics[width=1\columnwidth]{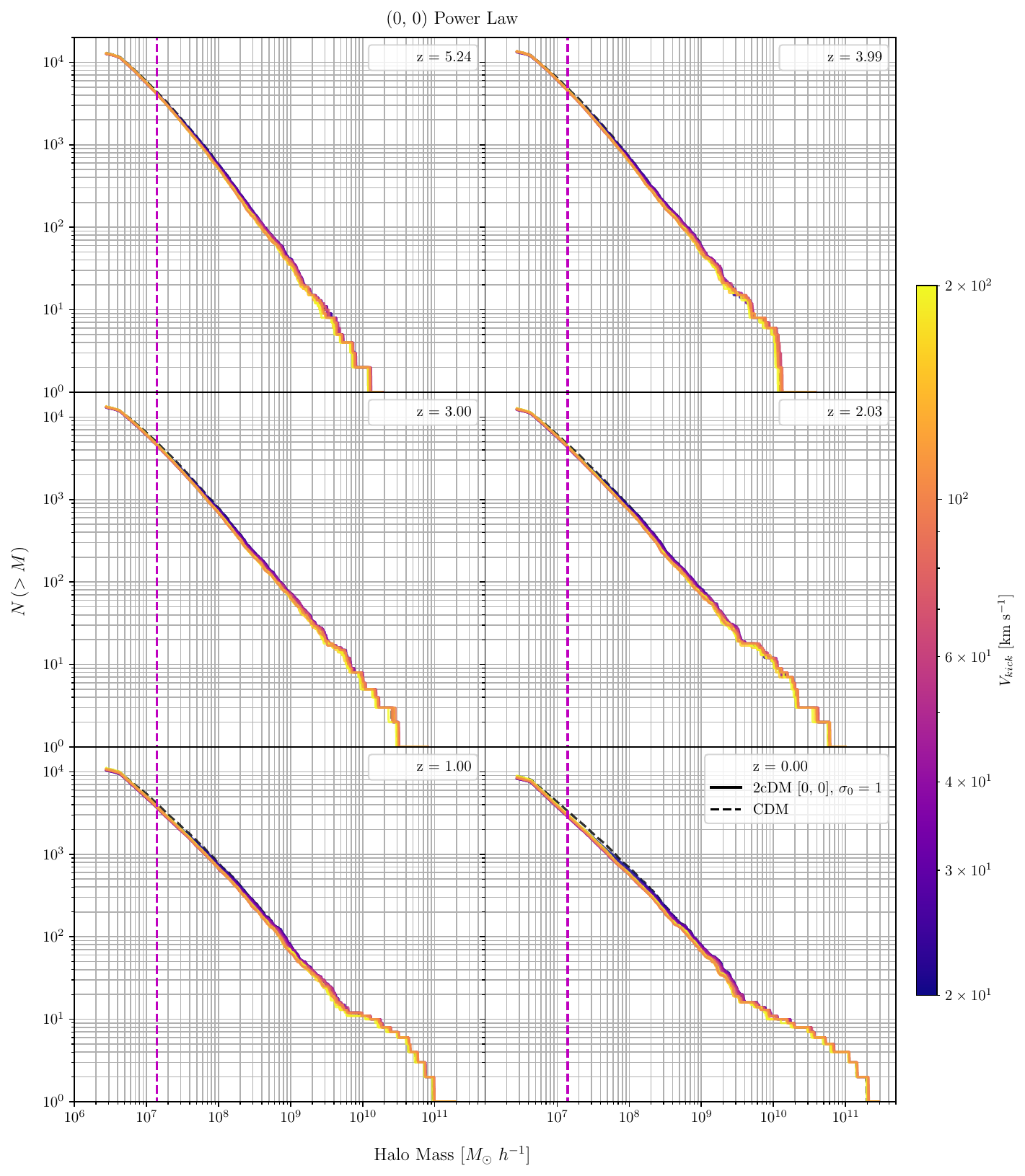}
\end{minipage}
\noindent\begin{minipage}[t]{1\columnwidth}
\includegraphics[width=1\columnwidth]{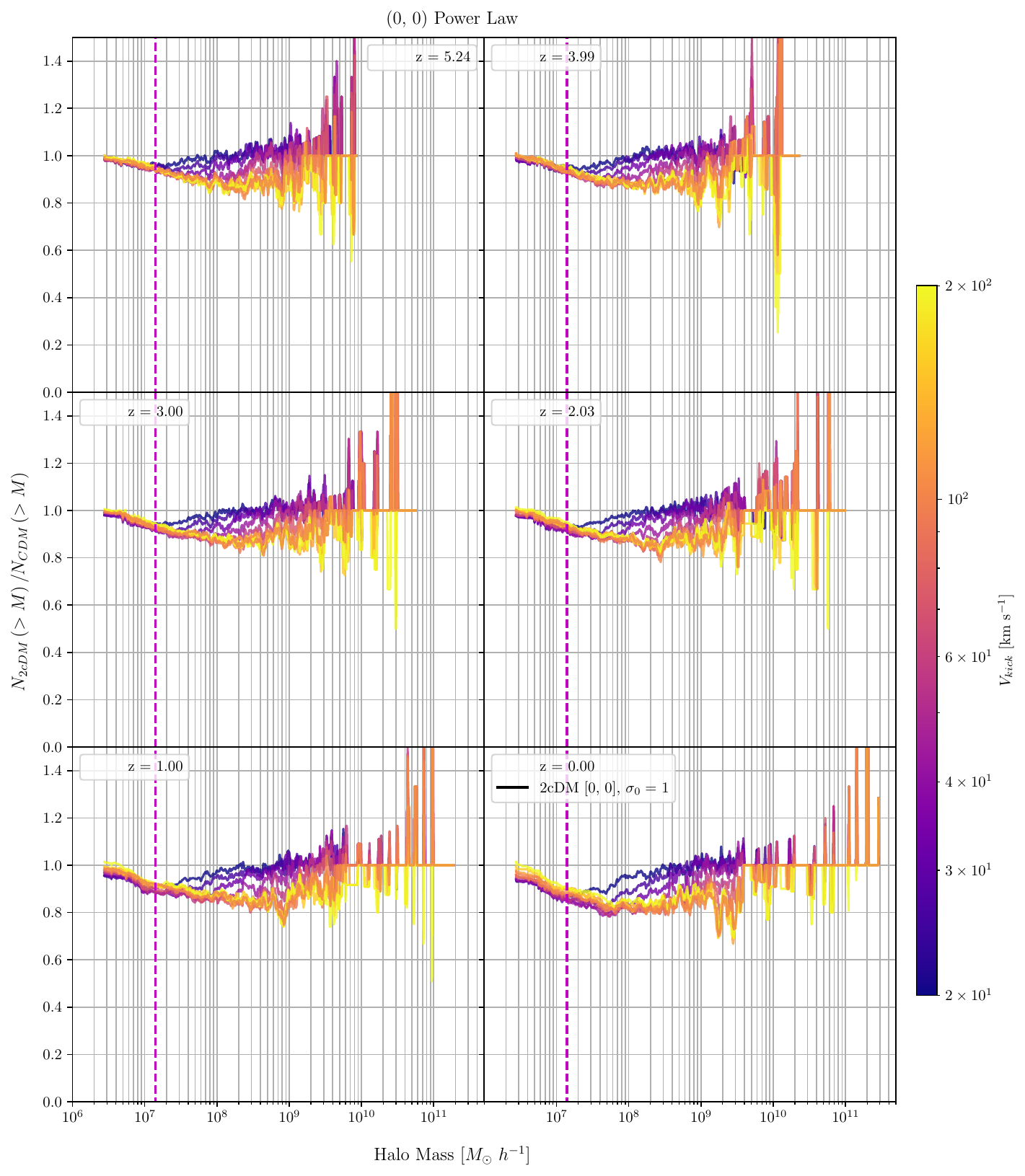}
\end{minipage}\\
\noindent\begin{minipage}[t]{1\columnwidth}
\includegraphics[width=1\columnwidth]{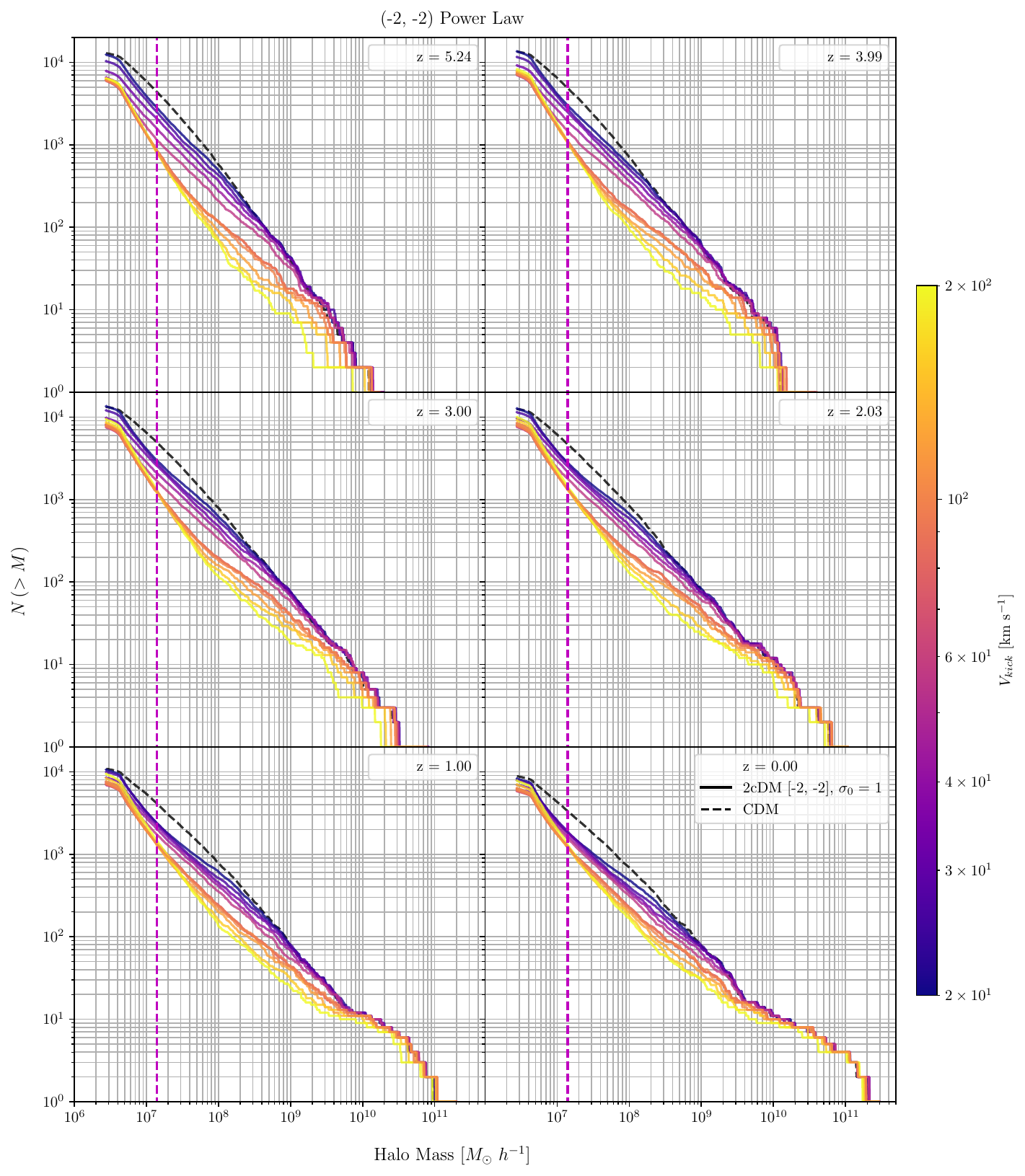}
\end{minipage}
\noindent\begin{minipage}[t]{1\columnwidth}
\includegraphics[width=1\columnwidth]{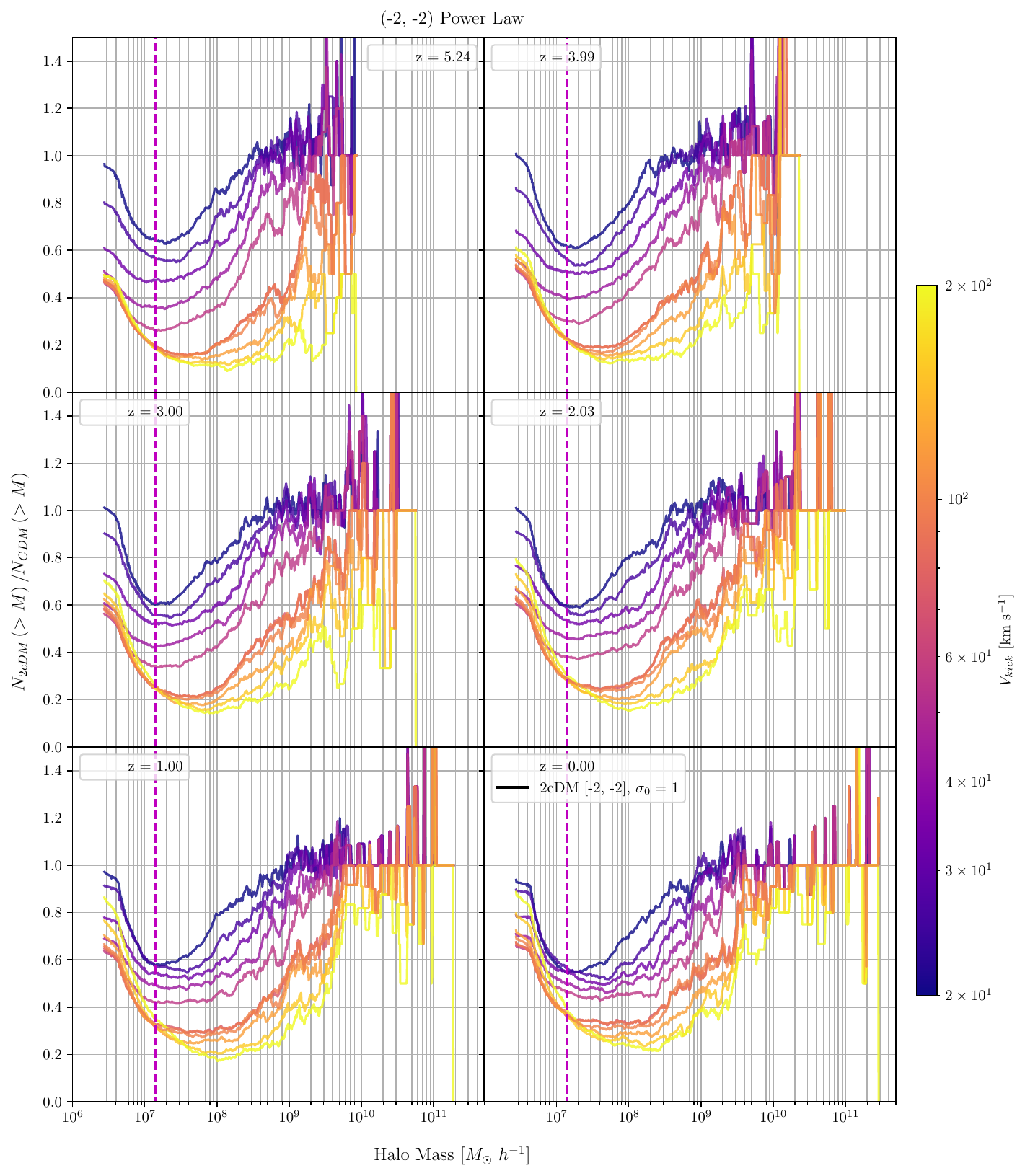}
\end{minipage}
\caption{(\textit{Left Column}) The halo mass functions for the 20 $V_{kick}$ variations, separated by power law. CDM is denoted by the dotted black line. 
Vertical lines denote the mass at which numerical effects can dominate.
We choose a cutoff of $100$ simulation particles.
Lines are coloured by the value of $V_{kick}$, with yellow representing larger values and blue representing smaller values. 
$\sigma_0$ is fixed at $1\cmg$ for these simulations.
(\textit{Right Column})
The ratio of 2cDM halo mass functions to the CDM halo mass function.
The main effect of increasing $V_{kick}$ is to increase the energy injected in a conversion,
thereby letting particles escape from halos more easily. This effect is clearly shown,
where larger $V_{kick}$ leads to the suppression of more small structure
and a shift in the peak amount of suppression to higher mass structures.
The effect is less pronounced in the \zz power law compared to the \mm power law.
For the \mm power law, as $z\rightarrow0$, the separation between CDM and the largest
$V_{kick}$ becomes smaller and smaller values of $V_{kick}$ begin to cluster
together.
\protect\label{fig:vk-mass}}
\end{figure*}

\begin{figure*}
\noindent\begin{minipage}[t]{1\columnwidth}
\includegraphics[width=1\columnwidth]{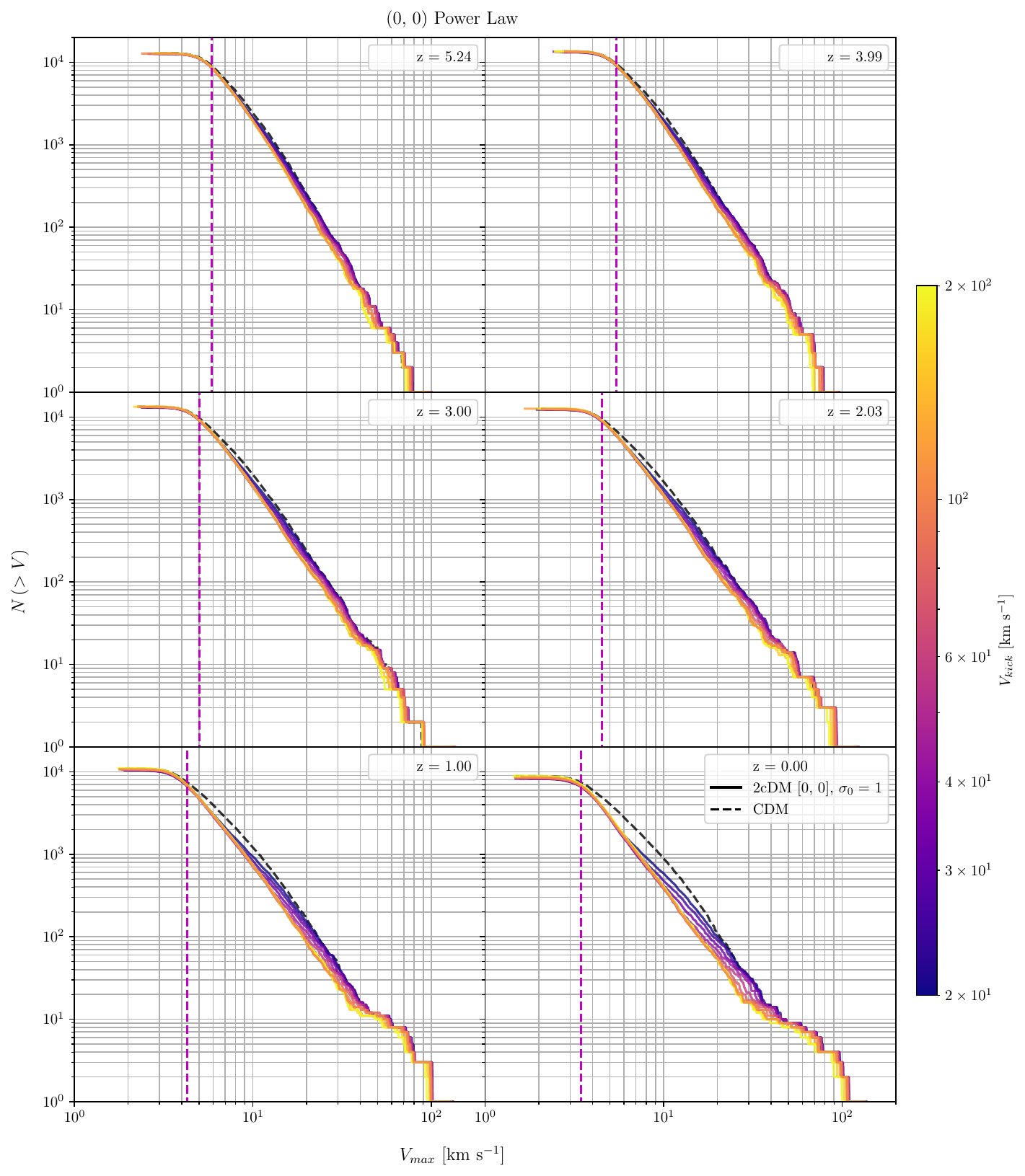}
\end{minipage}
\noindent\begin{minipage}[t]{1\columnwidth}
\includegraphics[width=1\columnwidth]{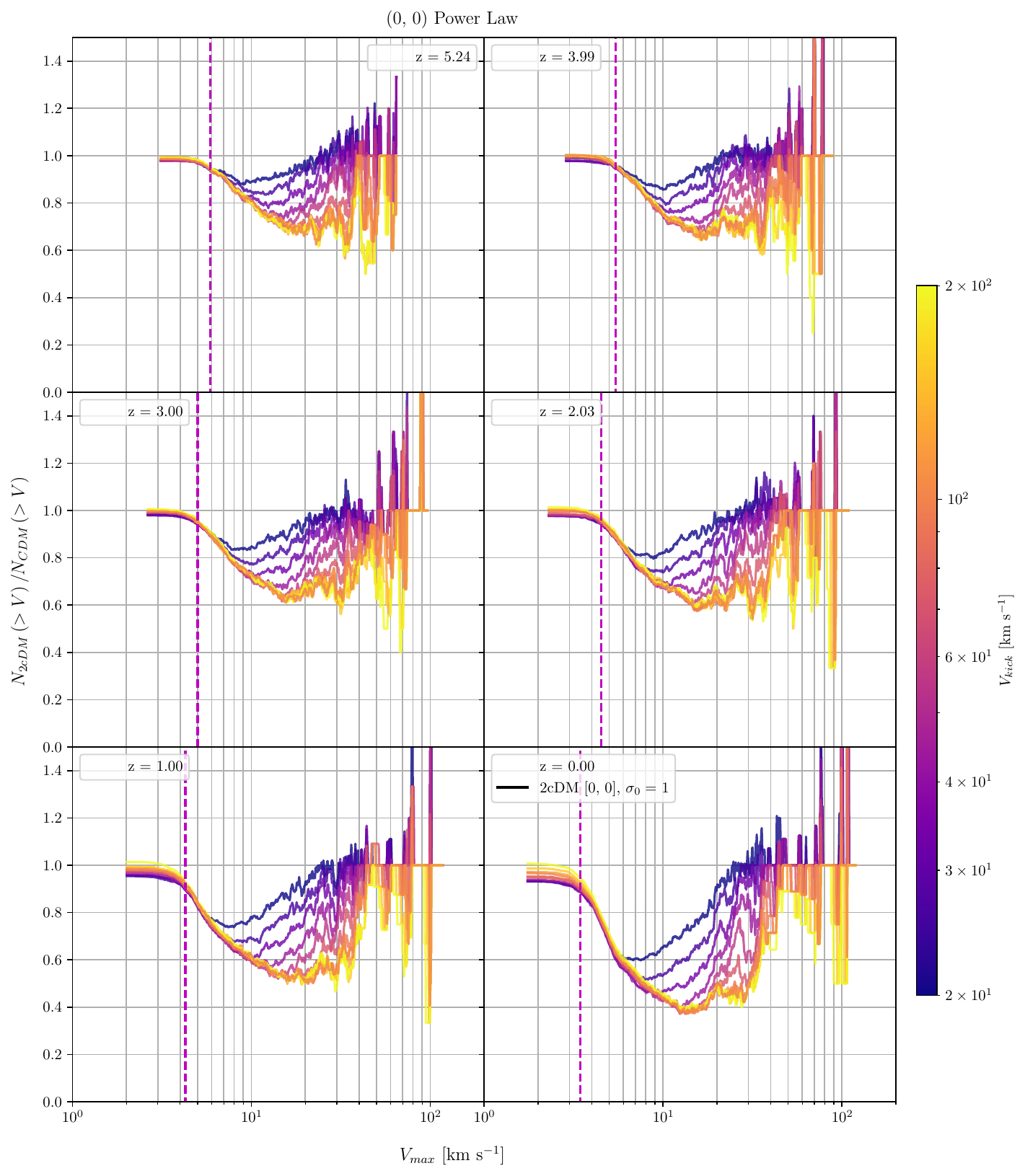}
\end{minipage}\\
\noindent\begin{minipage}[t]{1\columnwidth}
\includegraphics[width=1\columnwidth]{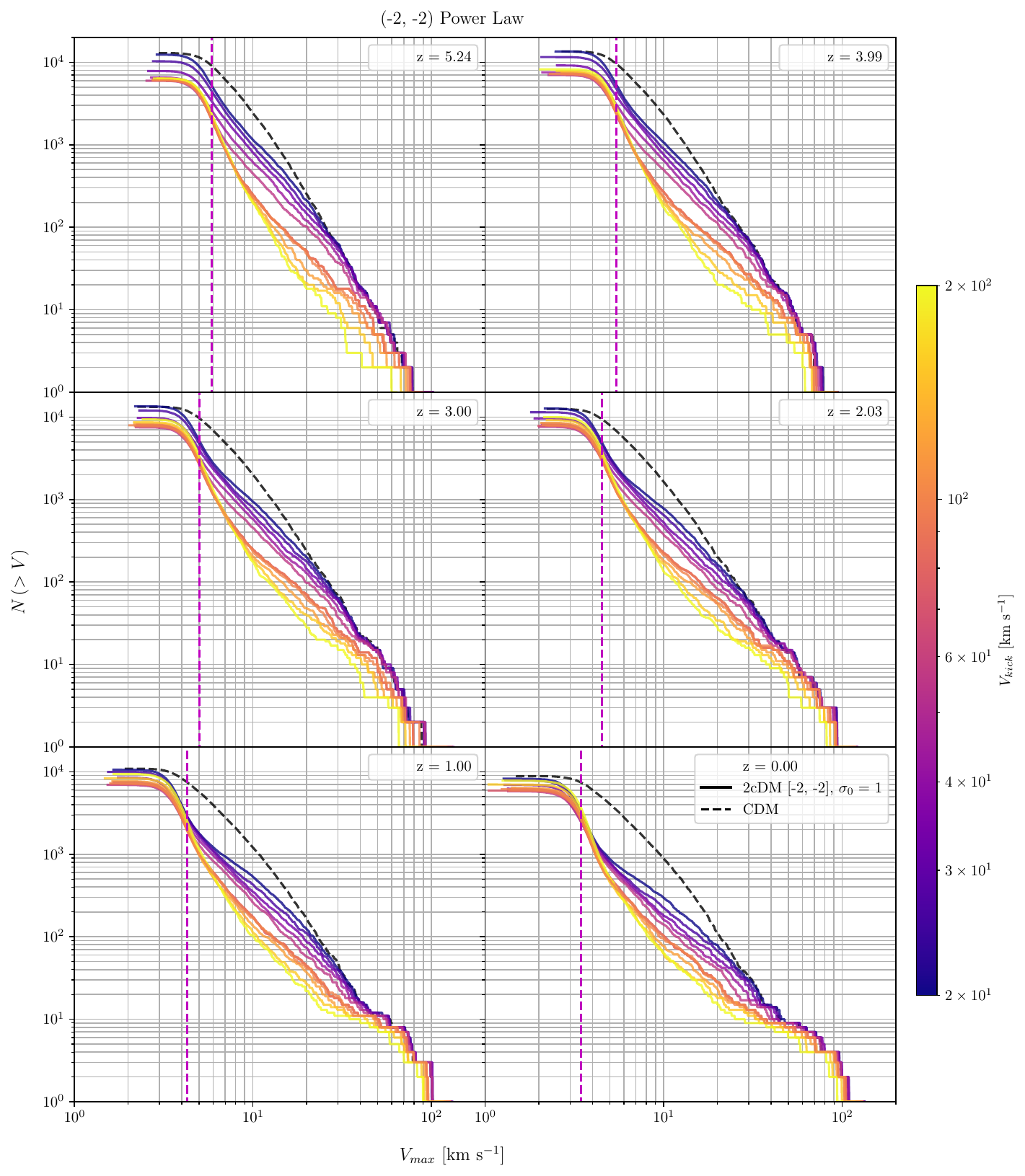}
\end{minipage}
\noindent\begin{minipage}[t]{1\columnwidth}
\includegraphics[width=1\columnwidth]{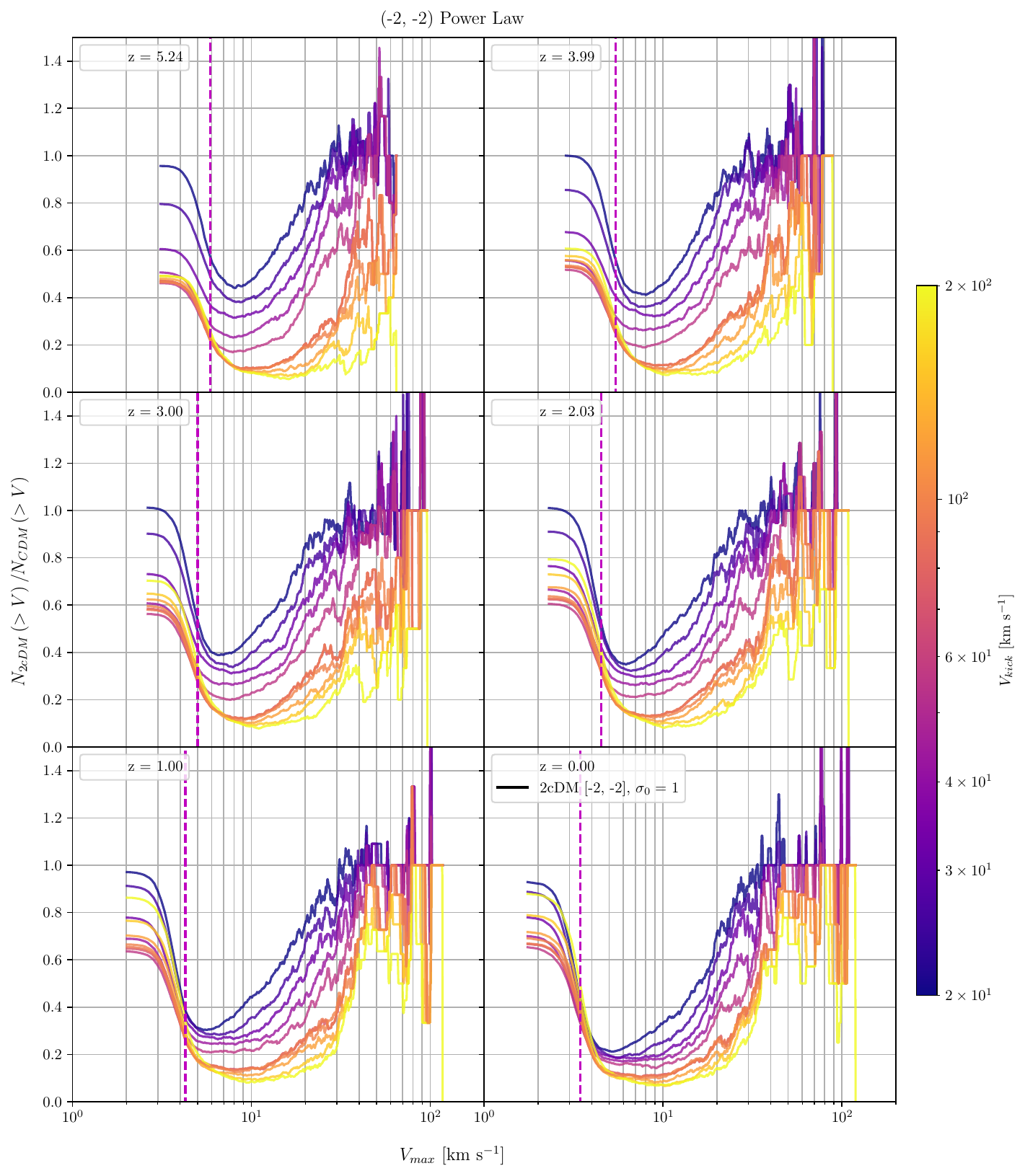}
\end{minipage}
\caption{The $V_{max}$ function for the $V_{kick}$ variation suite. Labelling 
and colouring are identical to Figure \ref{fig:vk-mass}. The effect of moving
peak suppression to higher mass structures is now seen in both power laws.
As expected, the \mm power law shows a much higher degree of suppression,
but the location of the peak is similar between both power laws.
In addition, we observe a similar clustering for the \mm power law
as in Figure \ref{fig:vk-mass} as $z\rightarrow0$.
\protect\label{fig:vk-vel}}
\end{figure*}

\begin{figure*}
\noindent\begin{minipage}[t]{1\columnwidth}
\includegraphics[width=1\columnwidth]{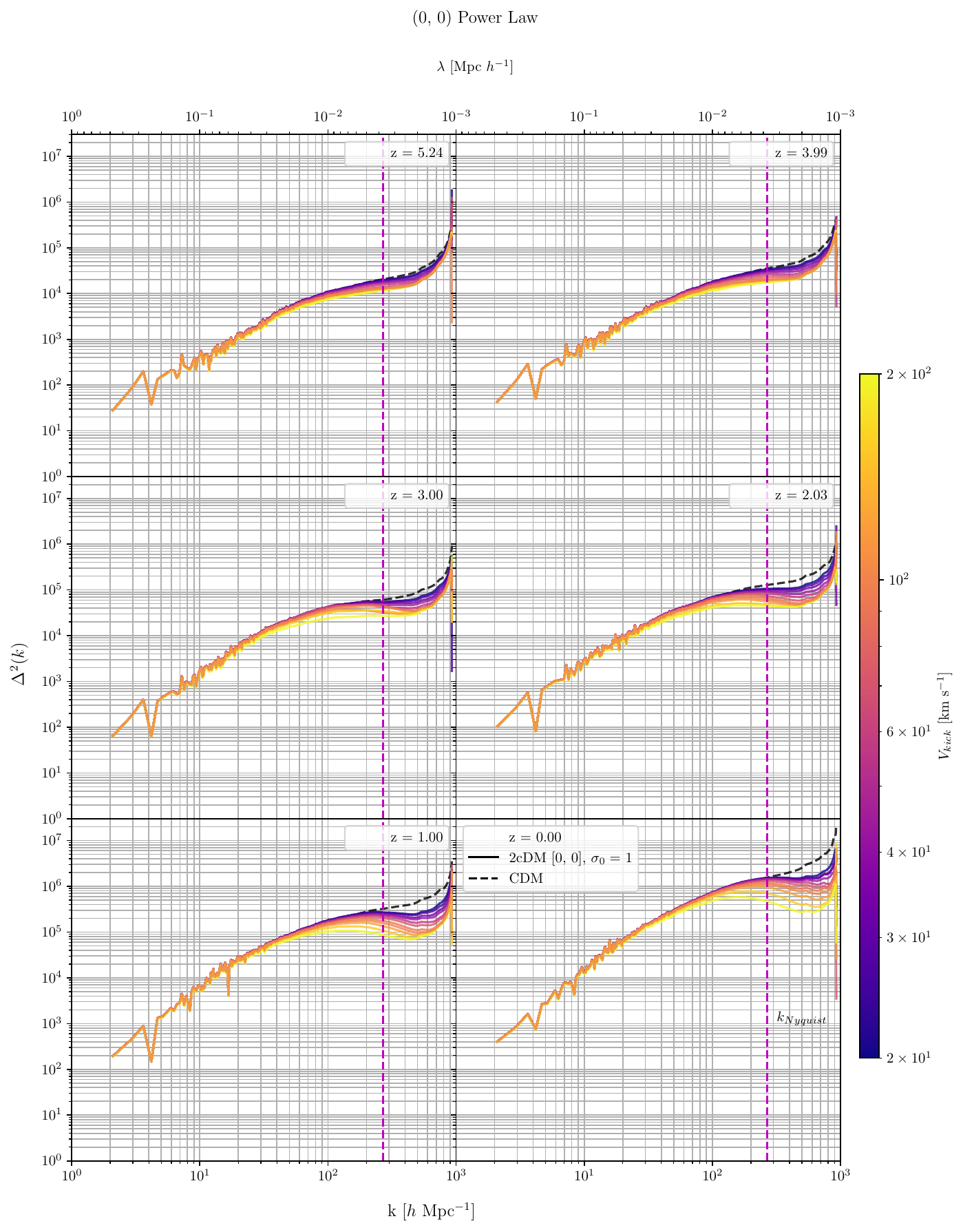}
\end{minipage}
\noindent\begin{minipage}[t]{1\columnwidth}
\includegraphics[width=1\columnwidth]{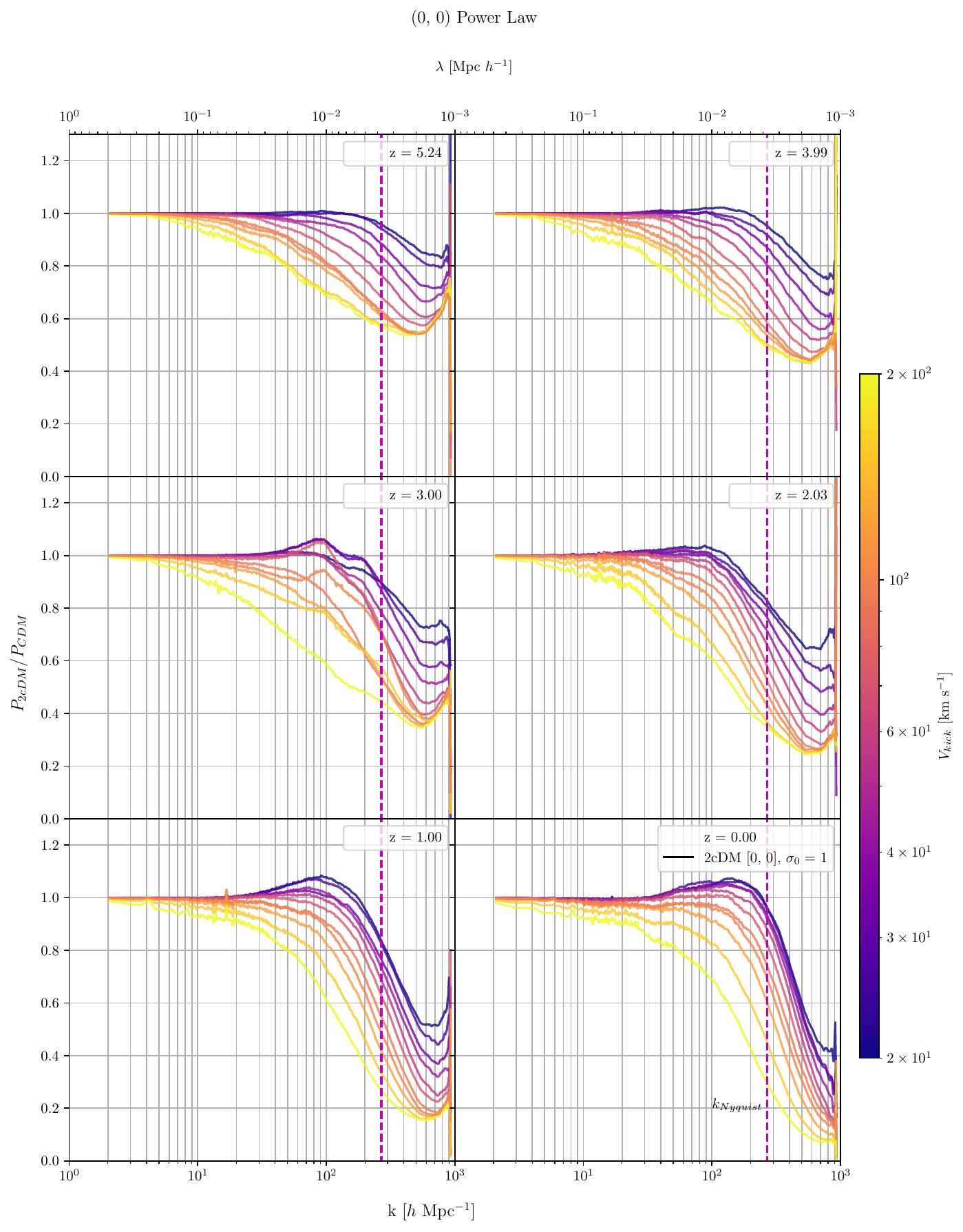}
\end{minipage}
\noindent\begin{minipage}[t]{1\columnwidth}
\includegraphics[width=1\columnwidth]{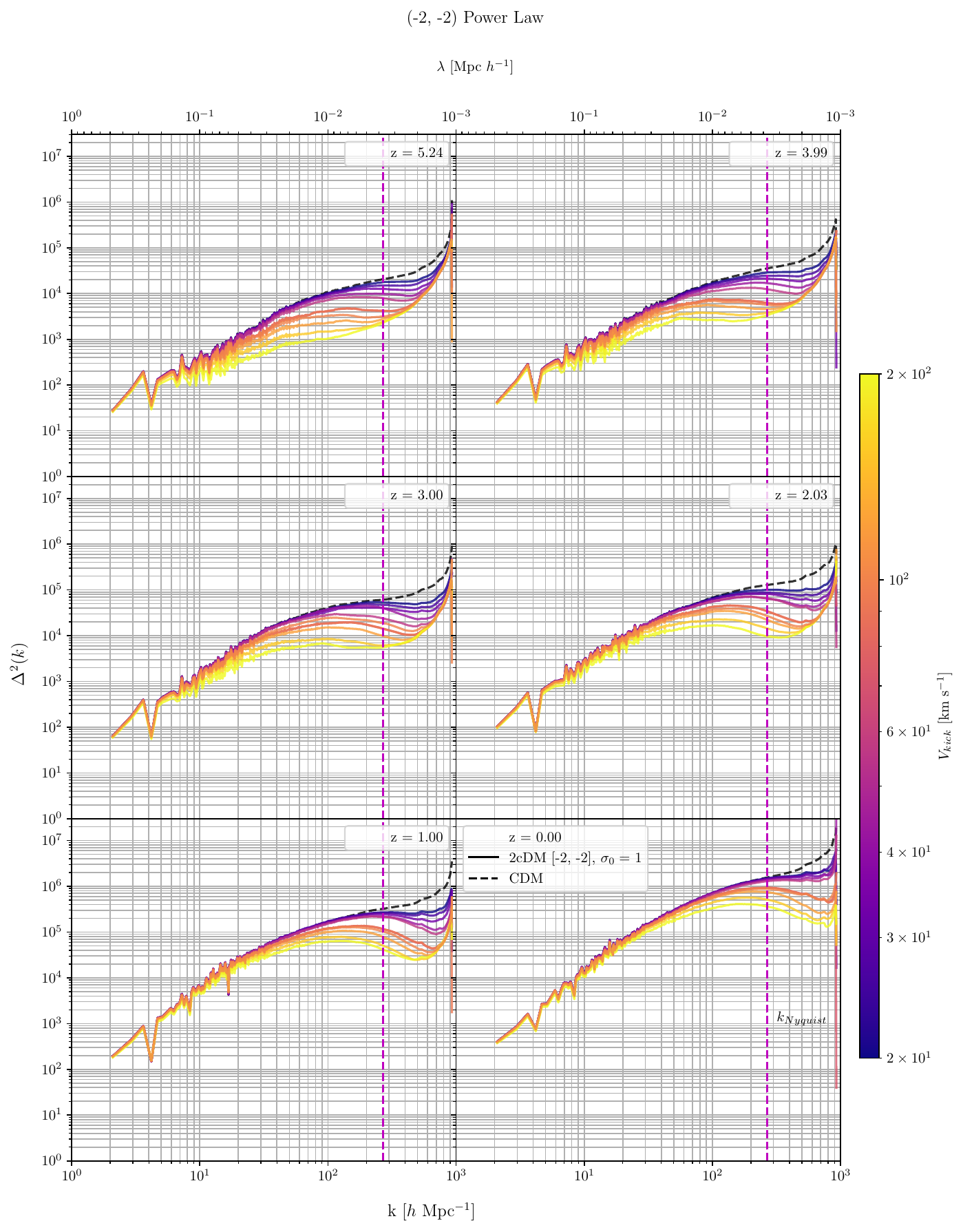}
\end{minipage}
\noindent\begin{minipage}[t]{1\columnwidth}
\includegraphics[width=1\columnwidth]{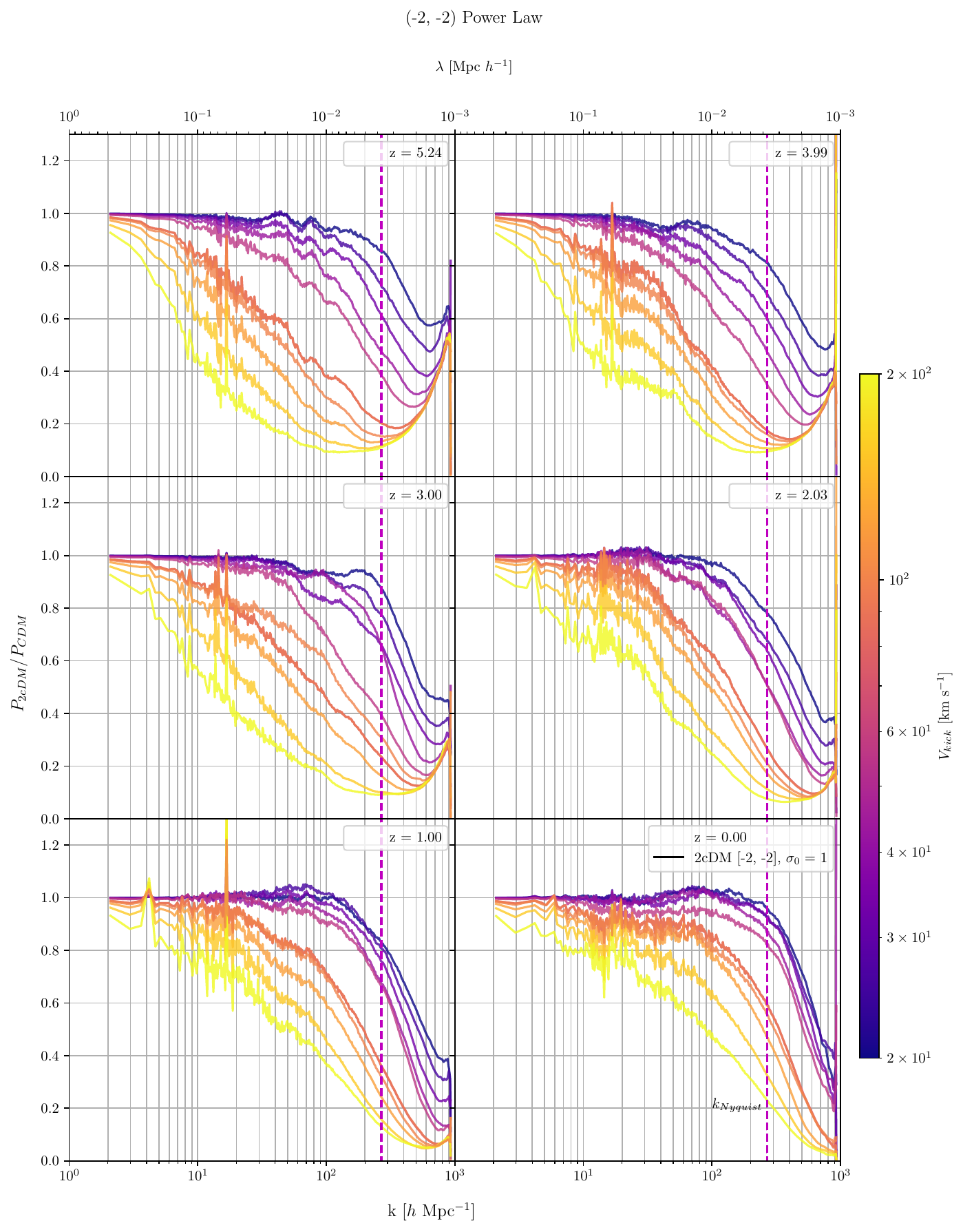}
\end{minipage}
\caption{The dimensionless power spectrum for the $V_{kick}$ variation suite. Labelling 
and colouring are identical to Figure \ref{fig:vk-mass}.
Vertical lines denote the Nyquist wavenumber for the simulations.
While the effects of the \zz power law are less pronounced in halo statistics,
the power spectra shows that even a modest $V_{kick}$ can produce $\sim20\%$ suppression
in the density distribution. As with the other statistics, the main effect of $V_{kick}$
is to move the onset of suppression towards larger structures.
The difference between CDM and the highest $V_{kick}$ reduces as
$z\rightarrow0$, similar to what is observed in Figure \ref{fig:vk-mass}.
\protect\label{fig:vk-pk}}
\end{figure*}

We vary $V_{kick}$ between $20$ and $200\kms$ in 10 logarithmic
steps while keeping $\sigma_{0}$ fixed to $1\cmg$. Including the
fiducial simulation with $10\kms$, we performed a total of 11 simulations
per power law. The results are shown in Figures \ref{fig:vk-mass}-\ref{fig:vk-pk}.

In Figure \ref{fig:vk-mass}, we present the suppression in the HMF. For the \zz power
law, structure suppression only occurs with $V_{kick}\gtrsim10\kms$,
where the degree of suppression is $\sim20\%$ relative to CDM at
a scale of $M_{halo}\sim10^{8}\,\mathrm{M_{\odot}}$. The \mm power law produces a
much higher degree of suppression at all $V_{kick}$ compared to the
\zz power law. The \mm power law illustrates the main effect of $V_{kick}$
more clearly: increasing $V_{kick}$ changes where peak suppression
occurs. As $V_{kick}$ increases, particles can escape more easily
from more massive systems, thereby increasing the halo mass at which
halo evaporation occurs.

This effect is most easily seen in the MCVF (Figure \ref{fig:vk-vel}). For both power
laws, it is clear that the effect of increasing $V_{kick}$ is to
shift the suppression peak over towards higher $V_{max}$, while retaining
a similar shape throughout. We also see that the effects on the MCVF
can be different from the effects on the HMF. The \zz power law has
a growing degree of suppression across redshift between $30-60\%$
at $V_{max}\sim11\kms$. The maximum degree of suppression for the
\mm power law is $\sim90\%$ at $V_{max}\sim10\kms$ across all redshift.

Power spectra are similarly affected (Figure \ref{fig:vk-pk}). 
As $V_{kick}$ increases, the break between CDM and 2cDM power
spectra moves to larger scales, i.e. towards larger $k$.
Simulations with small $V_{kick}$ have enhancement
or only small suppression above the Nyquist level.
In the highest $V_{kick}$ cases, the \zz power law shows suppression of up to 
$60\%$, while the \mm power law a higher degree of
suppression, up to $80\%$.

\subsubsection*{Variation of $\sigma_{0}$}

\begin{figure*}
\noindent\begin{minipage}[t]{1\columnwidth}
\includegraphics[width=1\columnwidth]{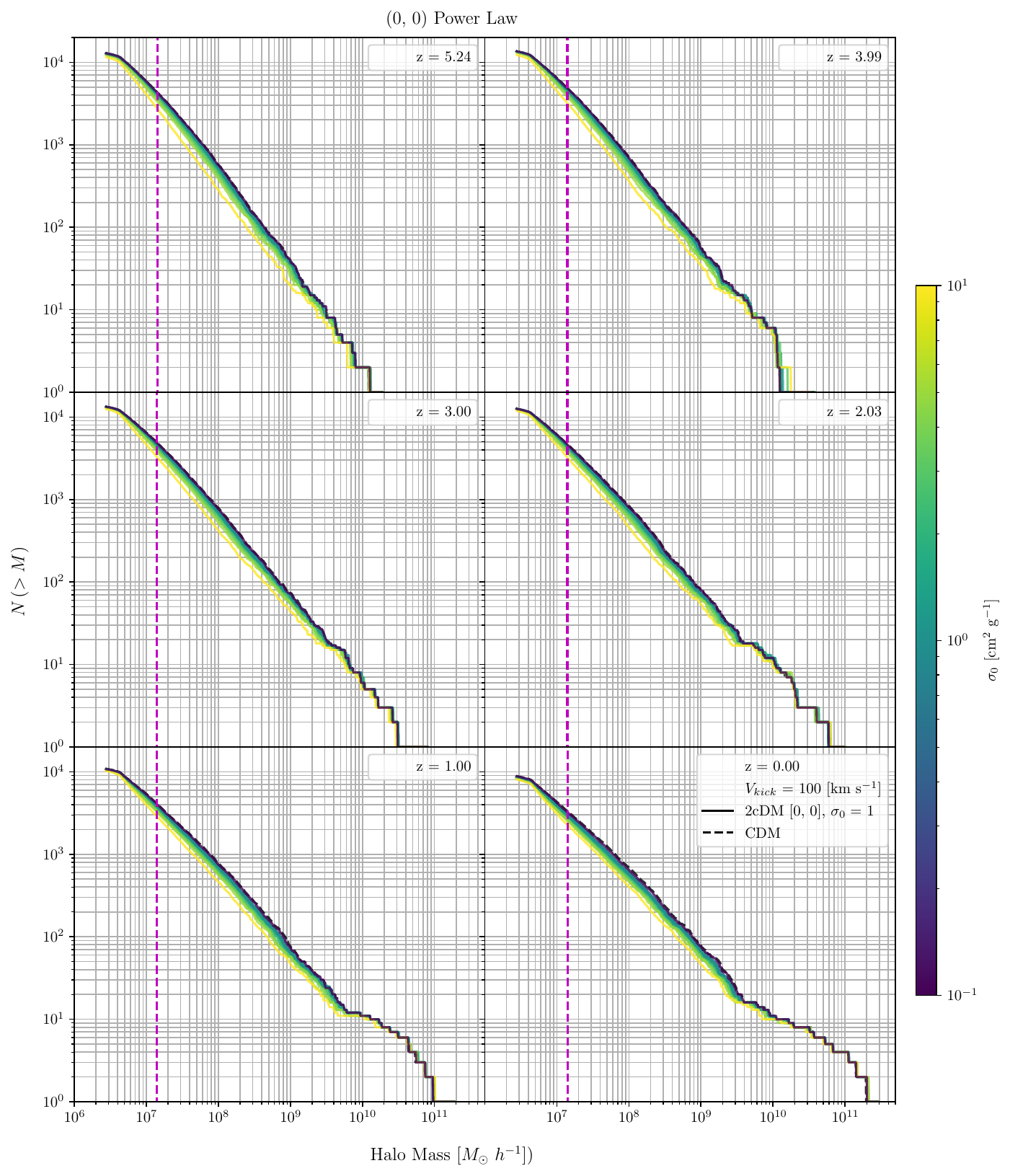}
\end{minipage}
\noindent\begin{minipage}[t]{1\columnwidth}
\includegraphics[width=1\columnwidth]{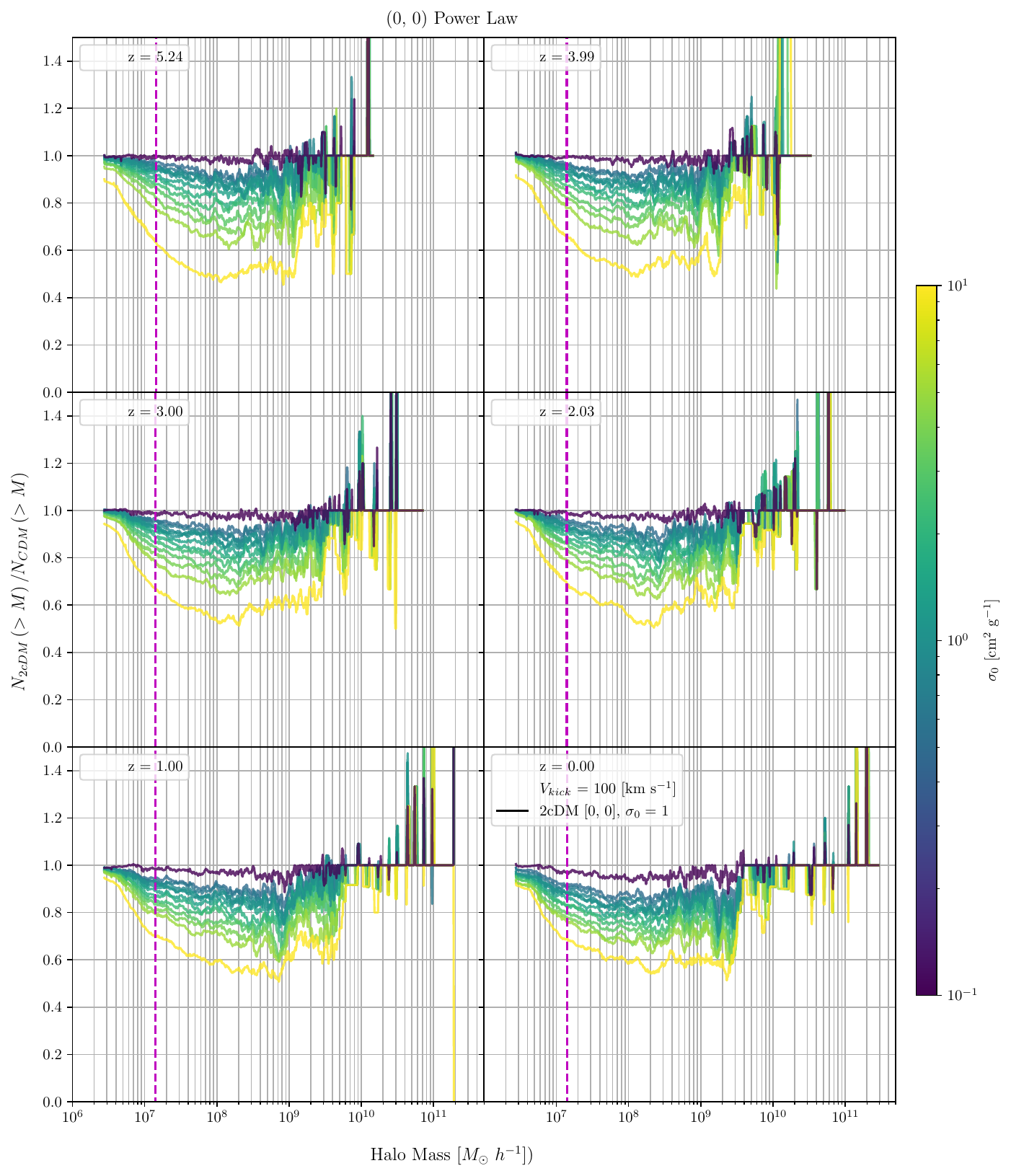}
\end{minipage}\\
\noindent\begin{minipage}[t]{1\columnwidth}
\includegraphics[width=1\columnwidth]{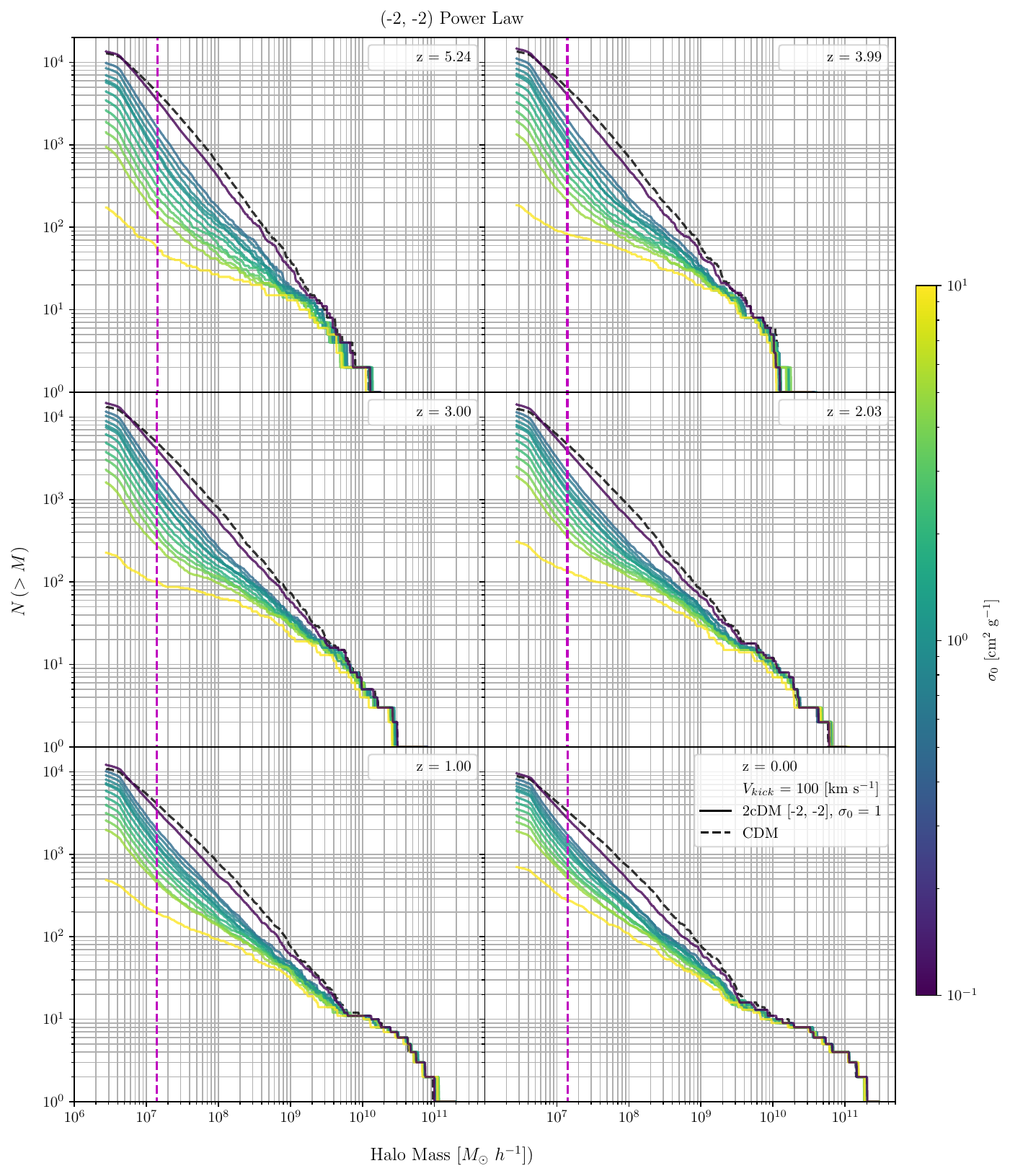}
\end{minipage}
\noindent\begin{minipage}[t]{1\columnwidth}
\includegraphics[width=1\columnwidth]{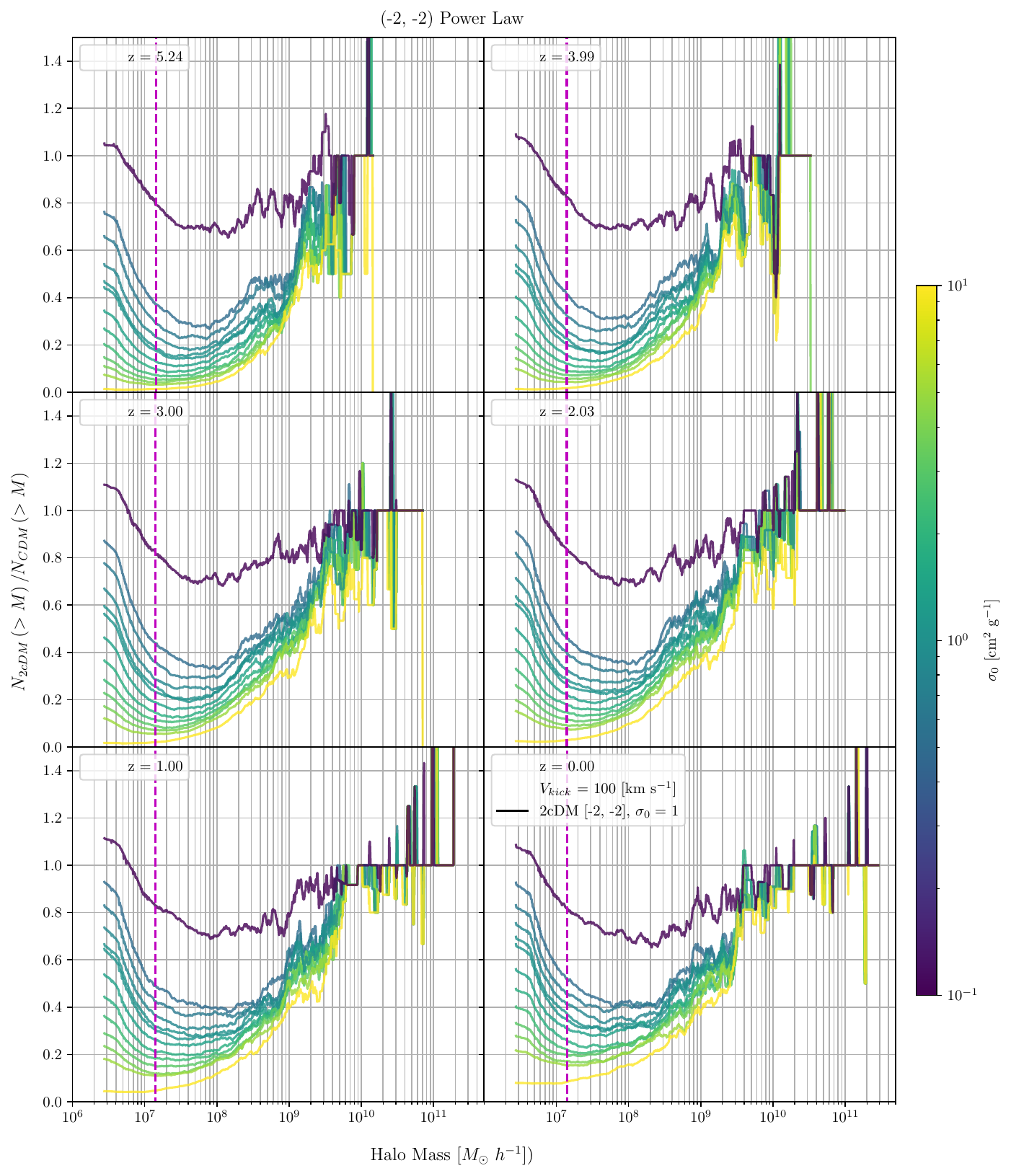}
\end{minipage}
\caption{(\textit{Left Column}) The halo mass functions for the 20 $\sigma_0$ variations, separated by power law. CDM is denoted by the dotted black line. 
Vertical lines denote the mass at which numerical effects can dominate.
We choose a cutoff of $100$ simulation particles.
Lines are coloured by the value of $\sigma_0$, with yellow representing larger values and blue representing smaller values.
$V_{kick}$ is fixed at $100\kms$ for these simulations.
(\textit{Right Column})
The ratio of 2cDM halo mass functions to the CDM halo mass function.
For both power laws, we see that the location of peak suppression remains the same.
Only the degree of suppression is increased. In other words, the overall
shape of the suppression curve remains the same, just deepened. This corresponds
to the higher collision rate from increasing $\sigma_0$.
\protect\label{fig:sig-mass}}
\end{figure*}

\begin{figure*}
\noindent\begin{minipage}[t]{1\columnwidth}
\includegraphics[width=1\columnwidth]{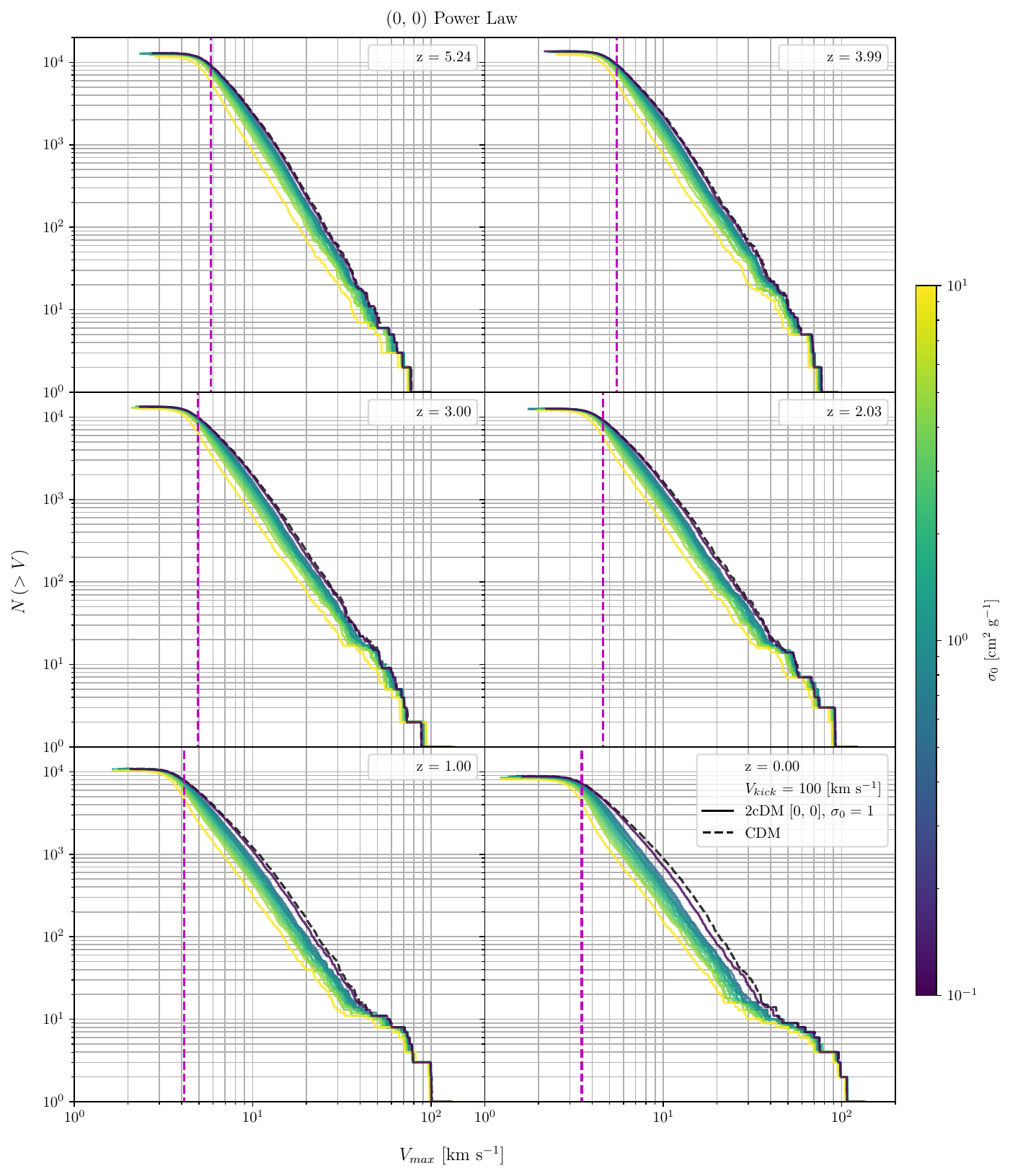}
\end{minipage}
\noindent\begin{minipage}[t]{1\columnwidth}
\includegraphics[width=1\columnwidth]{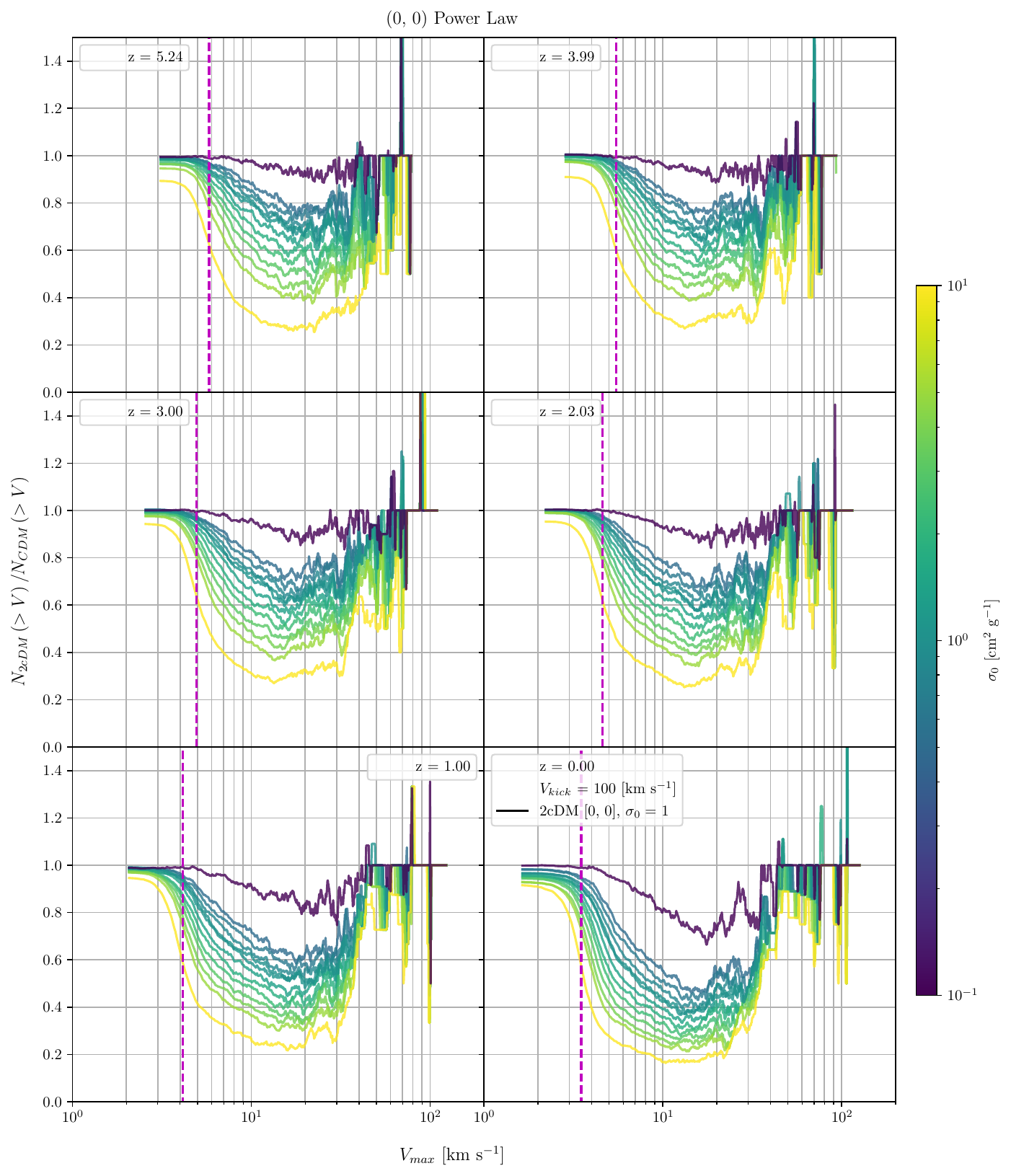}
\end{minipage}\\
\noindent\begin{minipage}[t]{1\columnwidth}
\includegraphics[width=1\columnwidth]{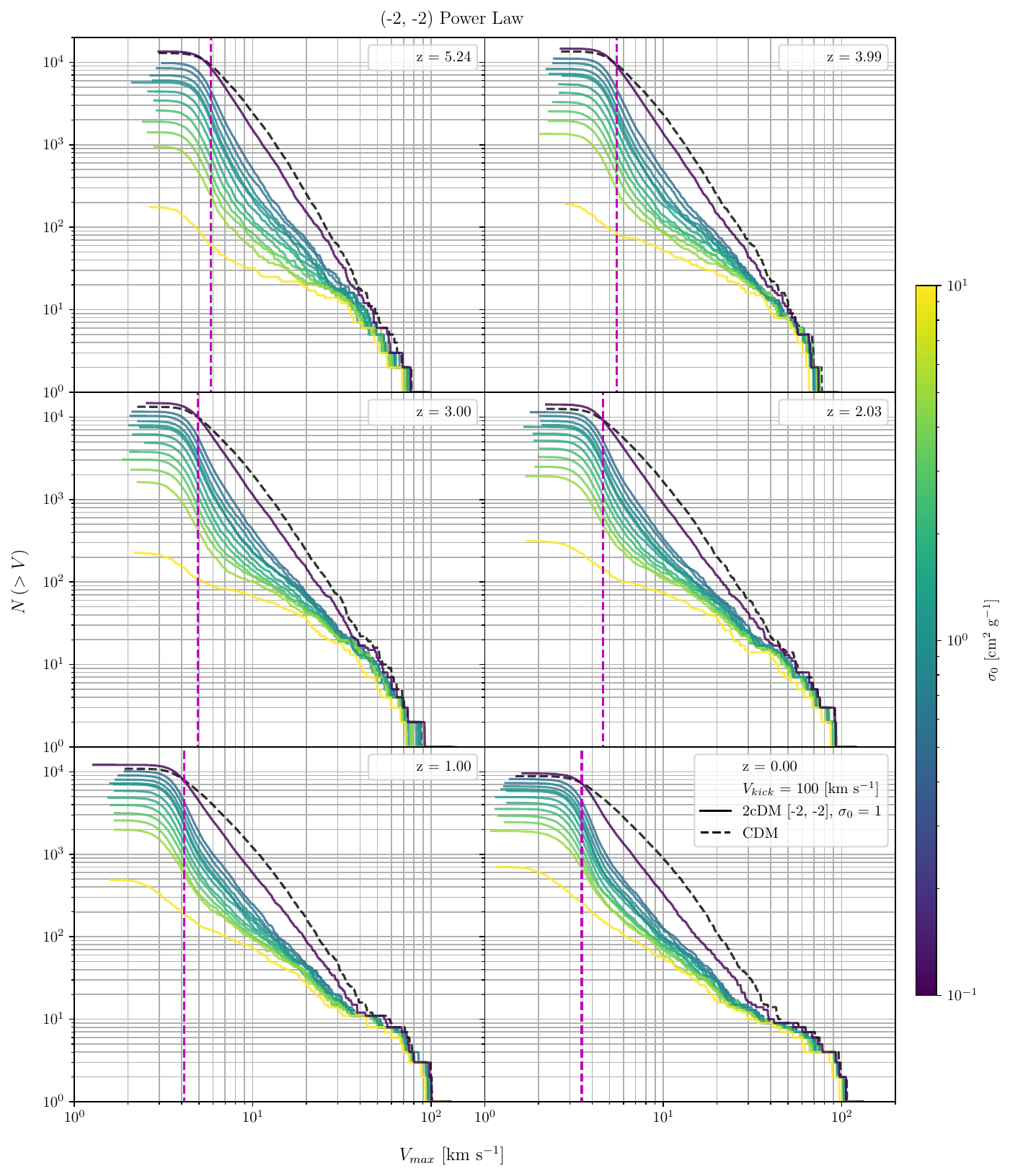}
\end{minipage}
\noindent\begin{minipage}[t]{1\columnwidth}
\includegraphics[width=1\columnwidth]{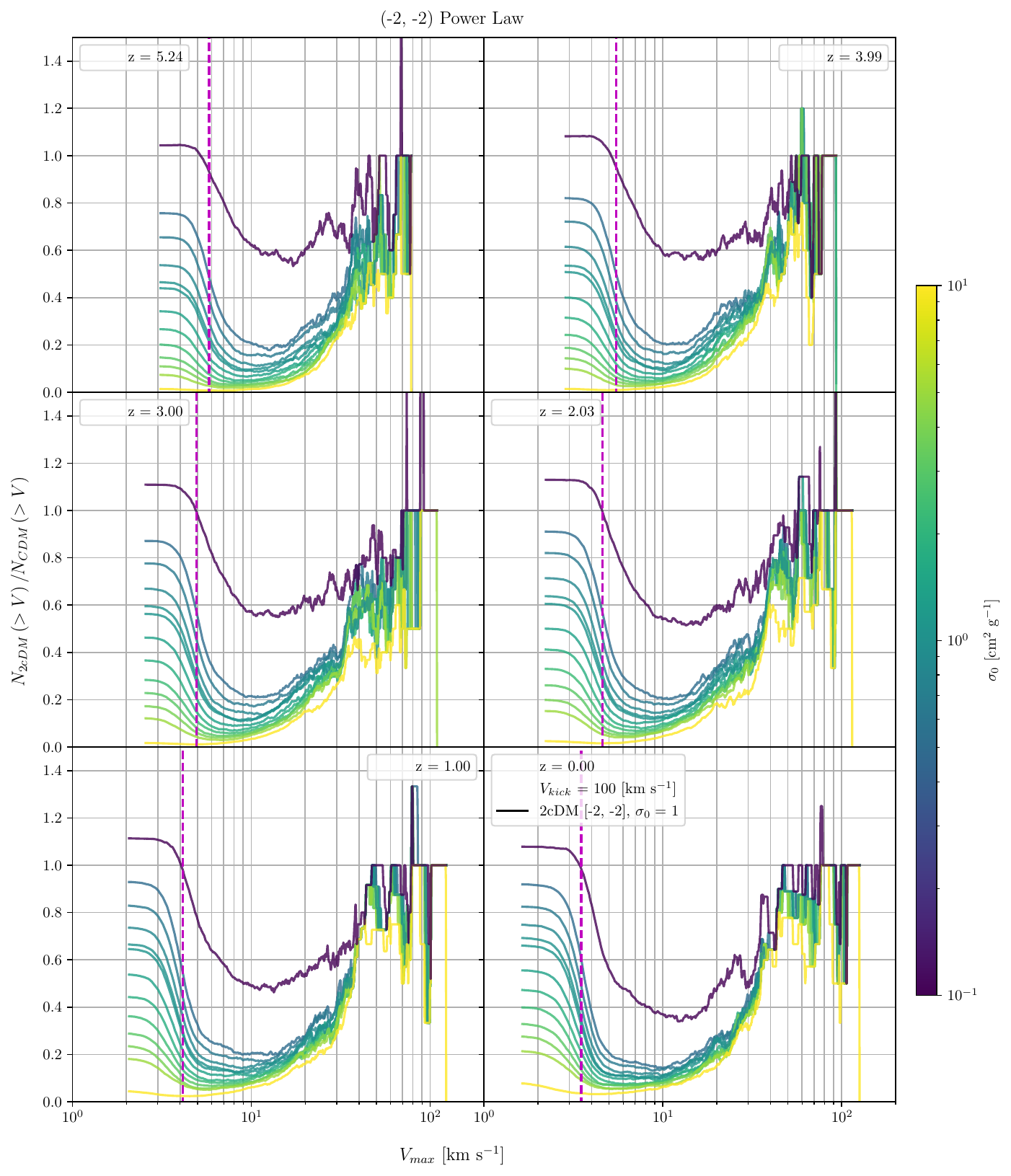}
\end{minipage}
\caption{The $V_{max}$ function for the $\sigma_0$ variation suite. Labelling 
and colouring are identical to Figure \ref{fig:sig-mass}. 
The same conclusion as in Figure \ref{fig:sig-mass} can be drawn here. The overall
shape of the suppression curve remains the same, just deepened.
\protect\label{fig:sig-vel}}
\end{figure*}

\begin{figure*}
\noindent\begin{minipage}[t]{1\columnwidth}
\includegraphics[width=1\columnwidth]{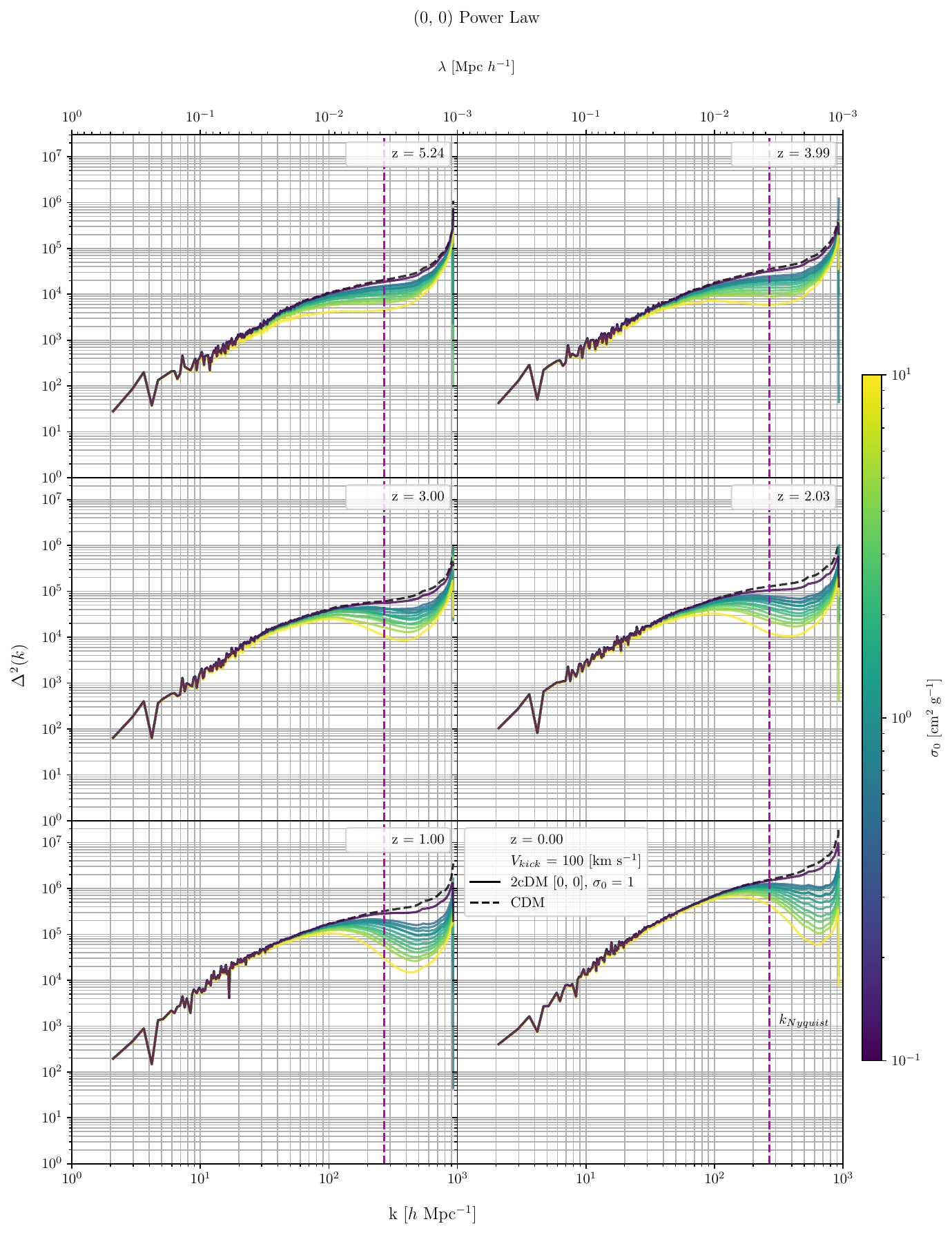}
\end{minipage}
\noindent\begin{minipage}[t]{1\columnwidth}
\includegraphics[width=1\columnwidth]{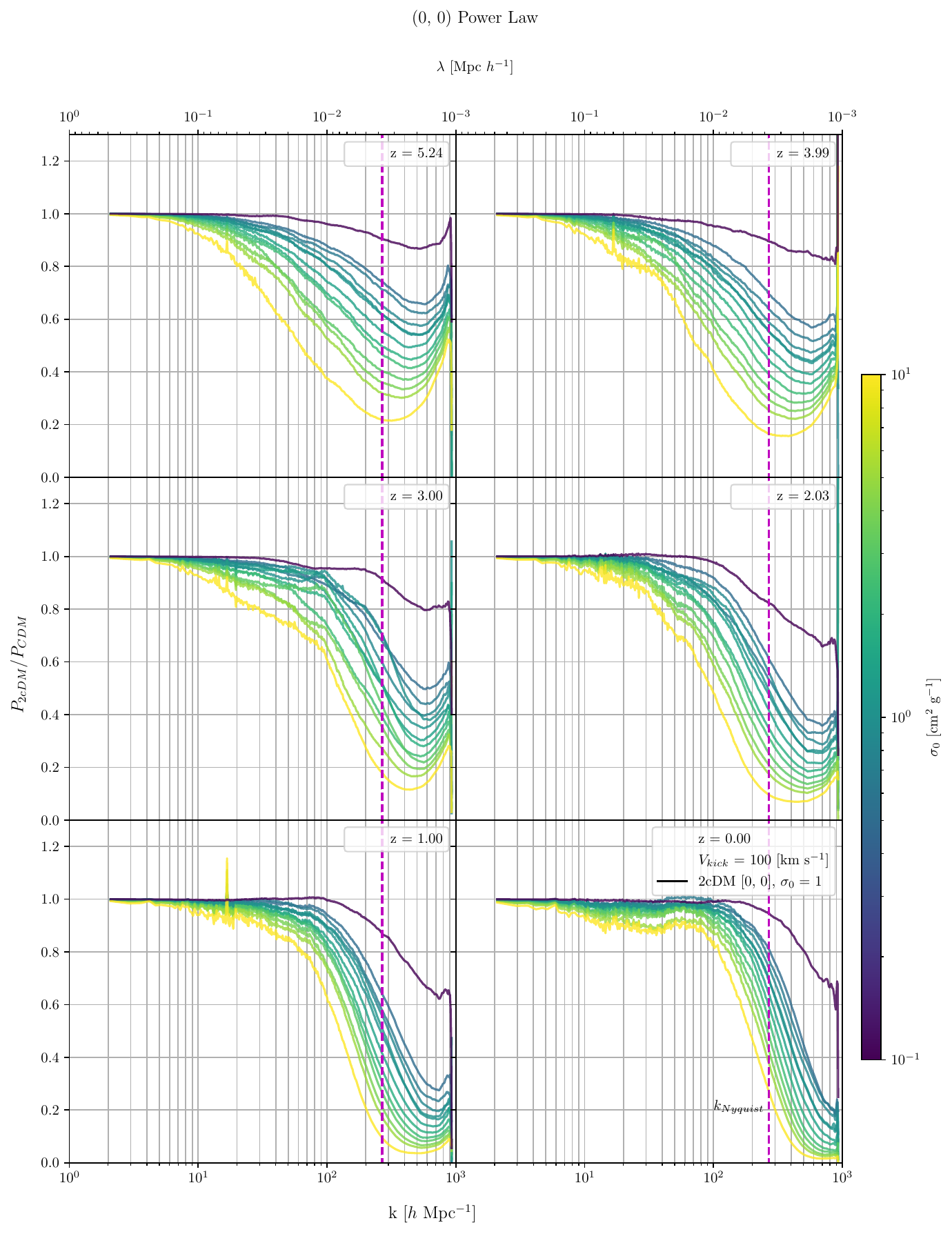}
\end{minipage}
\noindent\begin{minipage}[t]{1\columnwidth}
\includegraphics[width=1\columnwidth]{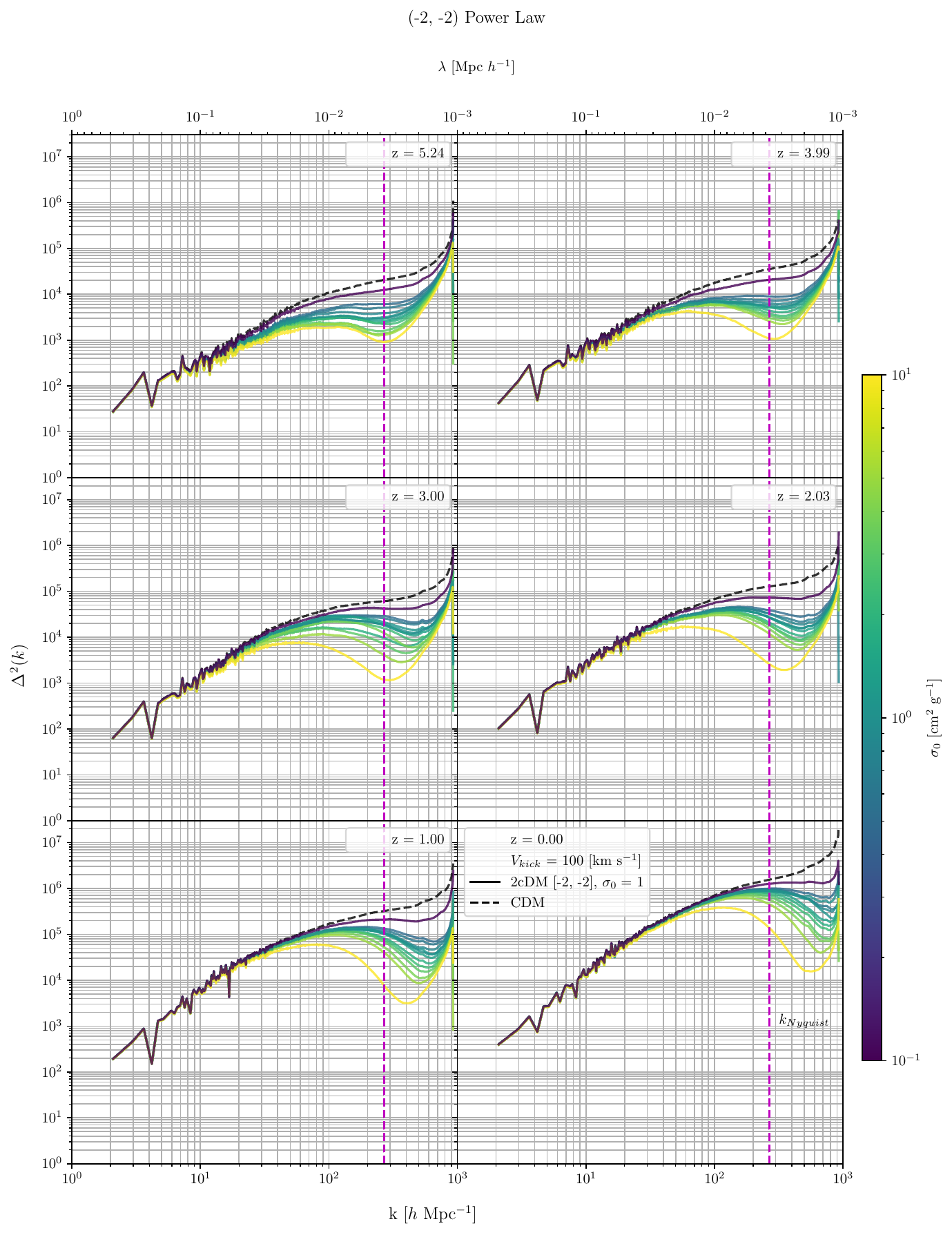}
\end{minipage}
\noindent\begin{minipage}[t]{1\columnwidth}
\includegraphics[width=1\columnwidth]{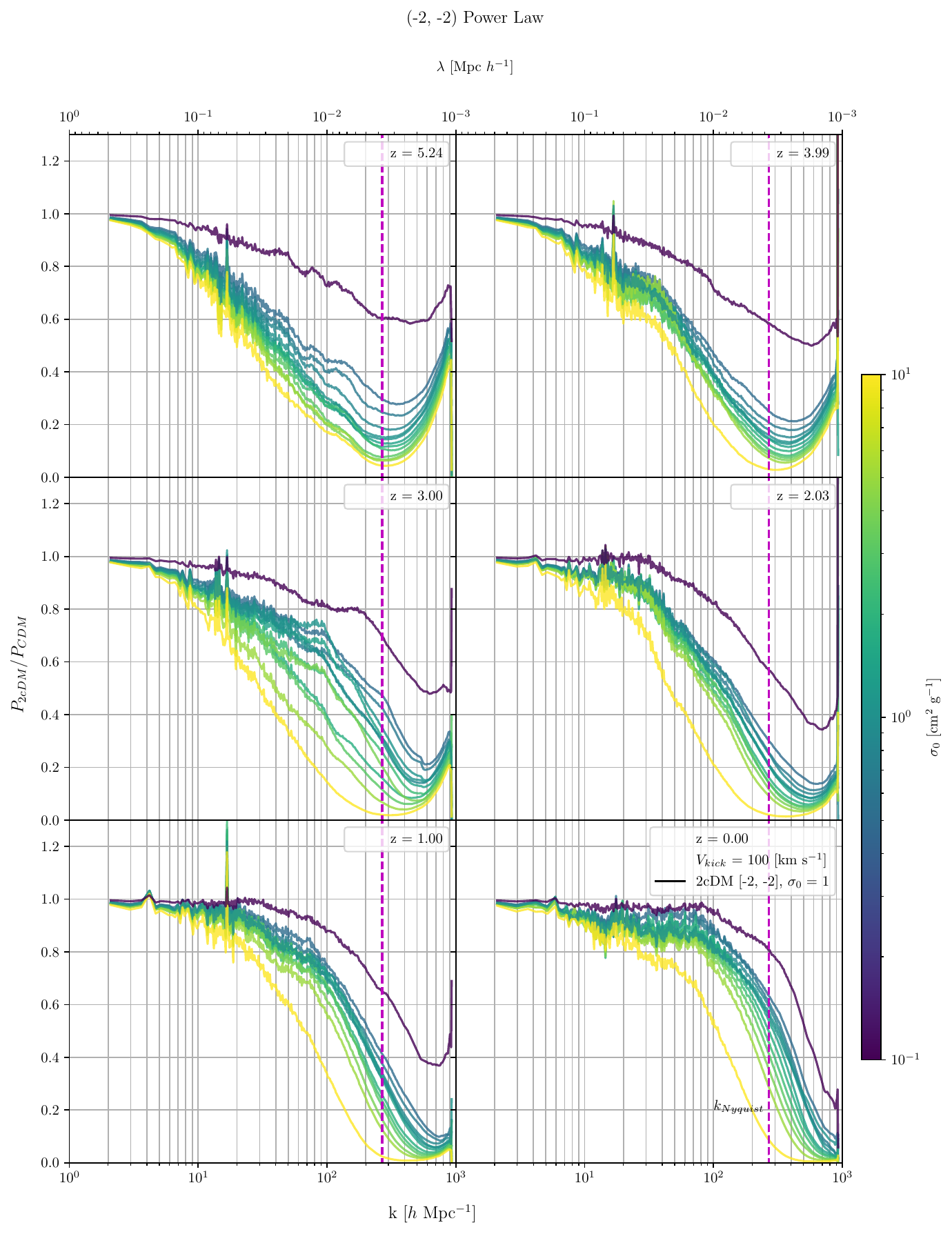}
\end{minipage}
\caption{The dimensionless power spectrum for the $\sigma_0$ variation suite. Labelling 
and colouring are identical to Figure \ref{fig:sig-mass}.
Vertical lines denote the Nyquist wavenumber for the simulations.
The \zz power law demonstrates that the location of the break with CDM
remains the same for all simulations and only the degree of suppression is
changed. It is more difficult to make this same conclusion for the \mm power law
because the break occurs at such large scale for $V_{kick}=100\kms$.
\protect\label{fig:sig-pk}}
\end{figure*}

We vary $\sigma_{0}$ between $0.5$ and $5\cmg$ in 10 logarithmic
steps keeping $V_{kick}$ fixed to $100\kms$. In addition to the
fiducial simulation with $\sigma_{0}=1\cmg$, as well as two additional
simulations with $\sigma_{0}=0.1,10\cmg$, we performed a total of
13 simulations per power law. 
The results are shown in Figures \ref{fig:sig-mass}-\ref{fig:sig-pk}. 

The HMF suppression is displayed in Figure \ref{fig:sig-mass}. With fixed $V_{kick}$,
we see that the scale at which suppression occurs remains consistent
between simulations and across redshift. Only the degree of suppression
increases with increasing $\sigma_{0}$. The \zz power law shows a consistent
amount of suppression peaking around $M_{halo}\sim10^9 \mathrm{M_{\odot}}$, with
a peak suppression of $\sim 40\%$. The \mm power law becomes exponentially
more collisional at small $V$ with increasing $\sigma_{0}$. In the most extreme
cases, there is a severe reduction in structure across all scales, due to the
difficulty in forming smaller seed structures to form larger ones. 
A similar story is shown
for the MCVF (Figure \ref{fig:sig-vel}) 
and $\Delta^{2}\left(k\right)$ (Figure \ref{fig:sig-pk}).

In all examples, the when compared to CDM the overall shape of each
curve remains similar, only deepening with increasing $\sigma_{0}$.
This makes intuitive sense. $\sigma_{0}$ is responsible for setting
the interaction rate, but $V_{kick}$ sets the amount of kinetic energy
injected, thus setting the scale at which structures become suppressed. 

\subsection{Hydrodynamical Simulations}

\begin{figure}
\includegraphics[width=1\columnwidth]{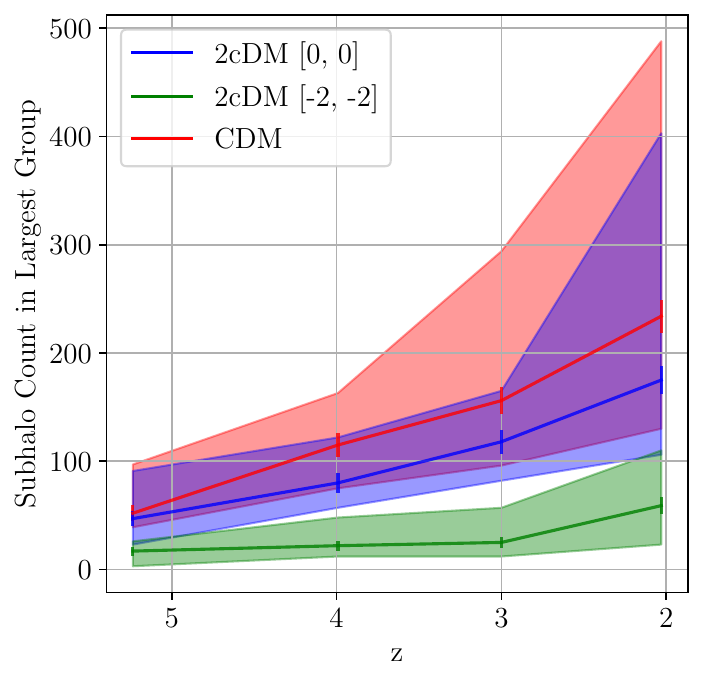}
\caption{Subhalo counts within the largest FoF group
for each simulation across redshift. 
Blue lines represent the \zz model,
green lines represent the \mm model, and red lines
represent CDM. Each line represents the average of 10 simulations.
Shaded regions denote the $10$th to $90$th percentile.
Error bars denote the Poisson counting error on the
halo counts.
Both \zz and \mm models form fewer subhalos than CDM,
the \mm model significantly so.
\protect\label{fig:var-halo-count}}
\end{figure}

\begin{figure*}
\noindent\begin{minipage}[t]{1\columnwidth}
\includegraphics[width=1\columnwidth]{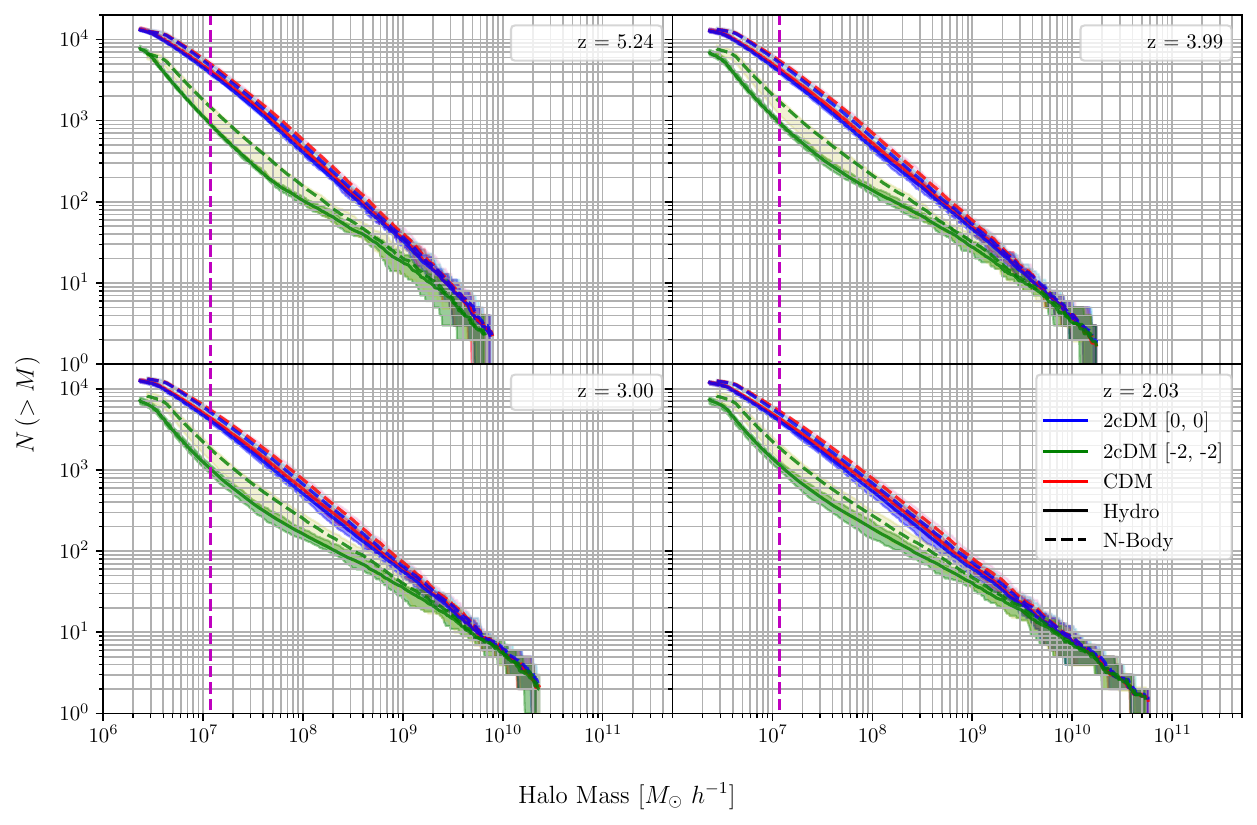}
\end{minipage}
\noindent\begin{minipage}[t]{1\columnwidth}
\includegraphics[width=1\columnwidth]{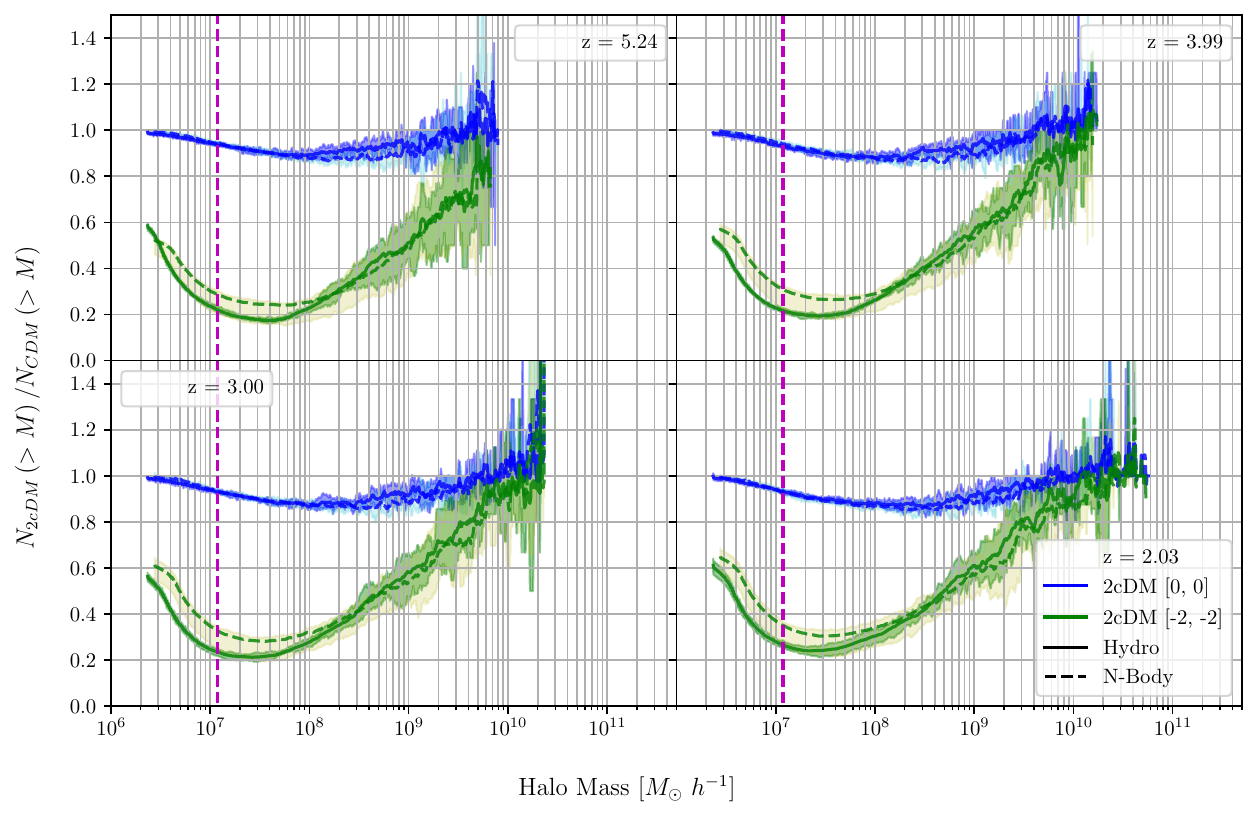}
\end{minipage}
\caption{(\textit{Left}) The halo mass functions for the fiducial suite of simulations
between $z\sim5-2$. 
Solid lines denote hydrodynamic simulations while dotted lines
denote DMO simulations. Blue lines represent the \zz model,
green lines represent the \mm model, and red lines
represent CDM. Each line represents the average of 10 simulations.
Shaded regions denote the $10$th to $90$th percentile.
Dashed vertical lines denote the mass resolution beneath which
numerical effects can dominate. We choose a cutoff of $100$ simulation particles.
Fiducial simulations have fixed 2cDM parameters $\sigma_0/m=1\cmg$ and $V_{kick}=100\kms$.
For clarity, shading for the DMO simulations are lighter than those for the 
hydrodynamical simulations.
The full DMO variations are shown in Figures \ref{fig:var-mass-dm} - \ref{fig:var-pk-dm}.
(\textit{Right}) Ratio of 2cDM halo mass functions to corresponding CDM halo mass functions.
Low mass structures are suppressed to a similar degree to that 
presented in Figures \ref{fig:vk-mass} and \ref{fig:sig-mass} for the given 2cDM parameters.
Baryons appear to provide slightly more suppression for the \mm power law, however
this is within the error bounds of the DMO simulations.
\protect\label{fig:var-mass-hy}}
\end{figure*}

\begin{figure*}
\noindent\begin{minipage}[t]{1\columnwidth}
\includegraphics[width=1\columnwidth]{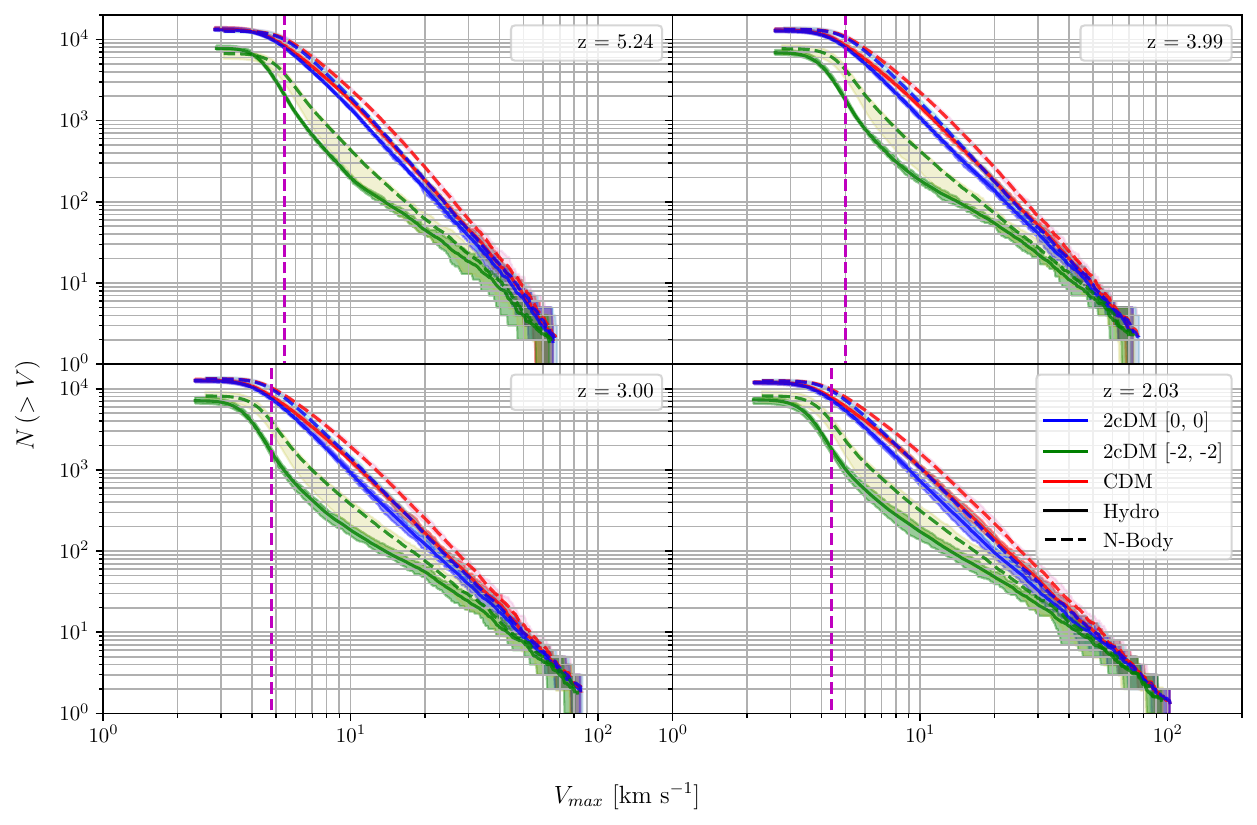}
\end{minipage}
\noindent\begin{minipage}[t]{1\columnwidth}
\includegraphics[width=1\columnwidth]{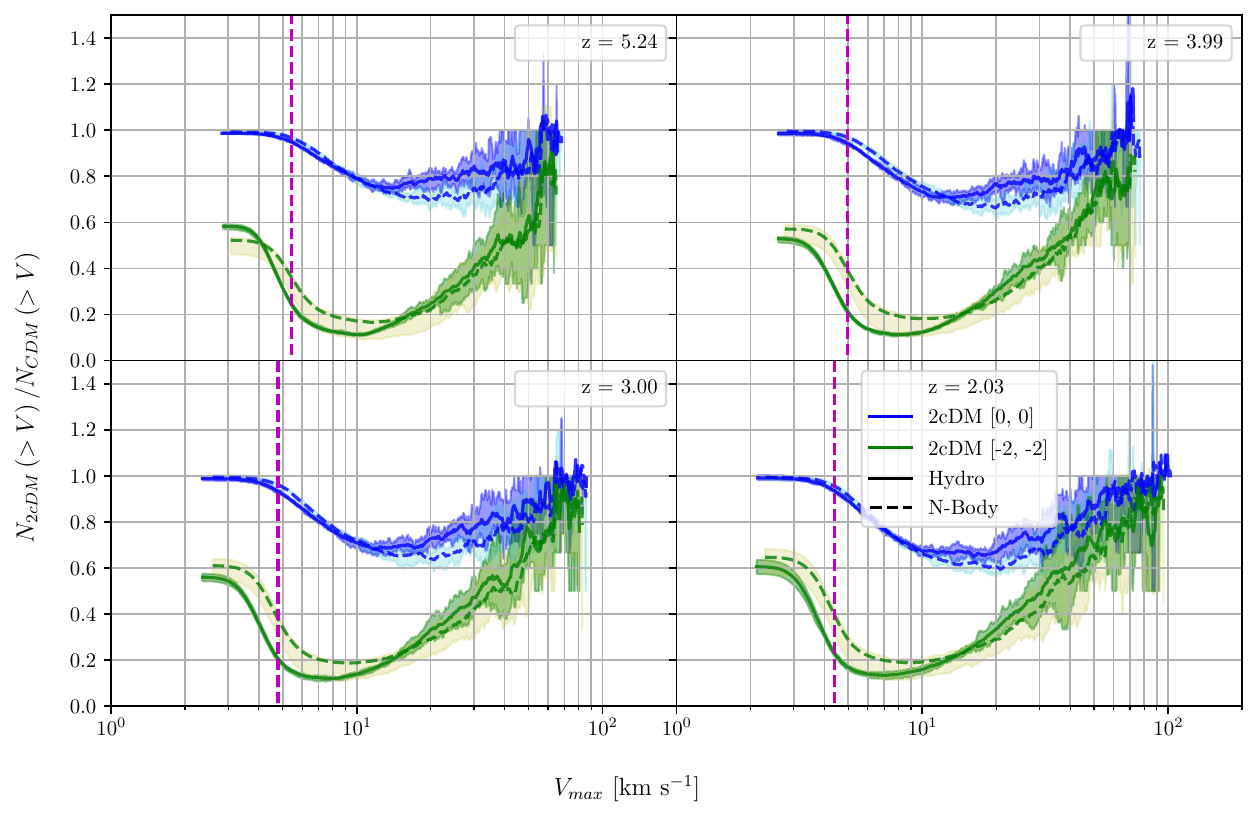}
\end{minipage}
\caption{The $V_{max}$ functions for the fiducial suite of simulations.
Colouring is identical to Figure \ref{fig:var-mass-hy}.
Low mass structures are suppressed to a similar degree to that 
presented in Figures \ref{fig:vk-vel} and \ref{fig:sig-vel} for the given 2cDM parameters.
Baryons provide significantly more suppression to systems with $V_{max}\lesssim10\kms$
for the \mm power law.
\protect\label{fig:var-vel-hy}}
\end{figure*}

\begin{figure*}
\noindent\begin{minipage}[t]{1\columnwidth}
\includegraphics[width=1\columnwidth]{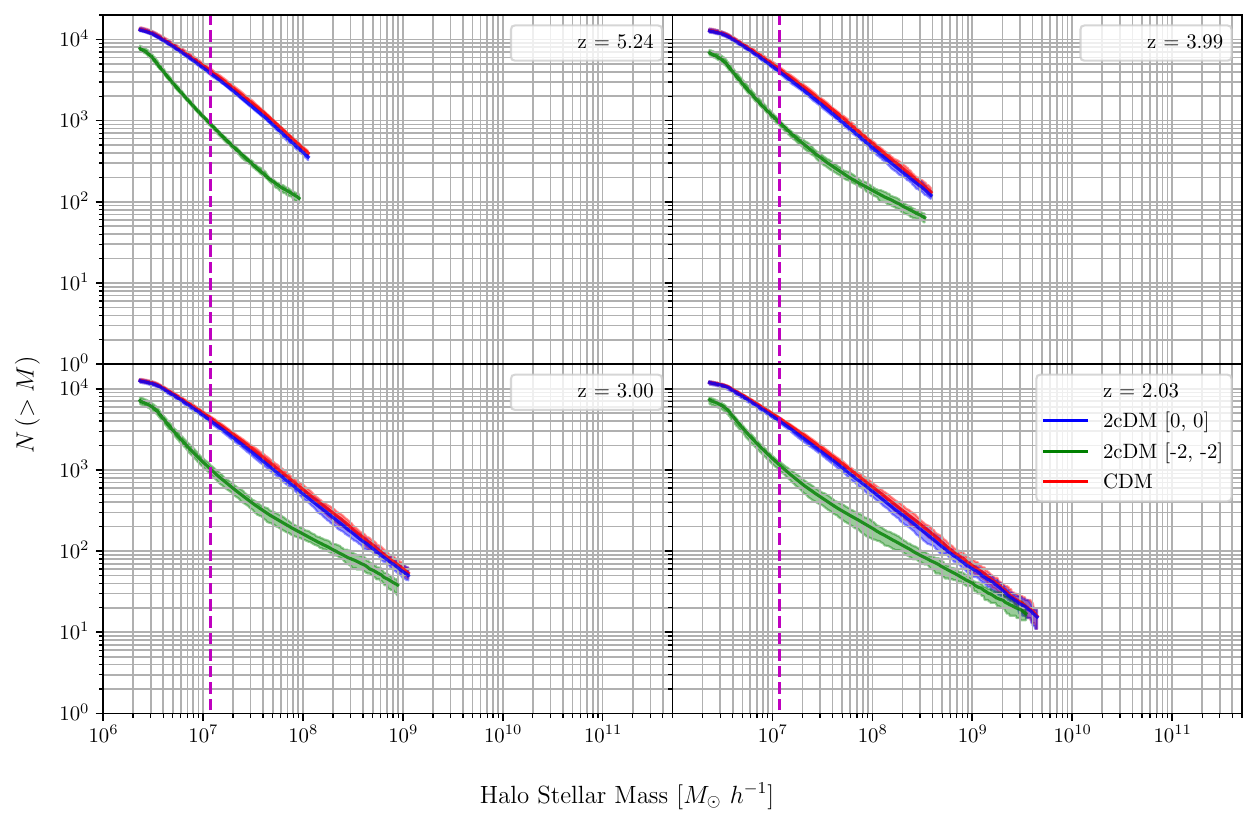}
\end{minipage}
\noindent\begin{minipage}[t]{1\columnwidth}
\includegraphics[width=1\columnwidth]{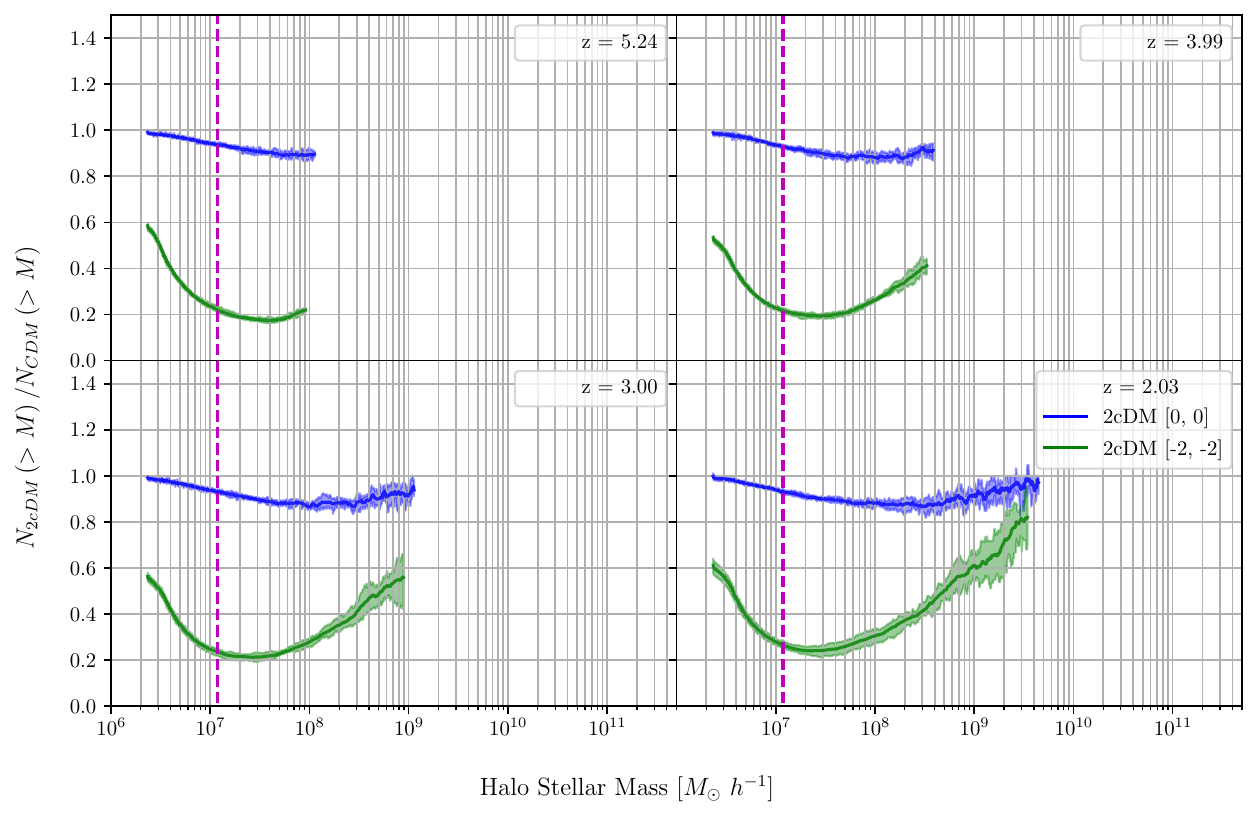}
\end{minipage}
\caption{The stellar halo mass functions for the fiducial suite of simulations
between $z\sim5-2$. 
Colouring is identical to Figure \ref{fig:var-mass-hy}.
The suppression in stellar mass follows the same trend
as the suppression in total halo mass.
In particular, the \mm model produces halos with up to $\sim80\%$
suppression in stellar mass relative to CDM.
\protect\label{fig:var-mass-stellar}}
\end{figure*}

\begin{figure*}
\noindent\begin{minipage}[t]{1\columnwidth}
\includegraphics[width=1\columnwidth]{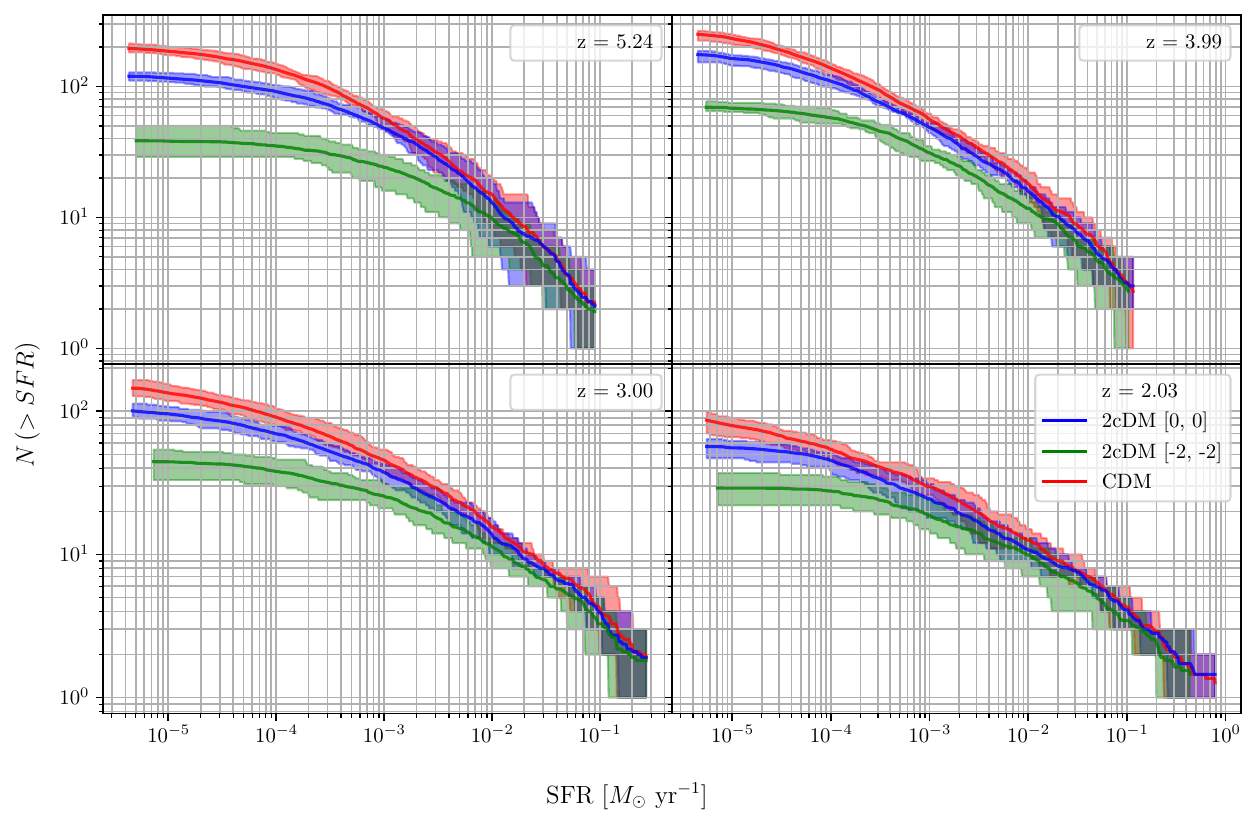}
\end{minipage}
\noindent\begin{minipage}[t]{1\columnwidth}
\includegraphics[width=1\columnwidth]{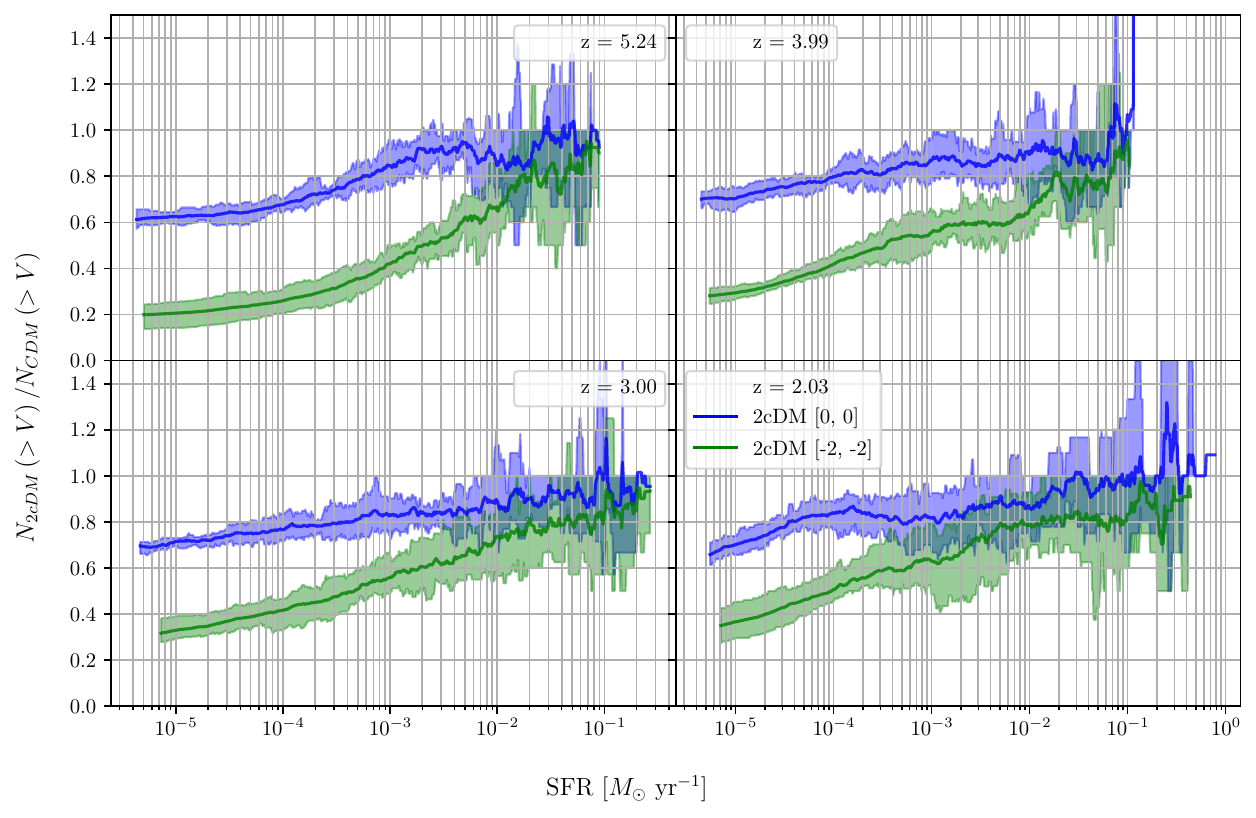}
\end{minipage}\caption{
(\textit{Left})
The distribution of SFRs for star-forming halos observed in the simulations.
Colouring is identical to Figure \ref{fig:var-mass-hy}.
Shading denotes the $10$th to $90$th percentile.
At high $z$, the \mm model produces significantly fewer star-forming halos
than either the \zz model or CDM.
(\textit{Right})
The ratio of the SFR distributions between 2cDM and CDM simulations.
\protect\label{fig:var-SFR-distribution}}
\end{figure*}

\begin{figure*}
\includegraphics[width=1\textwidth]{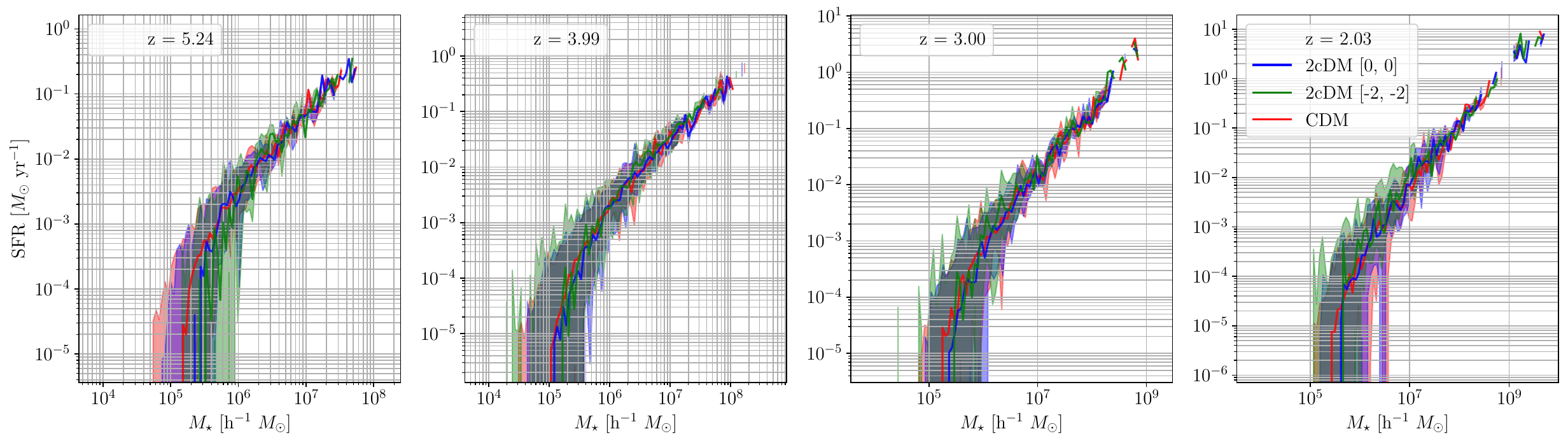}
\caption{The stellar mass - SFR relationship in the fiducial suite of simulations.
Colouring is identical to Figure \ref{fig:var-mass-hy}.
Shading indicates the $10$th to $90$th percentile.
All three simulation sets lie on top of each other up to the observed
scatter. It appears that the modified DM physics does not
affect the stellar mass - SFR relationship.
\protect\label{fig:var-SFR-mass}}
\end{figure*}

\begin{figure*}
\noindent\begin{minipage}[t]{1\columnwidth}
\includegraphics[width=1\columnwidth]{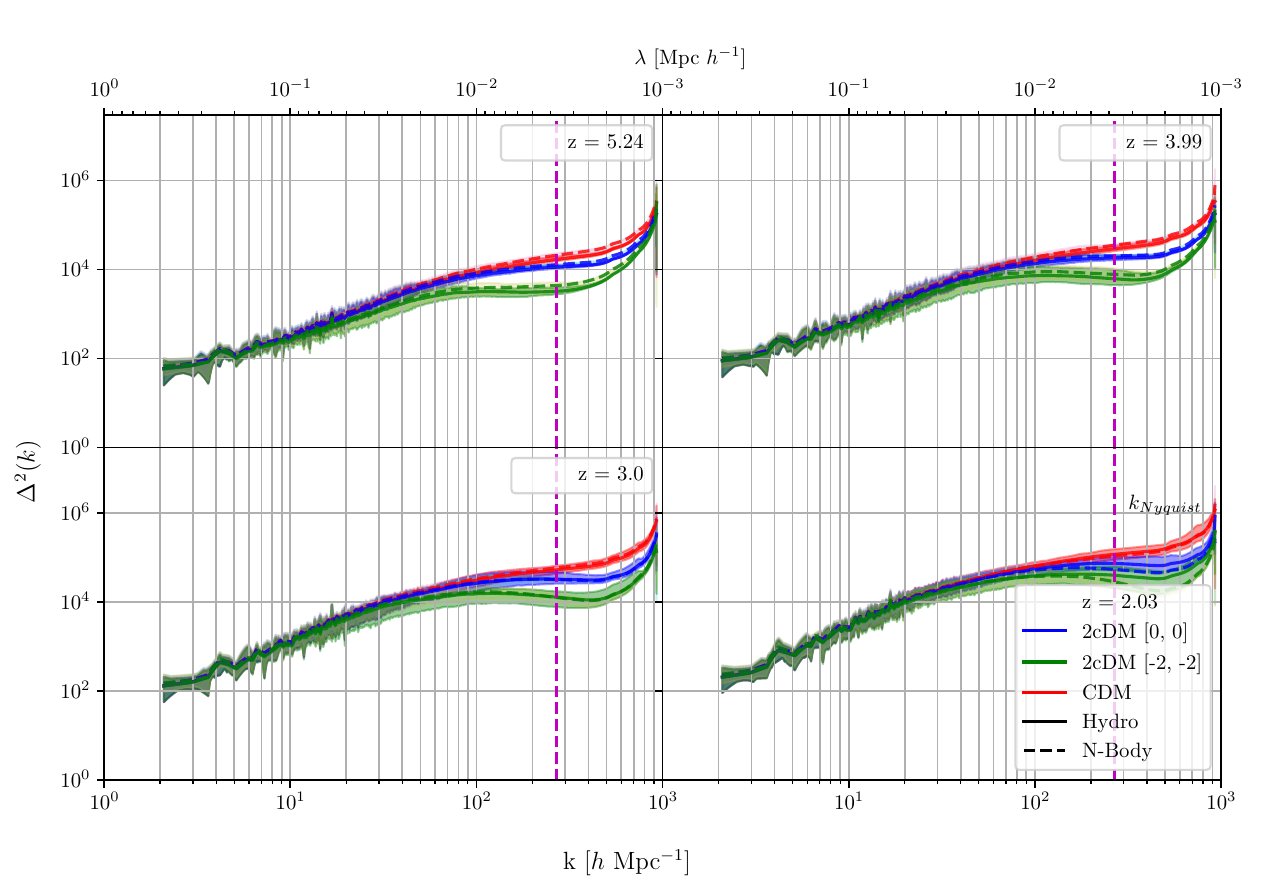}
\end{minipage}
\noindent\begin{minipage}[t]{1\columnwidth}
\includegraphics[width=1\columnwidth]{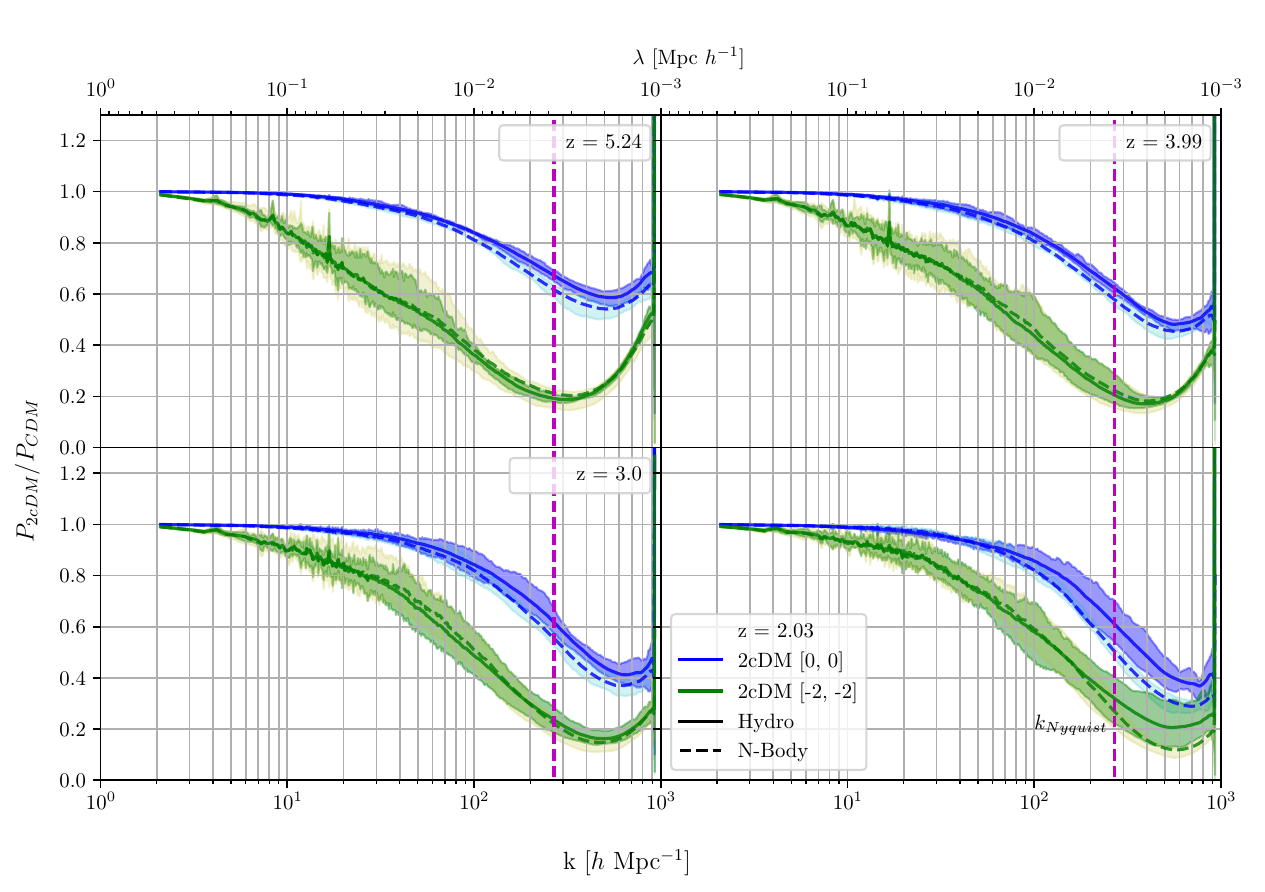}
\end{minipage}
\caption{The dimensionless power spectra for the fiducial suite of simulations.
Vertical lines denote the Nyquist wavenumber for the simulations. 
Colouring is identical to Figure \ref{fig:var-mass-hy}.
Both power laws exhibit significant suppression relative to CDM
at small scales across all redshifts. Results between $N$-body
and hydrodynamic simulations are similar, with the \zz power law
perhaps showing some enhancement relative to its $N$-body counterpart.
\protect\label{fig:var-pk-hy}}
\end{figure*}

\begin{figure}
\includegraphics[width=1\columnwidth]{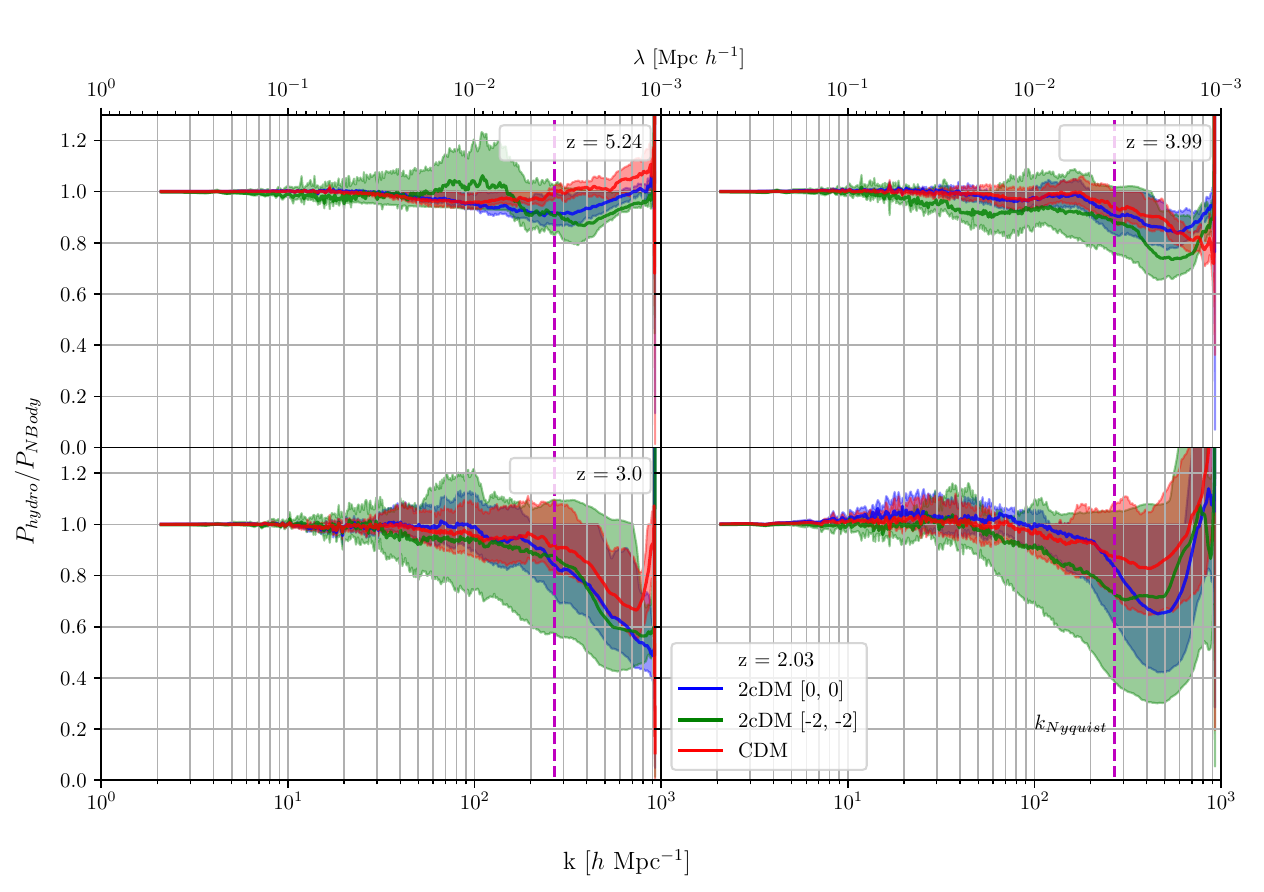}
\caption{The ratio between power spectra for hydrodynamic and $N$-body simulations.
Vertical lines denote the Nyquist wavenumber for the simulations. 
Colouring is identical to Figure \ref{fig:var-mass-hy}.
Deviations from unity indicate hydrodynamic simulations producing
different amounts of structure compared to $N$-body counterparts.
Shaded regions denote the $10$th to $90$th percentile.
Within the error bounds, all hydrodynamic simulations
produce similar levels of suppression at small scales,
though the deviation from unity is insignificant.
Suppression between CDM and 2cDM simulations can therefore be mainly attributed to the
modified DM physics.
\protect\label{fig:var-pk-nbody}}
\end{figure}

\begin{figure*}
\includegraphics[width=1\textwidth]{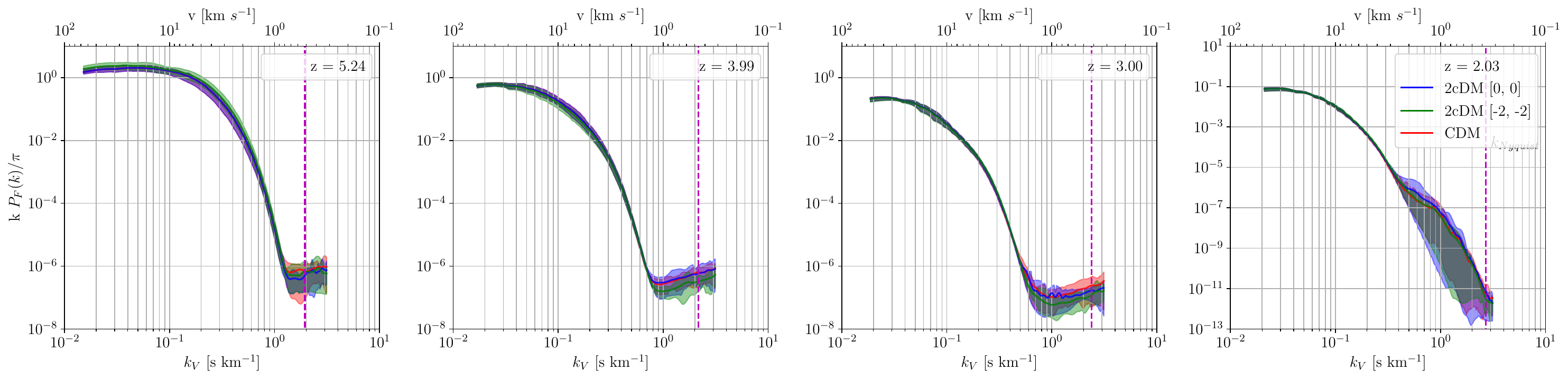}
\includegraphics[width=1\textwidth]{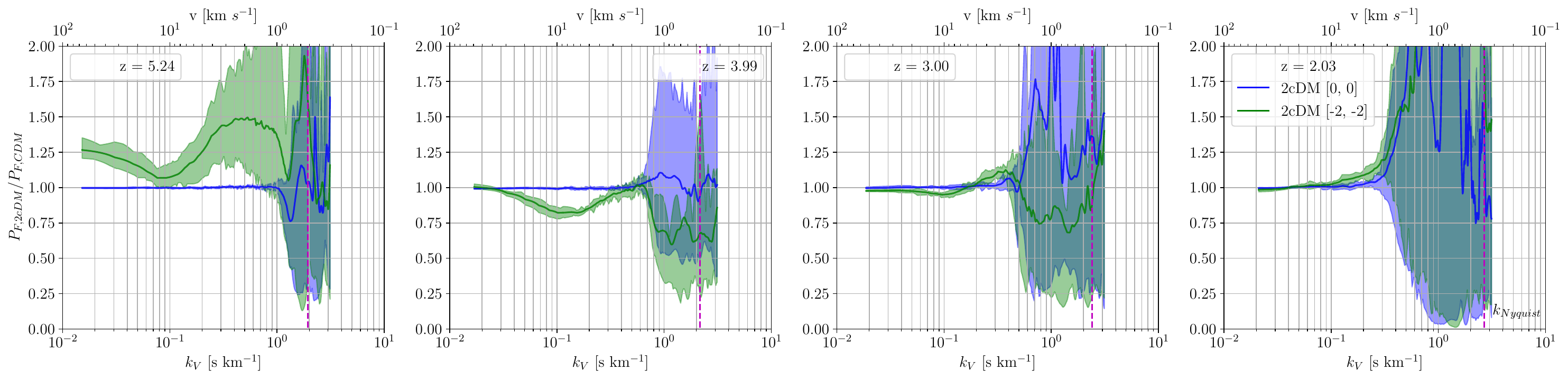}
\caption{
P1D generated from 
Lyman-$\alpha$ absorption in the hydrodynamic simulations.
Blue lines represent the \zz model,
green lines represent the \mm model, and red lines
represent CDM.
Shaded regions denote the $10$th to $90$th percentile. 
Vertical lines denote the velocity space Nyquist wavenumber for the simulations.
(\textit{Top}): The calculated P1D for each simulation type.
At small scales, the P1D are dominated by simulation-to-simulation variation,
which only grows at lower redshifts.
(\textit{Bottom}): The ratio of 2cDM P1D to CDM P1D.
No strong conclusions can be made about small scale suppression
for either model due to the large scatter.
\protect\label{fig:var-lyman}}
\end{figure*}

To demonstrate the resilience of results to cosmic variation, we perform
10 hydrodynamical simulations per power law with different initial
conditions. In addition to CDM and DMO counterparts, we performed
a total of 60 simulations. As a reminder, these fiducial simulations
use fixed 2cDM parameters of $\sigma_{0}=1\cmg$ and $V_{kick}=100\kms$.
We present our results in Figures \ref{fig:var-halo-count}-\ref{fig:var-lyman}.

To give a picture of how much substructure forms in the simulation,
we plot the number of subhalos formed within the largest FoF group
in Figure \ref{fig:var-halo-count}.
We remind that the subhalos identified by \textsc{Subfind}
typically correspond to galaxies hosted within a DM halo.
Both 2cDM models suppress the halo count relative to CDM,
even in the presence of baryons.
The effect is better seen in the other summary statistics.

Because the HMF (Figure \ref{fig:var-mass-hy}) and MCVF (Figure \ref{fig:var-vel-hy}) are 
cumulative distributions, we can directly compare between hydrodynamic and DMO
results. In both metrics, the \zz power law shows remarkably similar
results between hydrodynamic and DMO simulations.
With our sample size, we can say the results are consistent
up to cosmic variance.
The \mm power law have additional
$10\%$ suppression at scales $M_{halo}\lesssim10^{8}\,\mathrm{M_{\odot}}$
or $V_{max}\lesssim10\kms$ compared to the DMO simulations.
This difference is a most likely from baryonic feedback,
as it appears outside the variance bounds of the DMO simulation.

For hydrodynamic simulations, we calculate the HMF
for stellar mass (Figure \ref{fig:var-mass-stellar}).
The suppression in the halo count is similar to the suppression
for the full HMF. We highlight that star formation appears
to be severely suppressed in the \mm model
across all redshift, with
suppression up to $\sim80\%$ relative to CDM.

This suppression of star formation is also seen
in the SFR distribution (Figure \ref{fig:var-SFR-distribution}).
We remind that in these statistics, we only count
subhalos that have formed stars, that is with
nonzero stellar mass.
In both \zz and \mm models, the number of low SFR
systems are reduced relative to CDM before converging to CDM
levels at high SFR.
The \mm model much more severely affects star-forming halos, 
as the total number of star-forming halos is reduced up to
$\sim40\%$ relative to CDM at $z=3$.
Both models form fewer total star-forming halos than CDM,
which is consistent with 2cDM reducing the number
of low-mass halos.
In Figure \ref{fig:var-SFR-mass}, we plot the stellar mass - SFR relationship
with halos binned by stellar mass into $100$ logarithmically spaced bins.
The distributions for all simulations appear to lie on top of each other,
indicating that the modified DM does not significantly
affect the stellar mass - SFR relationship.

We plot the power spectra for the fiducial simulations in Figure \ref{fig:var-pk-hy}.
There is differences between the hydrodynamic and DMO suites for both power laws
and across all scales up to cosmic variance. Intuitively, this reflects the
reduced role baryonic feedback plays at high redshift.
We can further demonstrate that the effects are mainly due to the modified
dark physics.
In Figure \ref{fig:var-pk-nbody} we plot the ratio between the hydrodynamic and DMO
power spectra. In this comparison, effects from baryons are generally smaller, less than $20\%$ 
suppression at the Nyquist level, and are less consistent, seen in the much wider spread.

We present our calculations of the Lyman-$\alpha$ P1D in Figure \ref{fig:var-lyman}.
The \zz model does not significantly
affect the P1D relative to CDM at large scales.
At the same scales, the \mm model exhibits much larger variation
and can change between enhancement and suppression relative to CDM.
At small scales, the P1D are dominated by simulation-to-simulation variation.
These variations are emphasized in the ratios with CDM P1D.
As such, we cannot draw any strong conclusions about the effect of
2cDM on the P1D without performing further simulations.

\begin{figure*}
\noindent\begin{minipage}[t]{1\columnwidth}
\includegraphics[width=1\columnwidth]{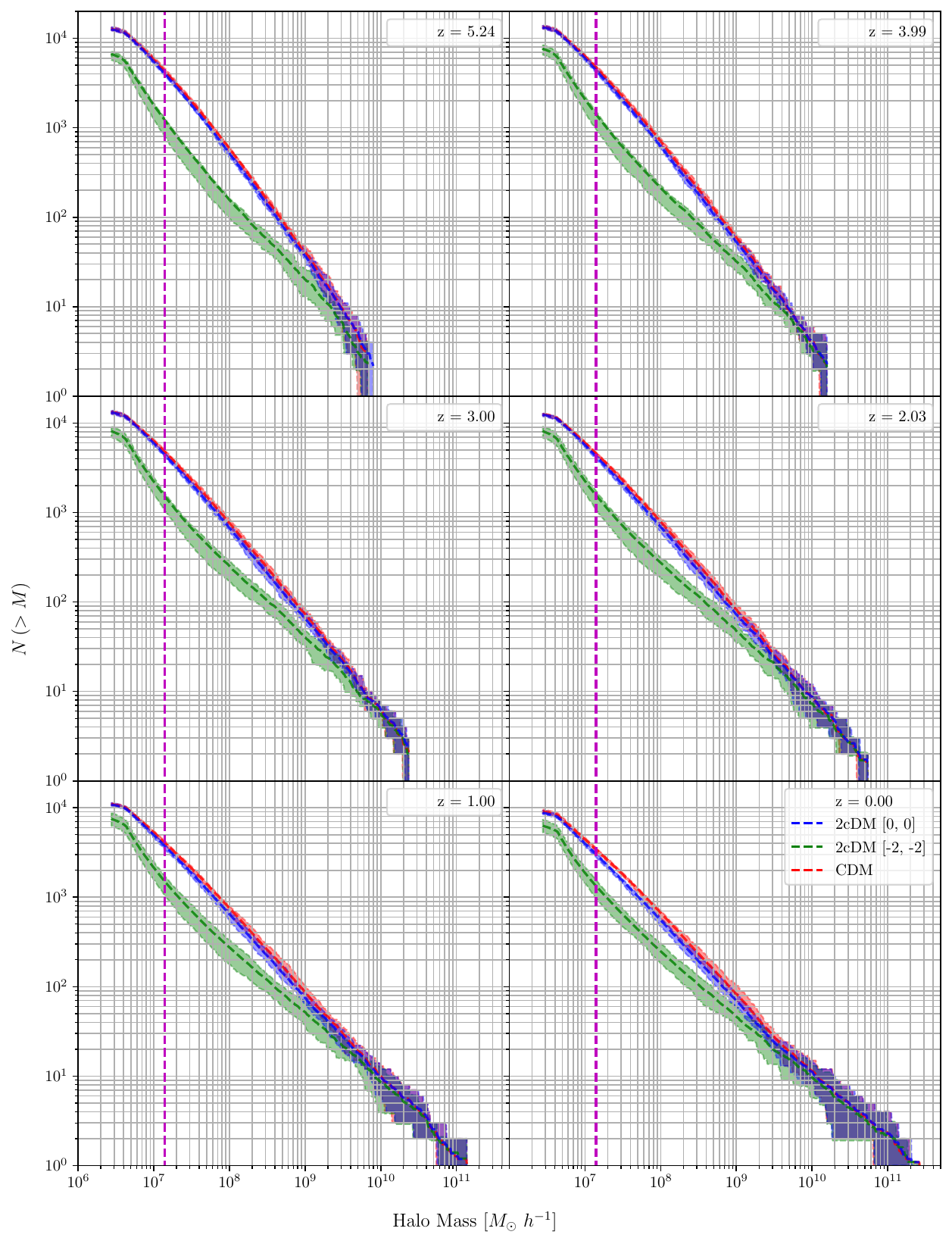}
\end{minipage}
\noindent\begin{minipage}[t]{1\columnwidth}
\includegraphics[width=1\columnwidth]{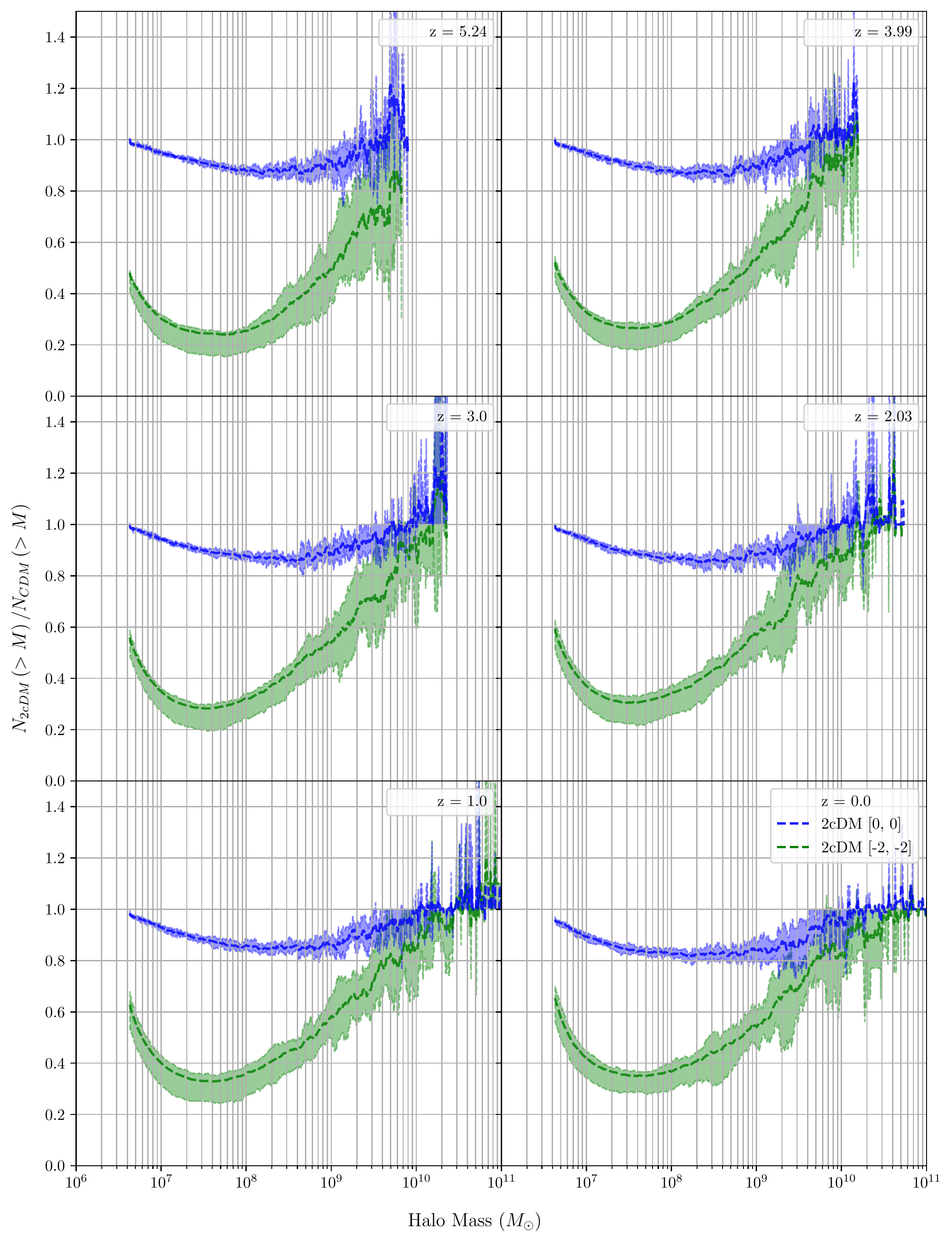}
\end{minipage}
\caption{(\textit{Left}) The halo mass function for the DMO suite of simulations
from $z\sim5-0$.
Dotted lines denote that these are DMO simulations. 
As with Figures \ref{fig:var-mass-hy}-\ref{fig:var-pk-hy},
blue lines represent the $\left(0,0\right)$ model,
green lines represent the $\left(-2,-2\right)$ model, and red lines
represent CDM. Each line represents the average of 10 simulations.
Shaded regions denote the $10$th to $90$th percentile.
Dashed vertical lines denote the mass resolution beneath which
numerical effects can dominate. We choose a cutoff of $100$ simulation particles.
Fiducial simulations have fixed 2cDM parameters $\sigma_0/m=1\cmg$ and $V_{kick}=100\kms$.
(\textit{Right}) Ratio of 2cDM halo mass functions to corresponding CDM halo mass functions.
Simulations with the \zz power law show low variation across all redshift while
the \mm power law shows variation of $\sim10\%$, biased towards more suppression.
Both power laws produce results consistent with the single parameter variations
presented in Section \ref{subsec:parameter-space}.
\protect\label{fig:var-mass-dm}}
\end{figure*}

\begin{figure*}
\noindent\begin{minipage}[t]{1\columnwidth}
\includegraphics[width=1\columnwidth]{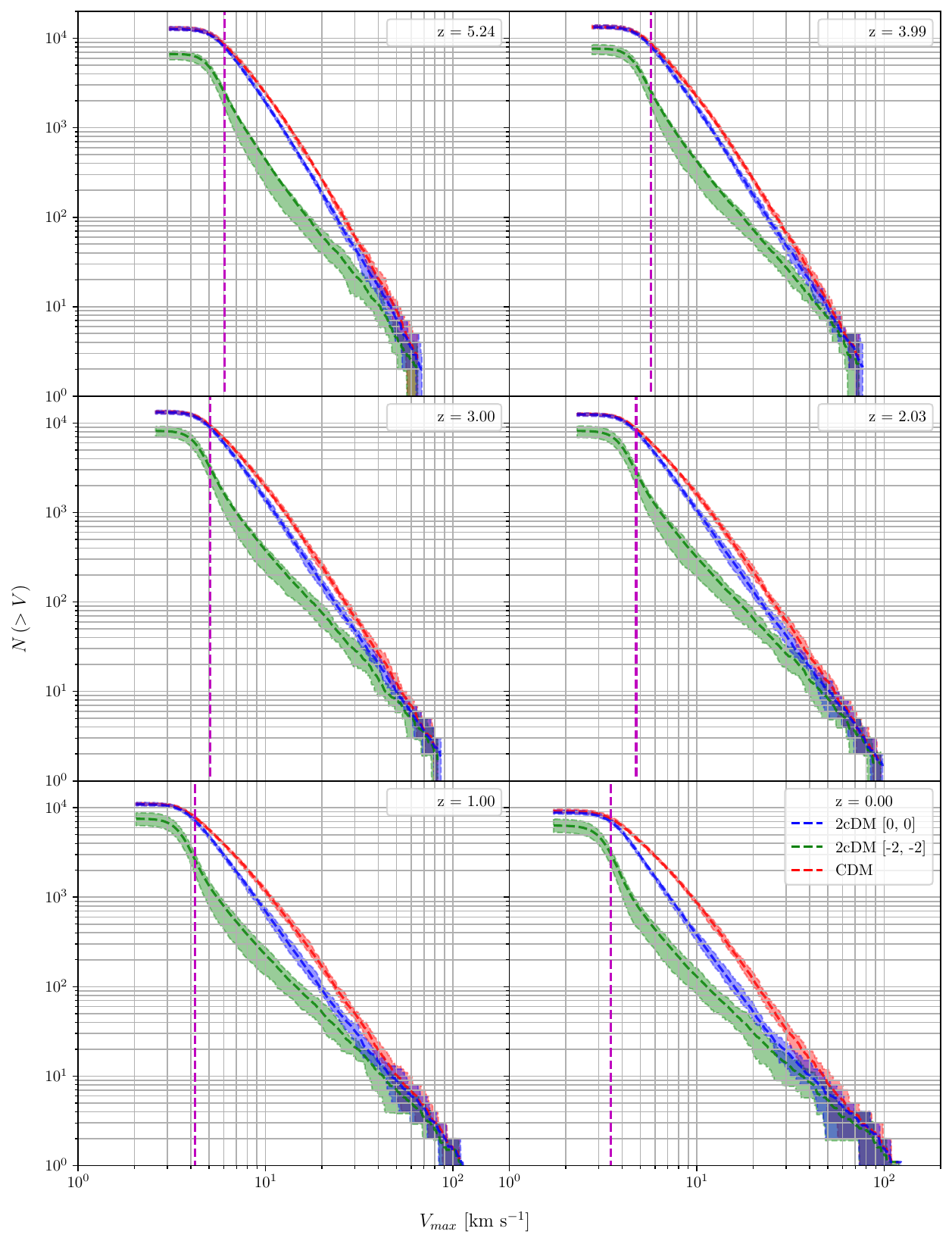}
\end{minipage}
\noindent\begin{minipage}[t]{1\columnwidth}
\includegraphics[width=1\columnwidth]{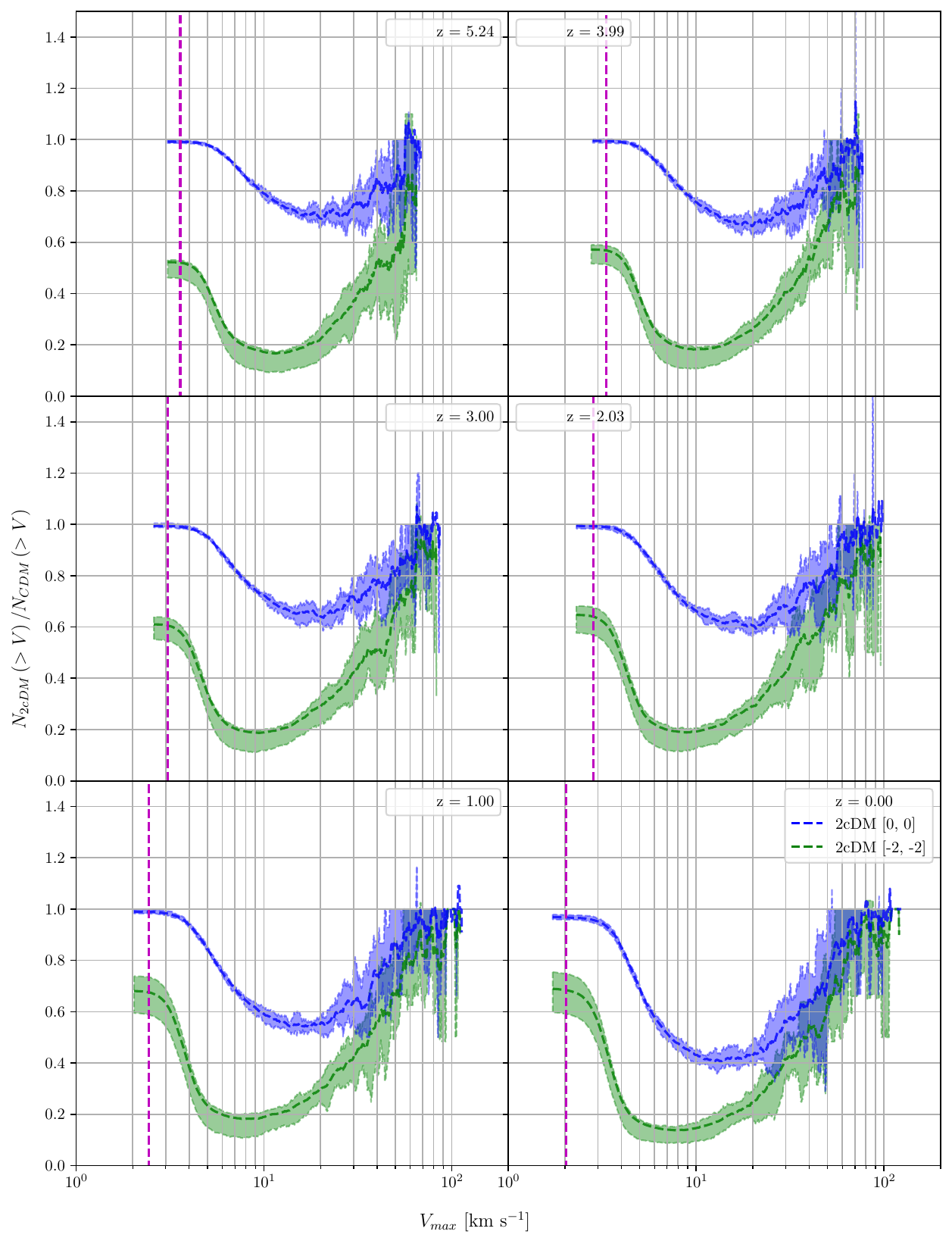}
\end{minipage}
\caption{The $V_{max}$ functions for the DMO suite of simulations. 
The variation in the $V_{max}$ functions is generally less than that of the
halo mass functions. 
Both power laws produce results consistent with the single parameter variations
presented in Section \ref{subsec:parameter-space}.
\protect\label{fig:var-vel-dm}}
\end{figure*}

\begin{figure*}
\noindent\begin{minipage}[t]{1\columnwidth}
\includegraphics[width=1\columnwidth]{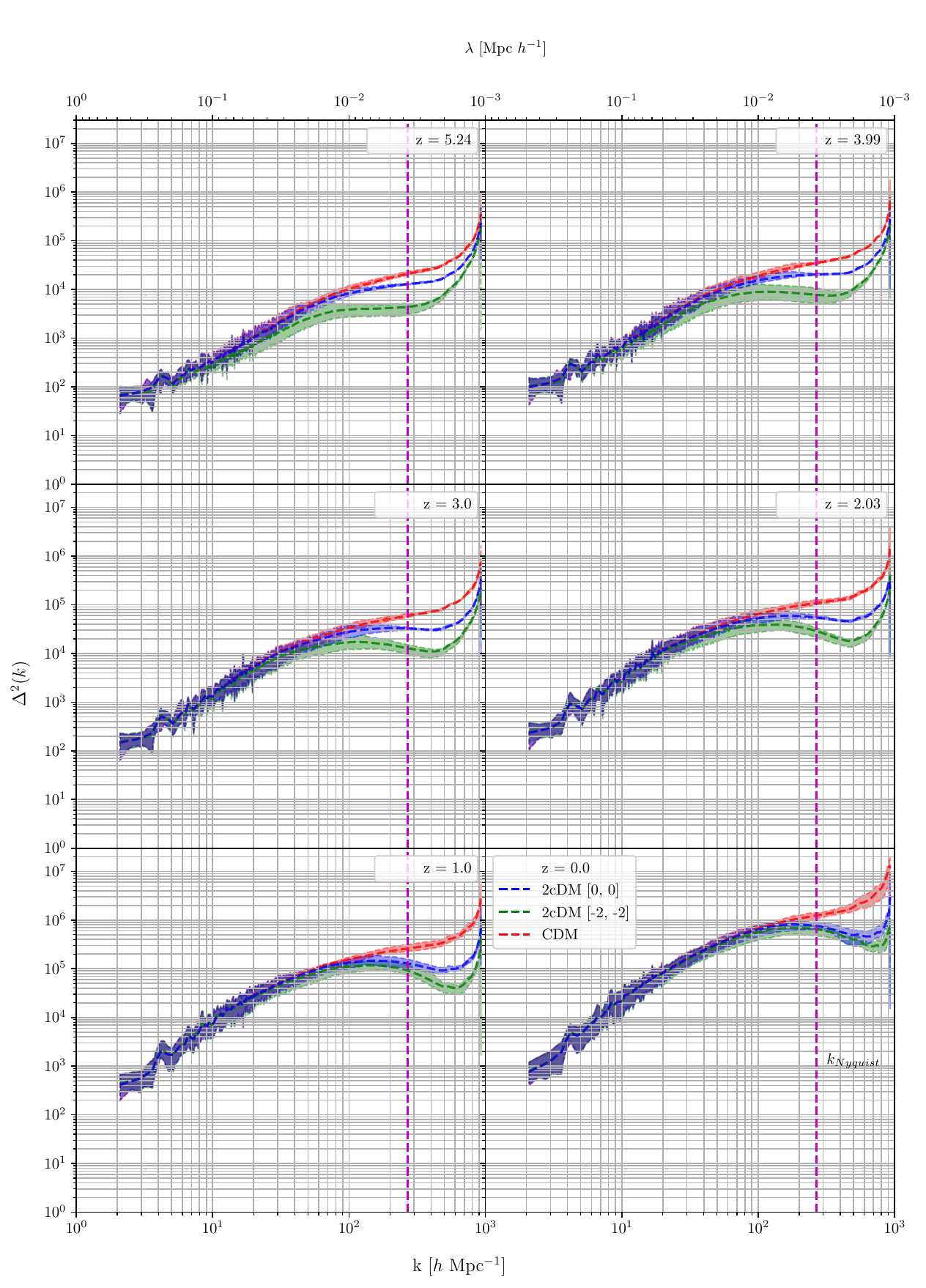}
\end{minipage}
\noindent\begin{minipage}[t]{1\columnwidth}
\includegraphics[width=1\columnwidth]{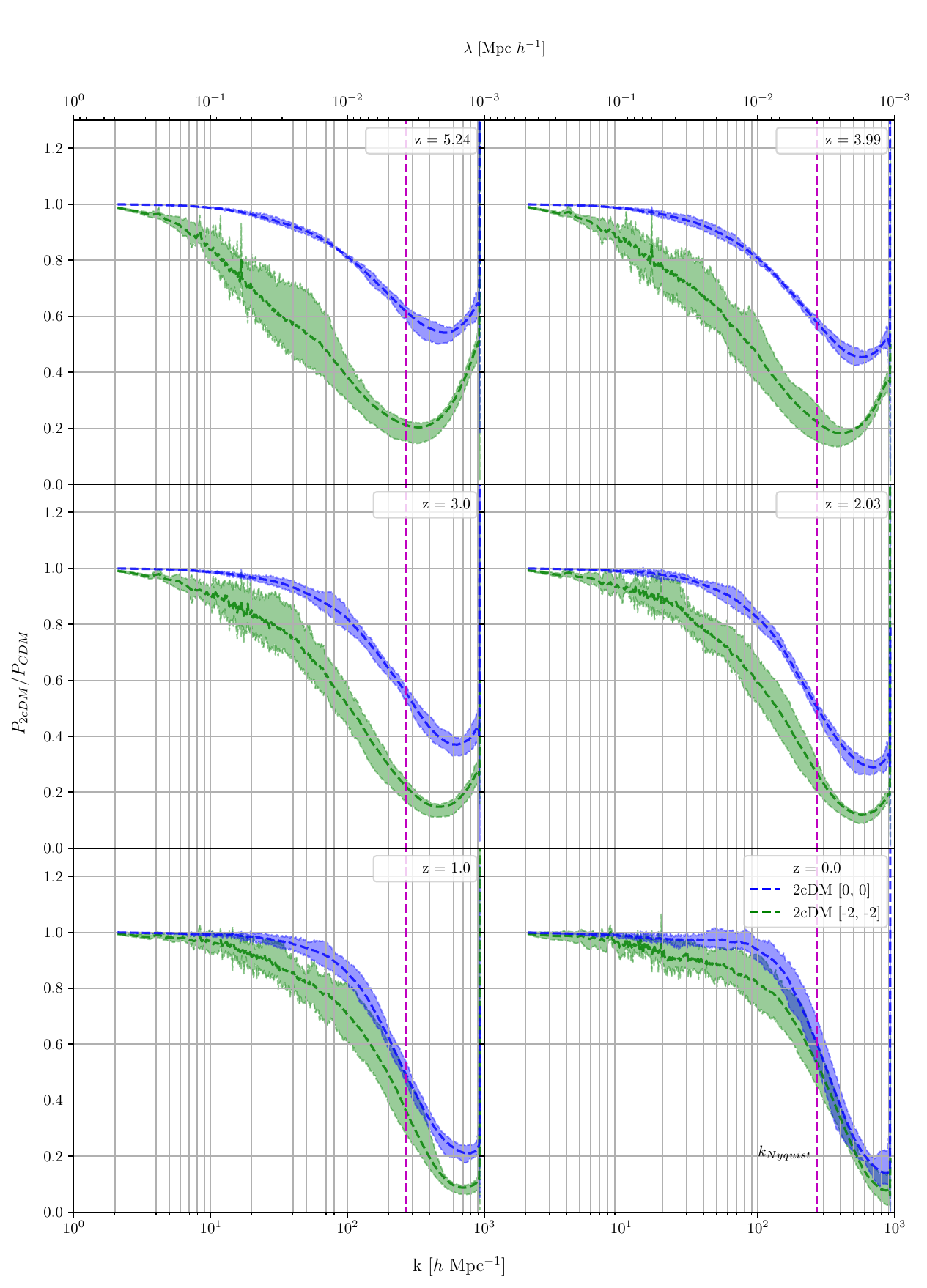}
\end{minipage}
\caption{The dimensionless power spectra for the DMO suite of simulations. Vertical lines denote the Nyquist wavenumber for the simulations.
Results as $z\rightarrow0$ should not be expected to hold in the hydrodynamical case,
as baryonic feedback becomes much more significant.
Both power laws produce results consistent with the single parameter variations
presented in Section \ref{subsec:parameter-space}.
\protect\label{fig:var-pk-dm}}
\end{figure*}

For completeness, we plot the analysed metrics for the DMO suites in Figures
\ref{fig:var-mass-dm}-\ref{fig:var-pk-dm}.
In DMO simulations, the suppression continues to $z=0$ with relatively low spread.
This is consistent with results reported by \cite{TodorokiI}, though extends
the analysis to a larger parameter range for the selected power laws.
We anticipate that the low redshift results will not hold in 
hydrodynamic simulations, as the effects of baryonic feedback 
become larger in this regime.
\section{Discussion\protect}
\label{sec:Discussion}

\subsection{Degeneracies Between 2cDM and Baryonic Effects}
Baryonic feedback is known to generally suppress power at scales $k\lesssim2 \text{Mpc }h^{-1}$
\citep{vanDaalen2011, Chisari2019, Schneider2019, vanDaalen2020}, 
and the effect is now well quantified across a range of cosmological and
astrophysical parameters in the CAMELS suites \citep{CAMELS, baryonSpread}.
As shown in Section \ref{sec:Results}, inelastic DM processes also provide suppression
at those scales, so the effects are potentially degenerate.
It is therefore crucial that we be able to distinguish baryonic effects from
those of dark physics.

In CAMELS, the observed suppression over the parameter range is well constrained
for IllustrisTNG, which we use in our simulations. In particular,
we use the fiducial set of IllustrisTNG astrophysical and cosmological parameters,
aligning ourselves with the CAMELS cosmic variance (CV) set of simulations.
At scales of $k\gtrsim2 \text{Mpc }h^{-1}$, the 
ratio of hydrodynamic to $N$-body power spectra in the
CAMELS TNG CV suite exhibits a "spoon" shape ,
where power is initially suppressed by up to $30\%$
before turning back towards $100\%$ \citep{CAMELS}.
This decrease in the ratio of hydrodynamic to $N$-body power spectra
indicates that hydrodynamic simulations tend to produce less
structure than equivalent $N$-body simulations at those scales.

In Figure \ref{fig:var-pk-nbody}, we see that 
comparing our hydrodynamic to $N$-body simulations
all simulations,
both CDM and 2cDM,
exhibit a small downturn at small scales,
but the effect is insignificant up to cosmic variance.
All hydrodynamic simulations appear to have similar amounts
of substructure suppression relative to their $N$-body counterparts,
indicating that the addition of baryons has a similar effect on
small structure formation in both DM models.
At $z\gtrsim2$, it appears that the dominant effect on the power
spectrum is therefore the modified dark physics.

However, degeneracies between baryonic physics and dark physics
are still possible. The CAMELS Latin Hypercube (LH) set of simulations
simultaneously varies astrophysical and cosmological parameters.
In the TNG LH suite, the degree of suppression on the same scales
is less constrained, with a maximum of $\sim60\%$ and minimum of $\sim10\%$,
which can overlap with some of the more severe 2cDM models.

Similar results are seen when comparing to other baryonic physics
prescriptions, such as SIMBA and ASTRID \citep{baryonSpread}. Under
fiducial sets of parameters, the suppression from 2cDM is unique,
but degeneracies appear once cosmological and astrophysical parameters
are varied.

SFRs may provide a method to break these degeneracies.
As seen in Figure \ref{fig:var-SFR-distribution},
an aggressive 2cDM model, such as the \mm model, significantly
reduces the total number of star-forming subhalos.
Consequently, this may affect certain summary statistics and observables,
such as the quenched fractions, at low redshift.
These will be studied at the completion of the simulations.

While it would be instructive to compare the other halo statistics -- the HMF,
MCVF, and SFR distribution -- to results from CAMELS, the present suite of simulations unfortunately cannot.
The choice of $L_{box}$ renders us unable to form halos of similar masses
to the CAMELS boxes. However, as discussed in Section \ref{subsec:2cDM-Model} and
shown in Figure \ref{fig:var-mass-box},
2cDM signatures are expected to stay at these small scales.
Future studies utilizing Milky Way-like zoom-in suites,
similar to those found in the DREAMS project \citep{DREAMS}
will help us to further understand the role astrophysical and
cosmological parameters play at these small scales and how
modified dark physics interacts with those parameters.

\subsection{Application to Other DM Models}
We now highlight potential applications of this 
computational approach to other DM models as well as its limitations.

\cite{BDMTheoryI} proposed a similar two-component inelastic model,
Boosted Dark Matter (BDM),
with one of the main differences being a large mass difference
and a primary annihilation process
(in contrast to our low mass difference and primary conversion process).
\cite{BDMPowerSpectra} explored
its early-time effects on the initial power spectrum.
\cite{BDMNBody} then used $N$-body
CDM simulations with modified initial conditions
to explore metrics similar to those presented in this work.
The full consequences of the BDM model,
including both early time and late time effects, can be readily
implemented into this framework if the mass degeneracy and
velocity dependent cross sections are known.
Despite the large mass difference, realistic results
could potentially be achieved through rare inelastic interactions
or a low initial excited state fraction.

Other, more complicated inelastic DM models are possible.
As we emphasized in Section \ref{sec:Methods}, 2cDM is a special
case of a generalized $N$-component flavour-mixed model. More interaction
channels will inevitably lead to higher computational cost, as well
as a higher probability of more-than-two particle collisions occurring
within a time step, violating the rare binary collision approximation that
we use. A highly-interacting DM acts more like a fluid,
requiring a hydrodynamic approach to obtain accurate results.
Work is already ongoing in adapting existing hydrodynamical codes
to model fluid-like DM.
\cite{ADMI} implement such a model, atomic DM,
in the \textsc{GIZMO} code.
This computational approach applies on much
different physical scales than the present framework.
It would be interesting to investigate at which scale
the binary collision approach becomes insufficient and
a fully hydrodynamical implementation becomes necessary,
and if a hybrid computational approach is practical
or desirable.
\section{Conclusions}
\label{sec:Conclusions}
We have presented the first suite of 2cDM simulations with IllustrisTNG physics.
Our results indicate that baryonic physics does not
significantly affect suppression due to 2cDM at $z\sim5-2$.
In this regime, suppression follows expectations from $N$-body simulations,
suggesting this modified dark physics can be distinguished
from degeneracies due to baryonic effects.
Degeneracies can still exist as $z\rightarrow0$.
A future publication will follow these results down to $z=0$
to analyse how
baryonic effects interact with modified DM physics deep into the nonlinear regime.
Other important observables, such as the radial density profiles for
individual halos, will also be included in this future publication.

We have also demonstrated how two power laws, \zz and \mm, behave under a wide
range of the 2cDM parameter space in $N$-body simulations.
Our fiducial $N$-body and hydrodynamic simulations are self-consistent
with these parameter space studies.
Effort is ongoing to perform a similar parameter space exploration
in hydrodynamic simulations.

It is known that uniform boxes struggle to simultaneously achieve a realistic
Milky Way-like environment while also probing the scales necessary to
observe 2cDM effects. Future studies will utilize zoom-in simulations
to explore 2cDM in Milky Way-like systems more thoroughly.

\section*{Acknowledgments}
We would like to thank the University of Kansas
Center for Research Computing
and the Massachusetts Institute of Technology
Office of Research Computing and Data
for use of their computing resources. 
We also thank the Referee for their
insightful comments.
RL and MM acknowledge the partial support by NSF through the grant PHY-2409249.
PT acknowledges support from NSF-AST 2346977 and the NSF-Simons AI Institute for Cosmic Origins which is supported by the National Science Foundation under Cooperative Agreement 2421782 and the Simons Foundation award MPS-AI-00010515.

We utilised the following software for analysis:
\begin{itemize}
\item \textsc{Python}: \cite{python}
\item \textsc{Matplotlib}: \cite{Matplotlib}
\item \textsc{NumPy}: \cite{Numpy}
\item \textsc{SciPy}: \cite{Scipy}
\end{itemize}
\section*{Data availability}
Data is available upon request.

%
%

%
%

%
%

%
%

\bibliographystyle{mnras}
\bibliography{bibliography}

\appendix
\section{Tests of Numerical Convergence}
\label{sec:Convergence}
\subsection*{Scaling With $L_{box}$}
\begin{figure*}
\noindent\begin{minipage}[t]{1\columnwidth}
\includegraphics[width=1\columnwidth]{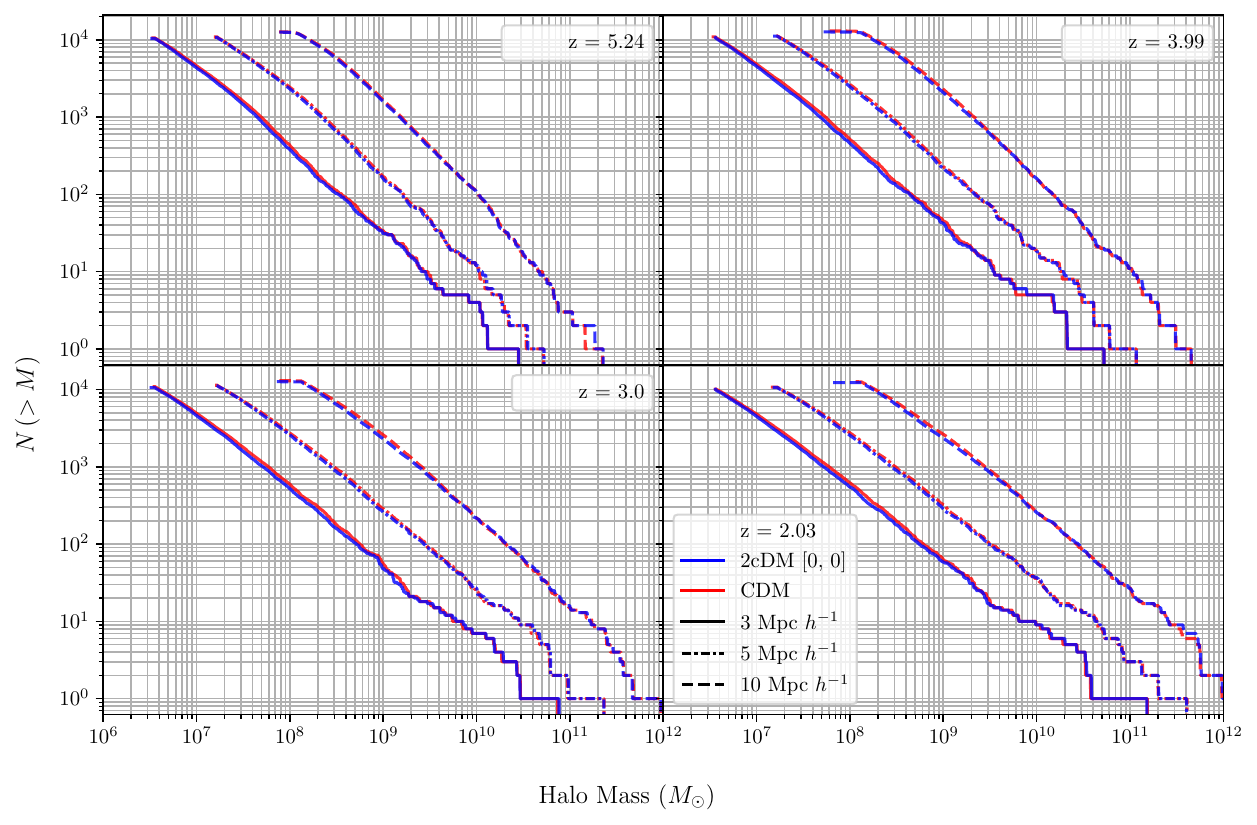}
\end{minipage}
\noindent\begin{minipage}[t]{1\columnwidth}
\includegraphics[width=1\columnwidth]{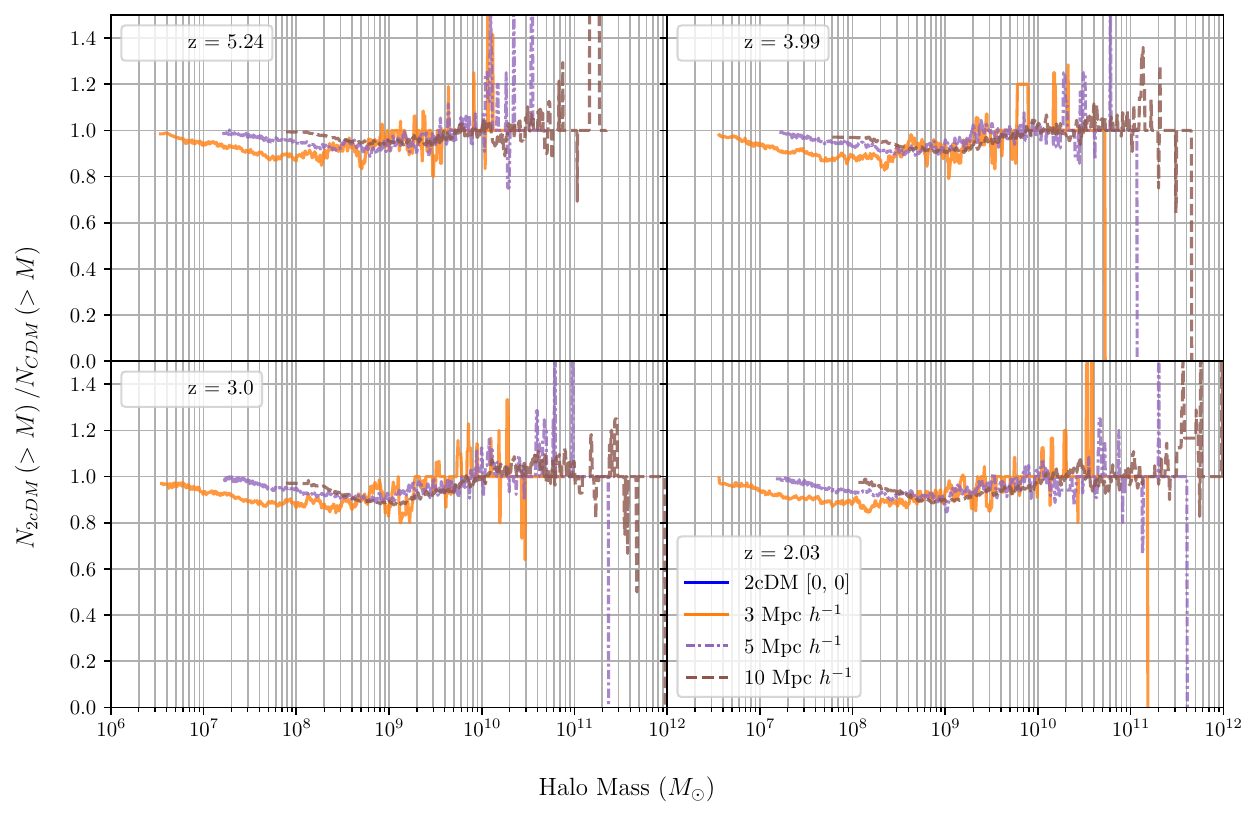}
\end{minipage}
\caption{(\textit{Left}) Comparing the halo mass function across simulations
with varying $L_{box}$. Blue lines represent 2cDM simulations, while red lines
represent CDM simulations. Solid lines correspond to the fiducial box size of 
$3 \text{Mpc }h^{-1}$. Dash-dotted and dashed lines correspond to box sizes of
$5 \text{Mpc }h^{-1}$ and $10 \text{Mpc }h^{-1}$ respectively.
All simulations presented here are hydrodynamic.
(\textit{Right}) The ratio between 2cDM and CDM halo mass functions.
Lines are coloured by box size.
\protect\label{fig:var-mass-box}}
\end{figure*}

\begin{figure*}
\noindent\begin{minipage}[t]{1\columnwidth}
\includegraphics[width=1\columnwidth]{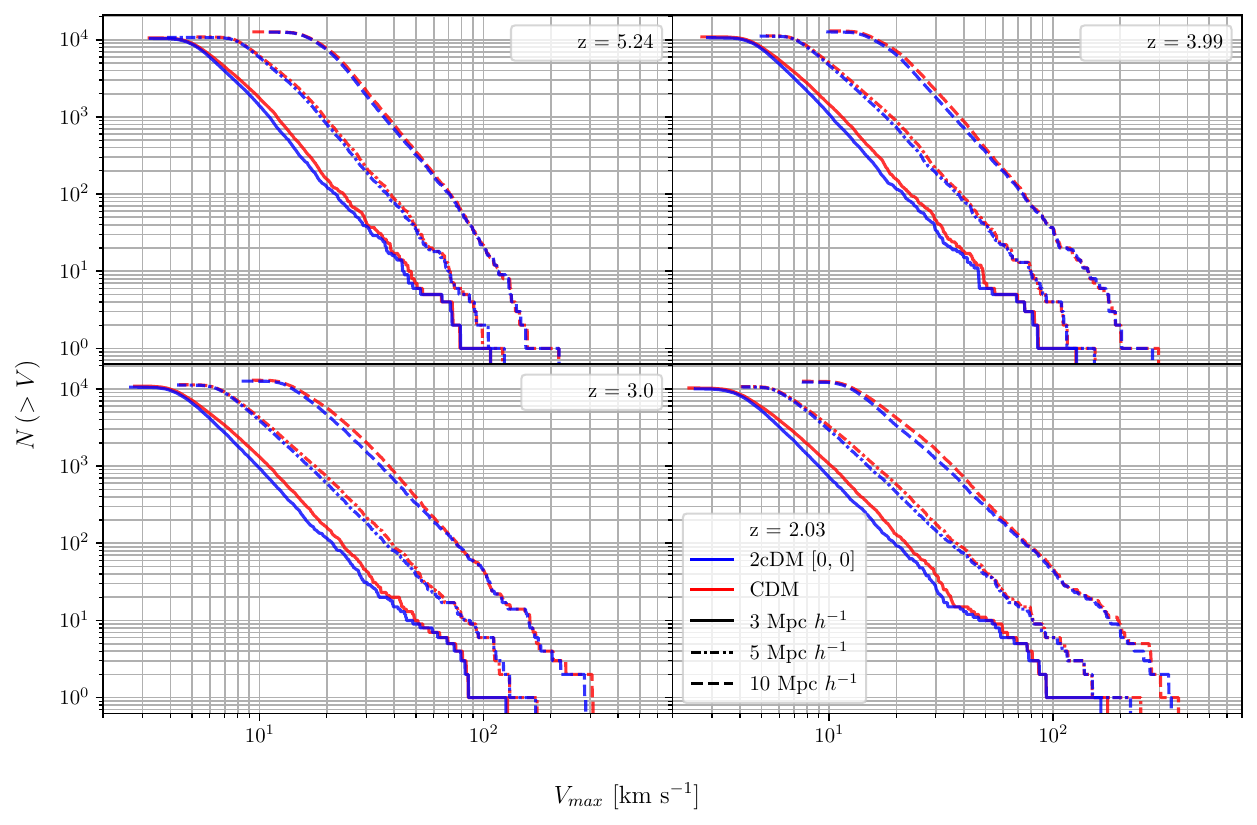}
\end{minipage}
\noindent\begin{minipage}[t]{1\columnwidth}
\includegraphics[width=1\columnwidth]{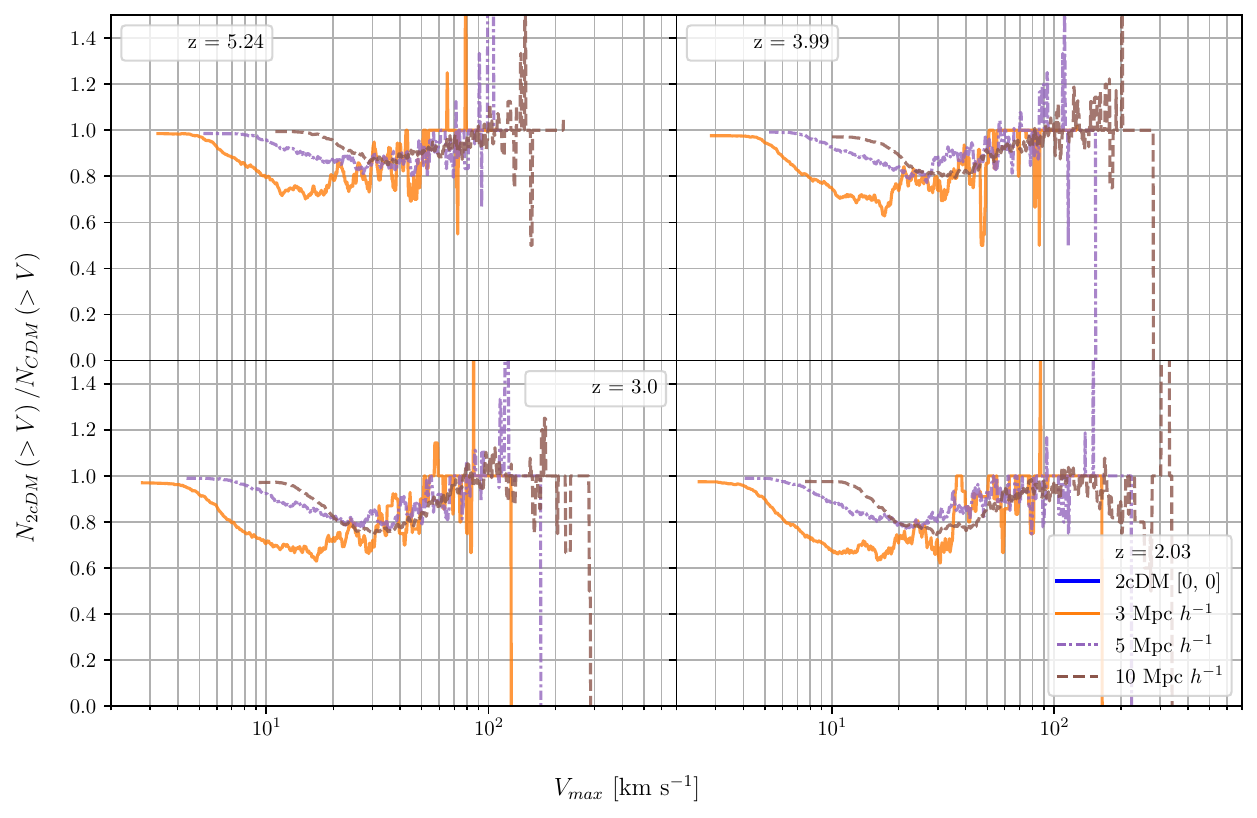}
\end{minipage}
\caption{The $V_{max}$ functions for the $L_{box}$ scaling test.
Colouring is identical to Figure \ref{fig:var-mass-box}.
\protect\label{fig:var-vel-box}}
\end{figure*}

\begin{figure*}
\noindent\begin{minipage}[t]{1\columnwidth}
\includegraphics[width=1\columnwidth]{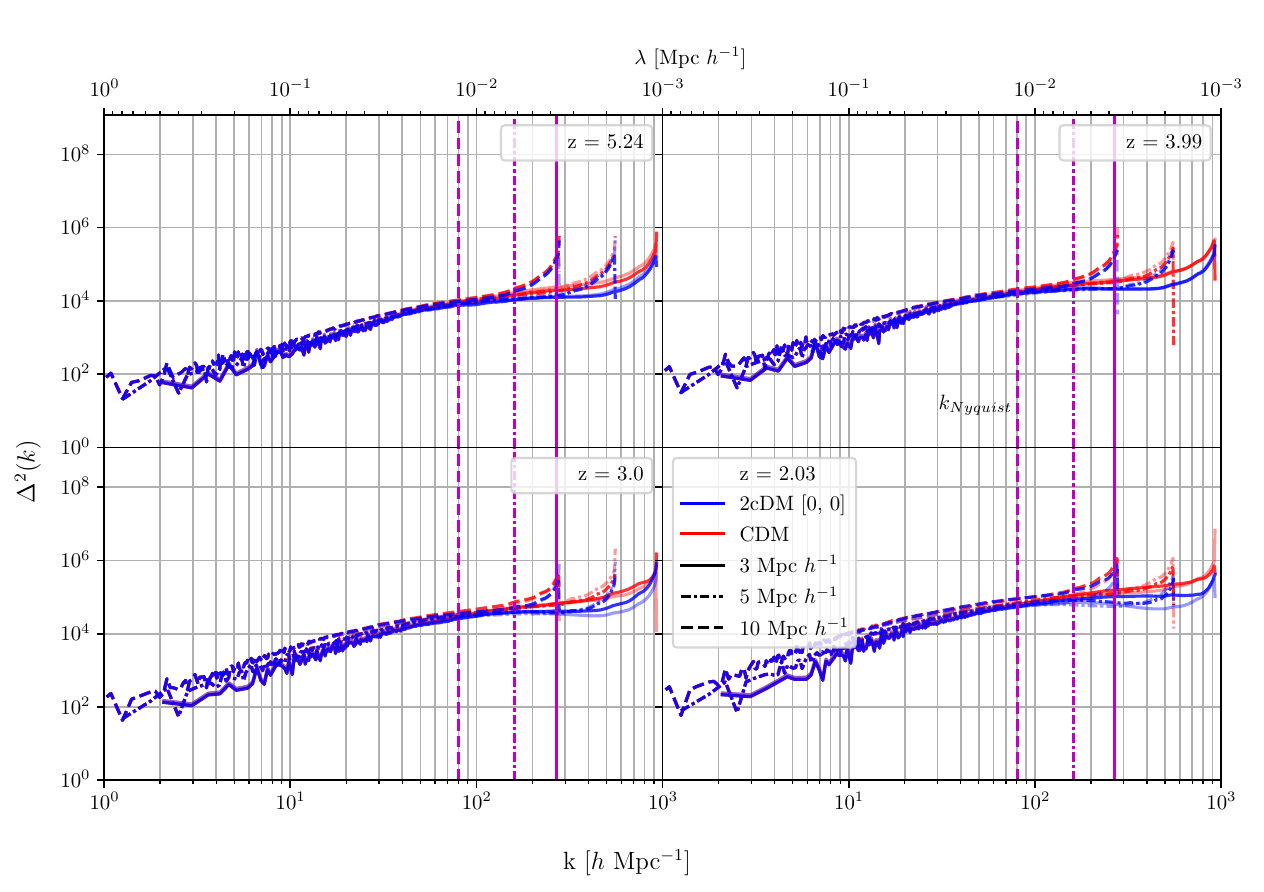}
\end{minipage}
\noindent\begin{minipage}[t]{1\columnwidth}
\includegraphics[width=1\columnwidth]{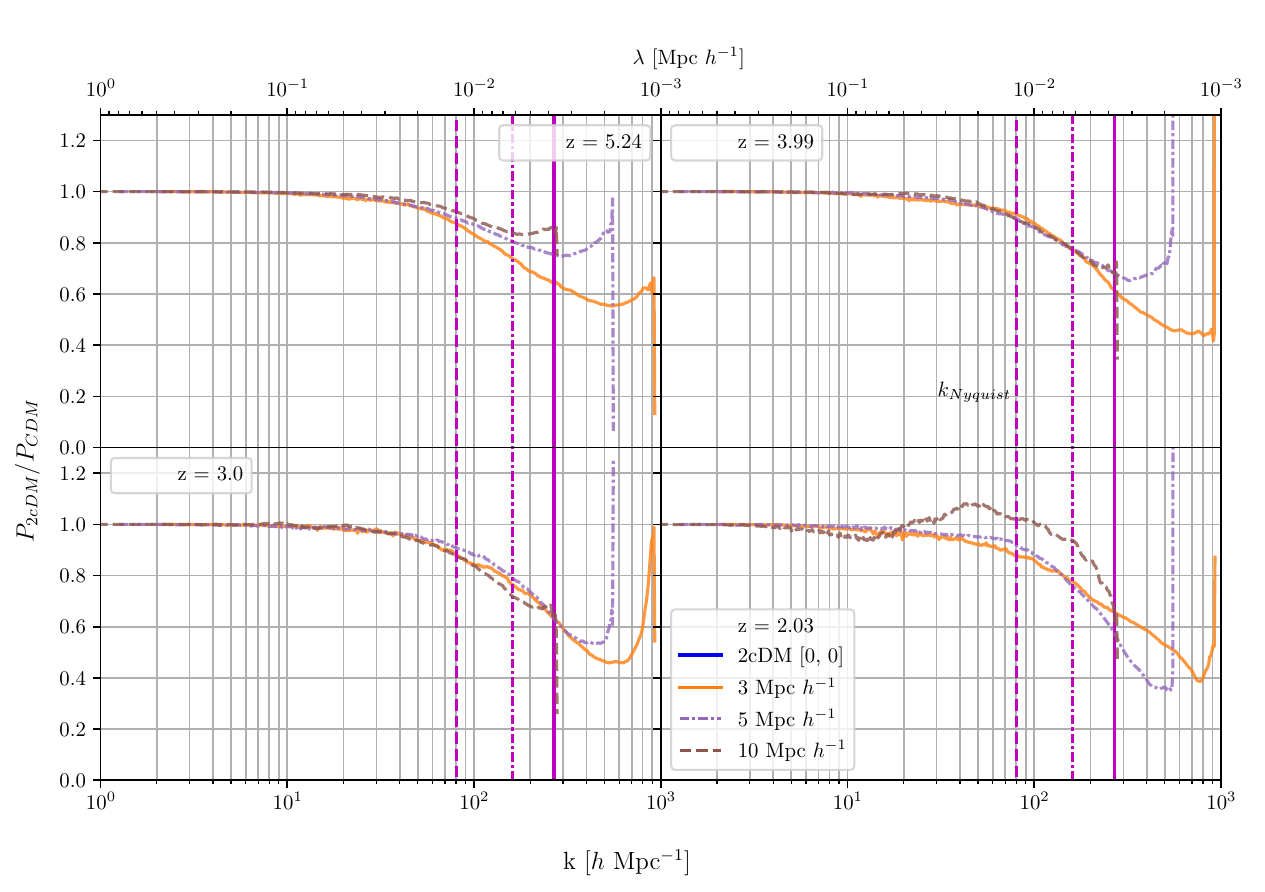}
\end{minipage}
\caption{The dimensionless power spectra for the $L_{box}$ scaling test.
Vertical lines denote the Nyquist wavenumber for the simulations. 
We note that the enhancement seen in the power spectra lie within the
error bounds of our fiducial simulations.
Colouring is identical to Figure \ref{fig:var-mass-box}.
Larger boxes at the same particle number cannot resolve smaller structures
as well due to coarser mass and force resolutions. Box size
does not appear to affect results significantly. Trends established
below the Nyquist level of the largest boxes are followed
well by the smaller boxes.
\protect\label{fig:var-pk-box}}
\end{figure*}

\begin{figure}
\includegraphics[width=1\columnwidth]{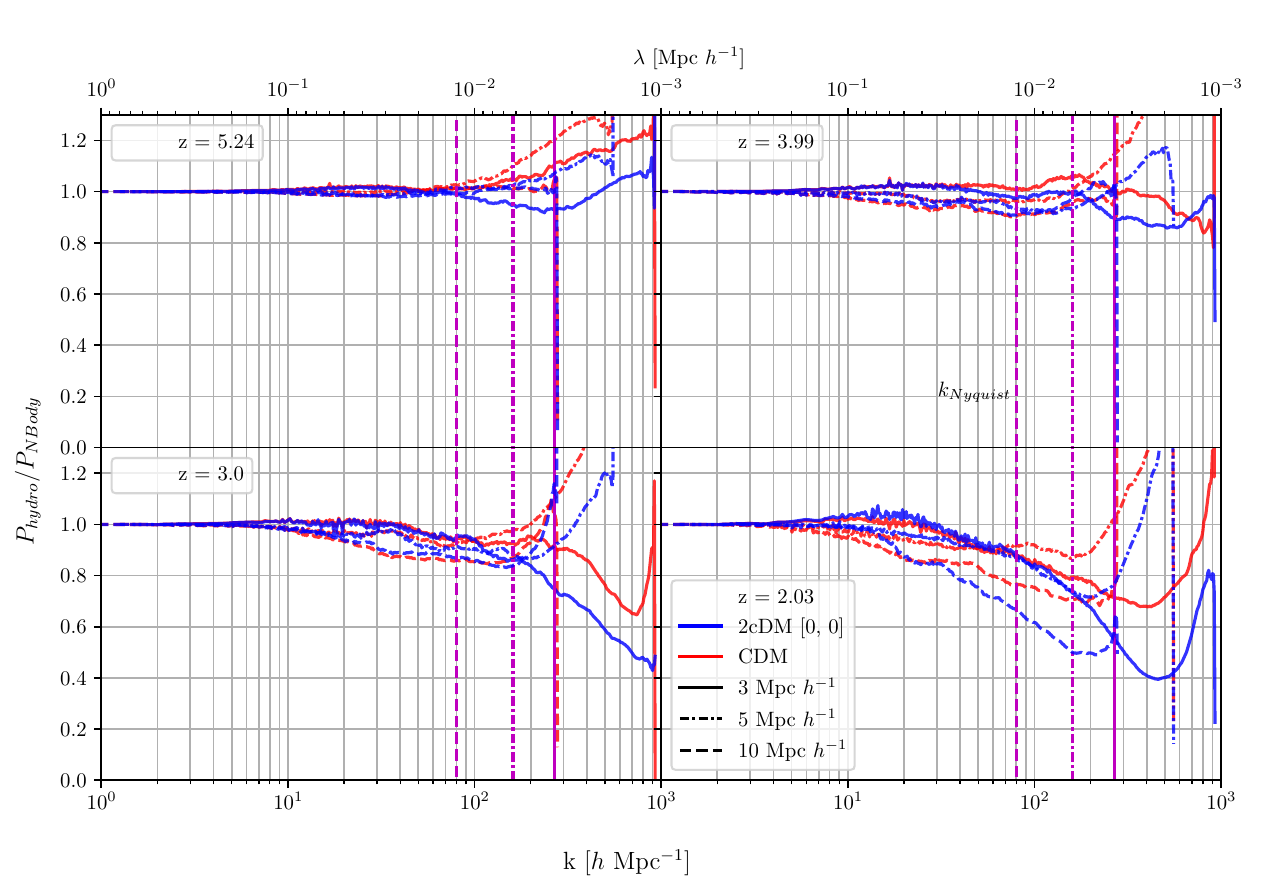}
\caption{The ratio between power spectra for hydrodynamic and $N$-body simulations
for the $L_{box}$ scaling test. Box size appears to have little effect
on the convergence of results on scales greater than the largest $\lambda_{Nyquist}$.
\protect\label{fig:var-box-pk-nbody}}
\end{figure}

We perform single simulations with CDM and the 2cDM \zz model with varying box sizes of
$L_{box}=3, 5, 10 \text{Mpc }h^{-1}$. The results for the same metrics discussed in 
Section \ref{sec:Results} are displayed in Figures \ref{fig:var-mass-box}-\ref{fig:var-box-pk-nbody}. Generally, the scale at which we see suppression relative to CDM is the same, though
the larger boxes are less sensitive to the small scales where effects are the strongest.

\subsection*{Scaling With $N_{part}$}
We perform single simulations with CDM and the 2cDM \zz model with varying particle numbers
of $N_{part}=128^{3}, 256^{3}, 512^{3}$. Results are only shown to $z=3$ due to the
computational and storage cost of the $512^{3}$ set of simulations. Results generally
align with each other, with the $128^{3}$ simulation exhibiting a high amount of noise
due to small number of particles.

%
%

%
%

\begin{figure*}
\noindent\begin{minipage}[t]{1\columnwidth}
\includegraphics[width=1\columnwidth]{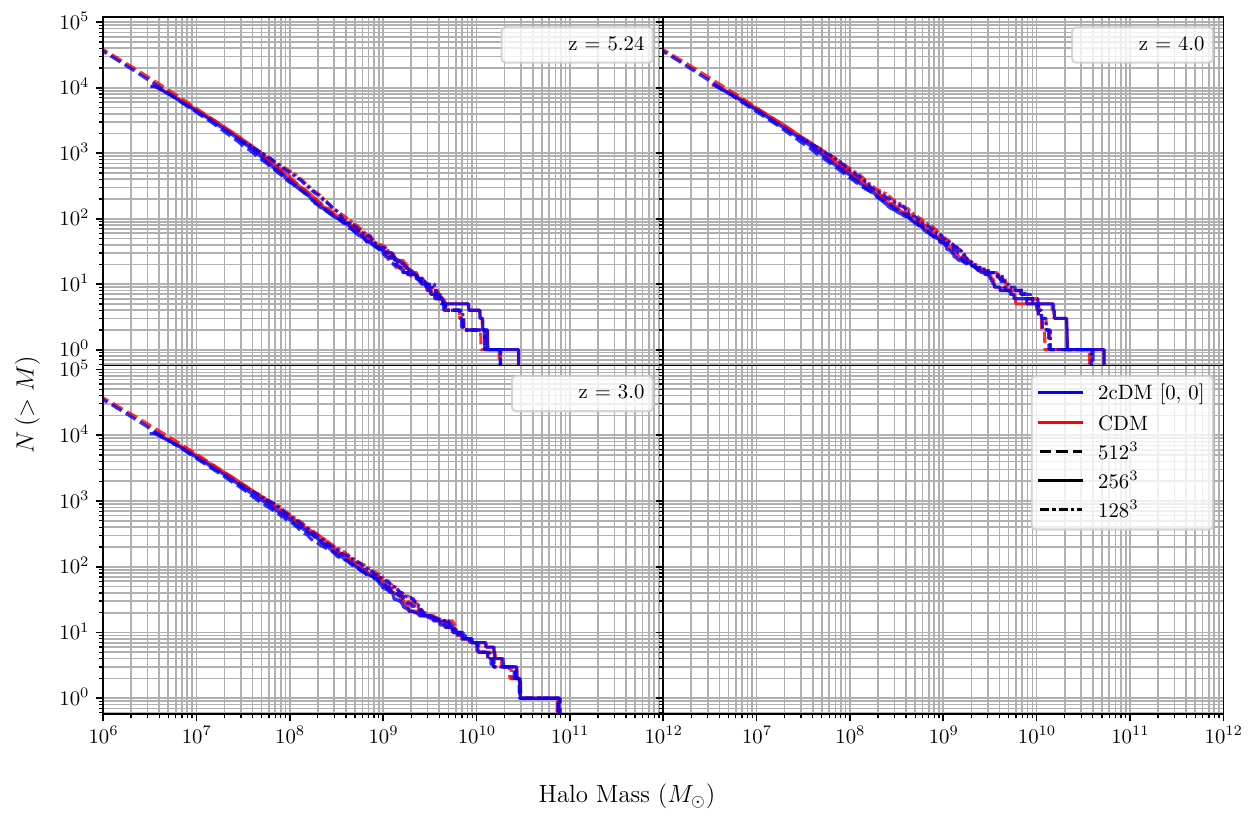}
\end{minipage}
\noindent\begin{minipage}[t]{1\columnwidth}
\includegraphics[width=1\columnwidth]{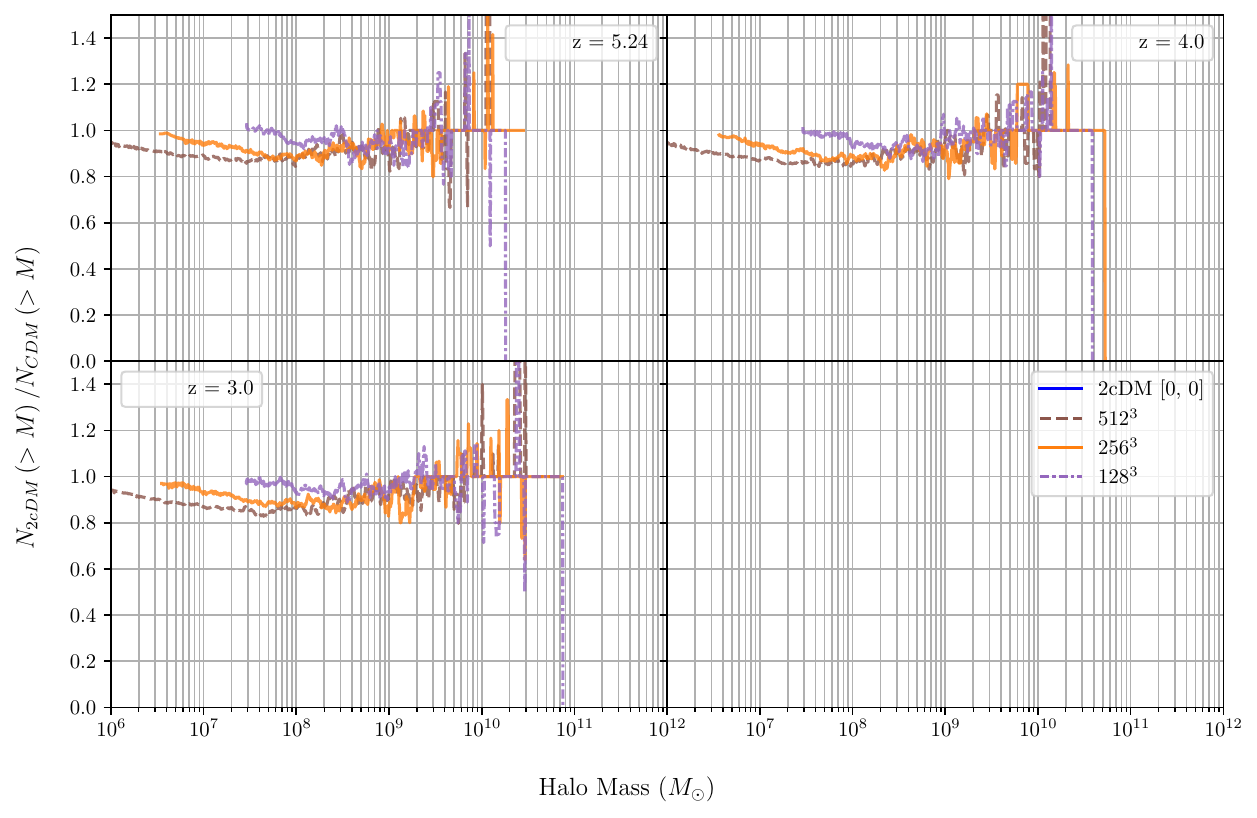}
\end{minipage}
\caption{(\textit{Left}) Comparing the halo mass function across simulations
with varying $N_{part}$. Blue lines represent 2cDM simulations, while red lines
represent CDM simulations. Solid lines correspond to the fiducial particle number of 
$256^{3}$. Dash-dotted and dashed lines correspond to particle numbers of
$128^{3}$ and $512^{3}$ respectively.
All simulations presented here are hydrodynamic.
(\textit{Right}) The ratio between 2cDM and CDM halo mass functions.
Lines are coloured by box size.
\protect\label{fig:var-mass-res}}
\end{figure*}

\begin{figure*}
\noindent\begin{minipage}[t]{1\columnwidth}
\includegraphics[width=1\columnwidth]{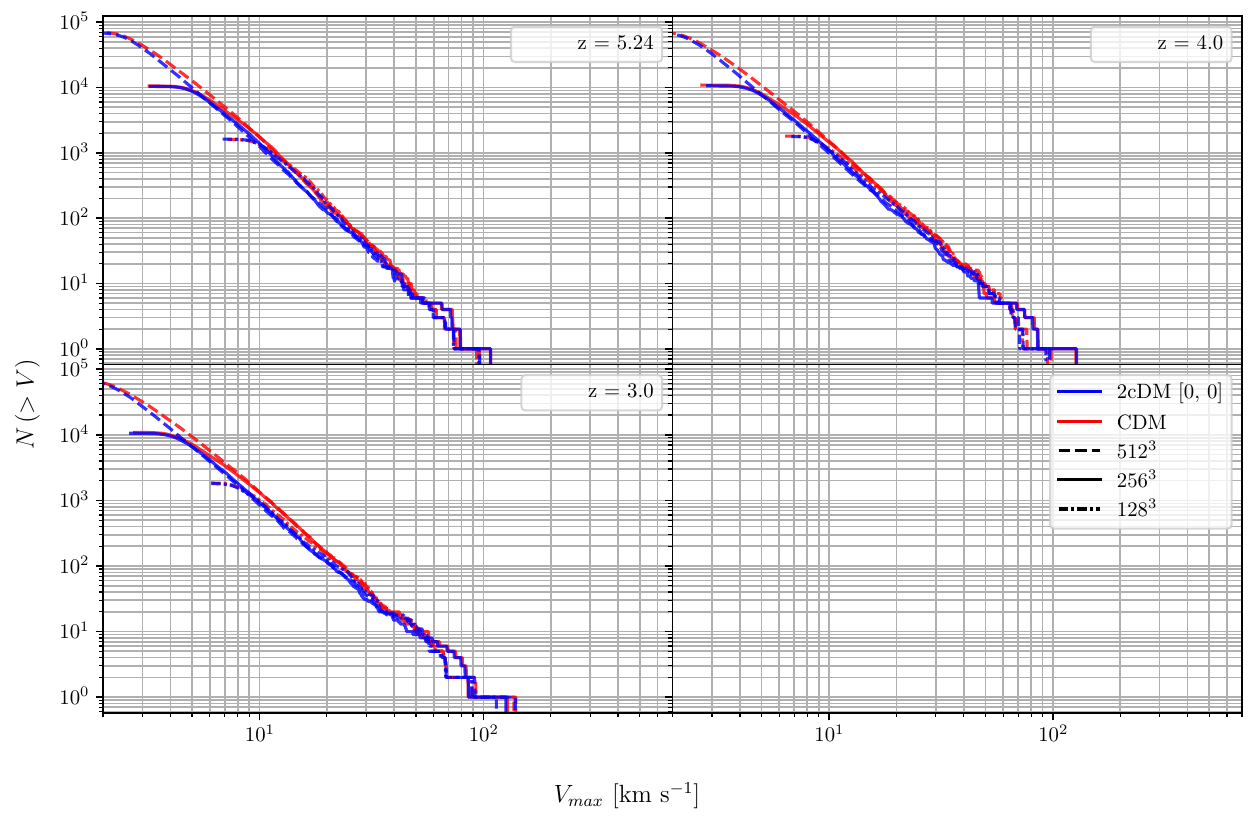}
\end{minipage}
\noindent\begin{minipage}[t]{1\columnwidth}
\includegraphics[width=1\columnwidth]{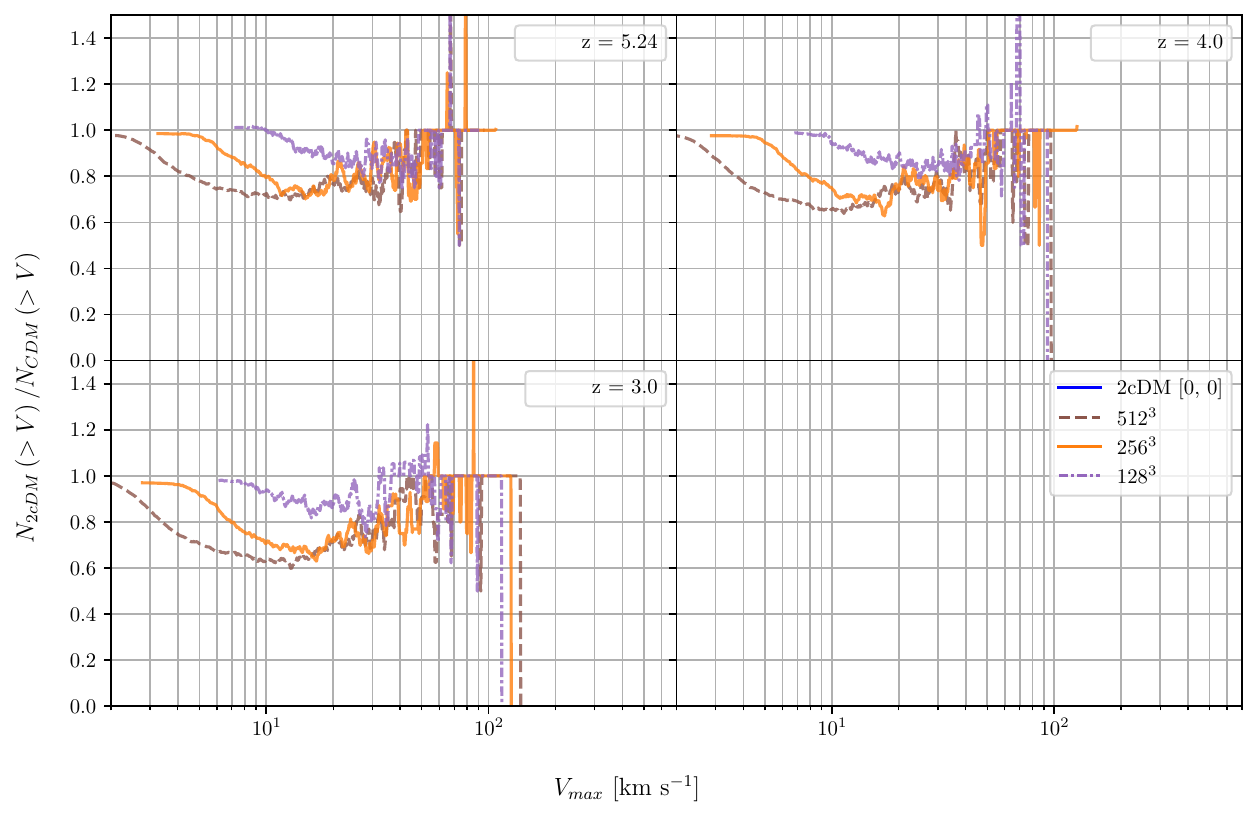}
\end{minipage}
\caption{The $V_{max}$ functions for the $N_{part}$ scaling test.
Colouring is identical to Figure \ref{fig:var-mass-res}.
\protect\label{fig:var-vel-res}}
\end{figure*}

\begin{figure*}
\noindent\begin{minipage}[t]{1\columnwidth}
\includegraphics[width=1\columnwidth]{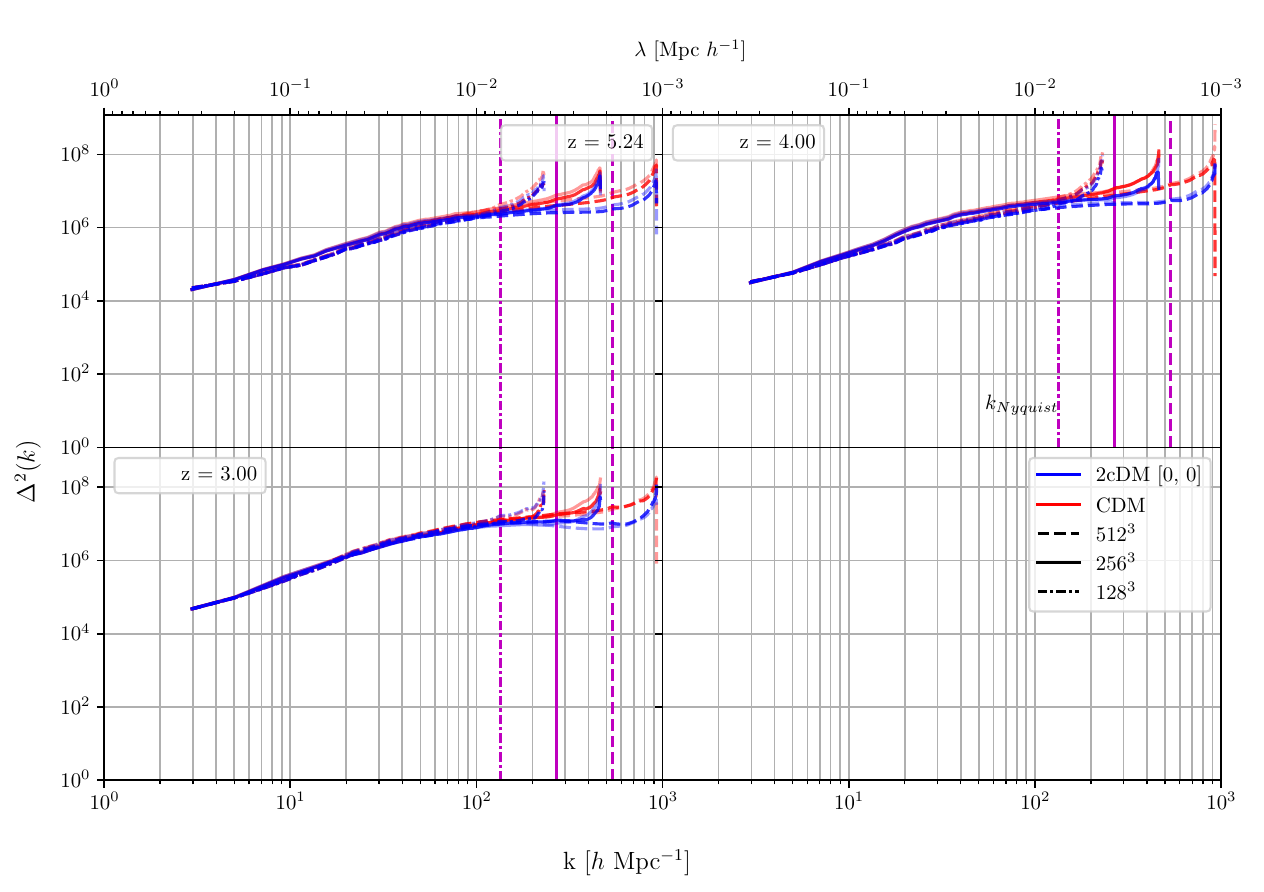}
\end{minipage}
\noindent\begin{minipage}[t]{1\columnwidth}
\includegraphics[width=1\columnwidth]{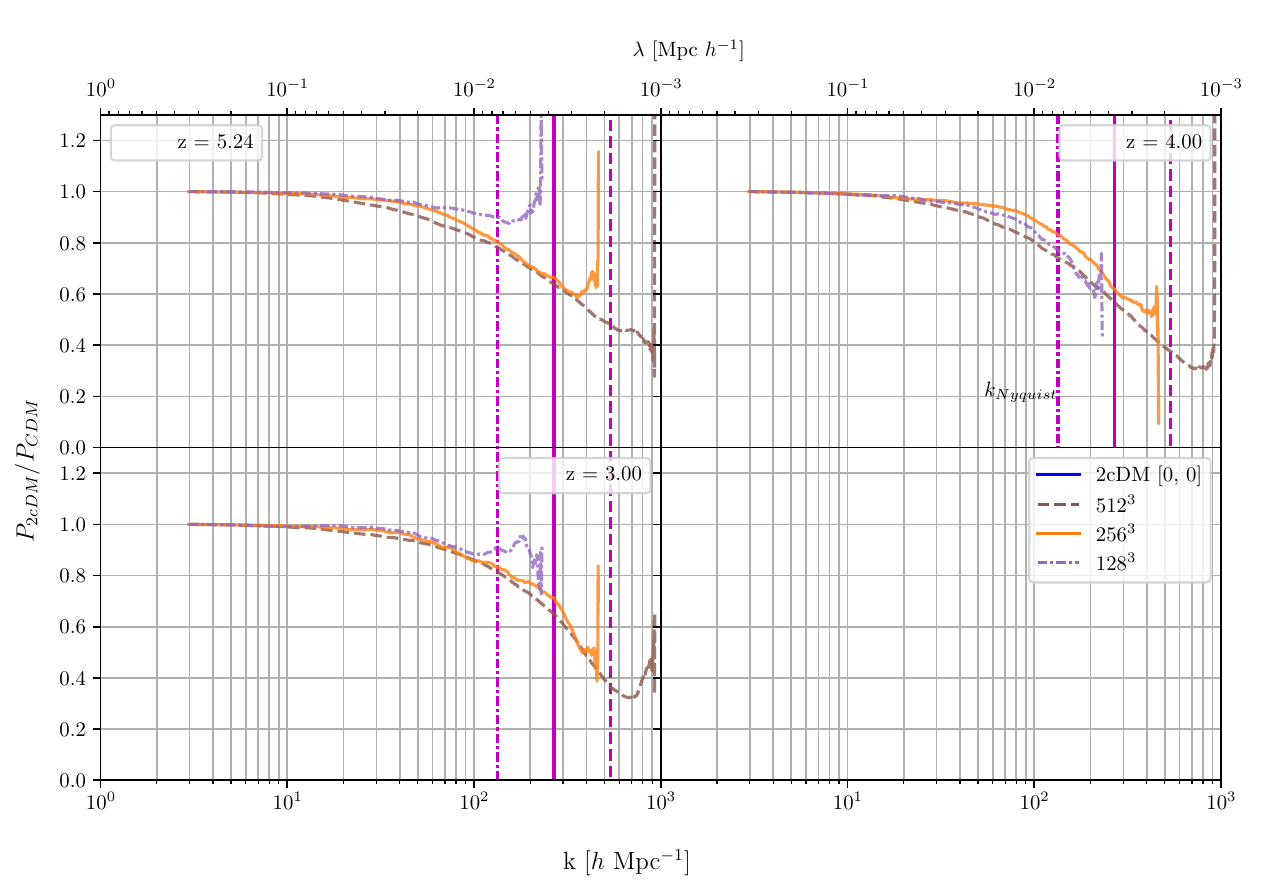}
\end{minipage}
\caption{The dimensionless power spectra for the $N_{part}$ scaling test.
Vertical lines denote the Nyquist wavenumber for the simulations. 
Colouring is identical to Figure \ref{fig:var-mass-res}.
Higher particle number leads to better resolved small structure.
The trend in the suppression that we see at lower resolutions continues
well at higher resolutions.
\protect\label{fig:var-pk-res}}
\end{figure*}

\begin{figure}
\includegraphics[width=1\columnwidth]{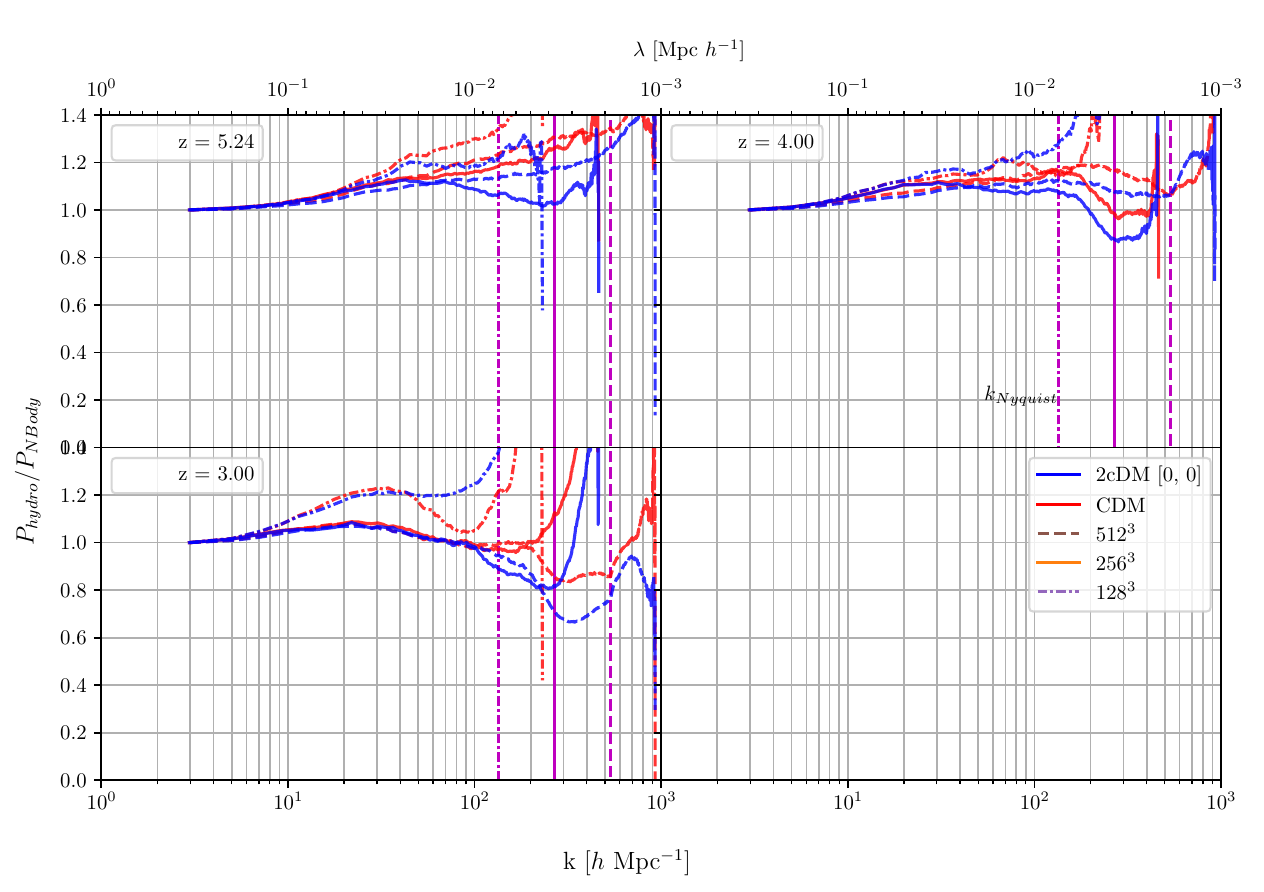}
\caption{The ratio between power spectra for hydrodynamic and $N$-body simulations
for the the $N_{part}$ scaling test. We note that the enhancement seen in the power
spectra lie within the error bounds of our fiducial simulations.
Lower gas particle numbers lead to worse convergence, suggesting
a minimum particle number for producing the correct baryonic effects.
\protect\label{fig:var-res-pk-nbody}}
\end{figure}

\label{lastpage}

\end{document}